\documentclass[preprint2]{emulateapj}
\usepackage[]{subfigure}

\def\kms{km\,s$^{-1}$}
\def\Ha{H{$\alpha$}}
\def\Hb{H{$\beta$}}

\def\ni{$^{56}$Ni}
\def\co{$^{56}$Co}
\def\fe{$^{56}$Fe}

\def\M{M$_{\odot}$}
\def\R{R$_{\odot}$}

\def\rks{PTF11rks}
\def\xk{SN 2011ke}
\def\css{SN 2011kf}
\def\fo{SN 2012il}
\def\paj{PTF10hgi}

\newcommand{\gps}{\ensuremath{g_{\rm P1}}}
\newcommand{\rps}{\ensuremath{r_{\rm P1}}}
\newcommand{\ips}{\ensuremath{i_{\rm P1}}}
\newcommand{\zps}{\ensuremath{z_{\rm P1}}}
\newcommand{\yps}{\ensuremath{y_{\rm P1}}}

\newcommand{\grizy}{\gps\rps\ips\zps\yps}

\shorttitle{SL-SNe Ic}
\shortauthors{Inserra et al.}

\begin{document}


\title{Super Luminous Ic Supernovae: catching a magnetar by the tail}


\author{
C. Inserra\altaffilmark{1}, 
S. J. Smartt\altaffilmark{1},
A. Jerkstrand\altaffilmark{1},
 S. Valenti\altaffilmark{2,3},
   M. Fraser\altaffilmark{1},
D. Wright\altaffilmark{1},
K. Smith\altaffilmark{1}, 
  T.-W. Chen\altaffilmark{1},
  R. Kotak\altaffilmark{1}, 
  A. Pastorello\altaffilmark{4},
M. Nicholl\altaffilmark{1}, 
 F. Bresolin\altaffilmark{5},
R. P. Kudritzki\altaffilmark{5},
   S. Benetti\altaffilmark{4}, 
M. T. Botticella\altaffilmark{6},
W. S. Burgett\altaffilmark{5},
K. C. Chambers\altaffilmark{5},
  M. Ergon\altaffilmark{7}, 
  H. Flewelling\altaffilmark{5},
  J. P. U. Fynbo\altaffilmark{8}, 
S. Geier\altaffilmark{8,9},
K. W. Hodapp\altaffilmark{5},
D. A. Howell\altaffilmark{2,3}, 
M. Huber\altaffilmark{5},
N. Kaiser\altaffilmark{5},
G. Leloudas\altaffilmark{10,8}, 
L. Magill\altaffilmark{1}, 
E. A. Magnier\altaffilmark{5},
M. G. McCrumm\altaffilmark{1}, 
N. Metcalfe\altaffilmark{11},
P. A. Price\altaffilmark{5},
A. Rest\altaffilmark{12},
J. Sollerman\altaffilmark{7},
W. Sweeney\altaffilmark{5},
 F. Taddia\altaffilmark{7}, 
S. Taubenberger\altaffilmark{13},
J. L. Tonry\altaffilmark{5},
R. J. Wainscoat\altaffilmark{5}, 
C. Waters\altaffilmark{5},
and
D. Young\altaffilmark{1}.
}

\altaffiltext{1}{Astrophysics Research Centre, School of Mathematics and Physics, Queens University
  Belfast, Belfast BT7 1NN, UK; c.inserra@qub.ac.uk}
\altaffiltext{2}{Las Cumbres Observatory Global Telescope Network, 6740 Cortona Dr., Suite 102 Goleta, Ca 93117}
\altaffiltext{3}{Department of Physics, University of California, Santa Barbara, Broida Hall, Mail Code 9530, Santa Barbara, CA 93106-9530, USA}
\altaffiltext{4}{INAF, Osservatorio Astronomico di Padova, vicolo dell'Osservatorio 5, 35122, Padova, Italy}
\altaffiltext{5}{University of Hawaii, Institute of Astronomy, 2680 Woodlawn Drive, Honolulu, Hawaii 96822 USA}
\altaffiltext{6}{INAF - Osservatorio astronomico di Capodimonte, Salita Moiariello 16, I- 80131 Napoli, Italy}
\altaffiltext{7}{The Oskar Klein Centre, Department of Astronomy, AlbaNova, Stockholm University, 10691 Stockholm, Sweden}
\altaffiltext{8}{Dark Cosmology Centre, Niels Bohr Institute, University of Copenhagen, 2100 Copenhagen, Denmark}
\altaffiltext{9}{Nordic Optical Telescope, Apartado 474, E-38700 Santa Cruz de La Palma, Spain}
\altaffiltext{10}{The Oskar Klein Centre, Department of Physics, Stockholm University, 10691 Stockholm, Sweden}
\altaffiltext{11}{Department of Physics, Durham University, South Road, Durham DH1 3LE}
\altaffiltext{12}{Space Telescope Science Institute, 3700 San Martin Drive, Baltimore, MD 21218, USA}
\altaffiltext{13}{Max-Planck-Institut f\"ur Astrophysik, Karl-Schwarzschild-Str. 1, 85741 Garching, Germany}

\begin{abstract}

We report extensive observational data for five of the lowest redshift
Super-Luminous Type Ic Supernovae (SL-SNe Ic) discovered to date, 
namely \paj, SN2011ke, \rks, SN2011kf and SN2012il.
Photometric
imaging of the transients at +50 to +230 days after peak combined with
host galaxy subtraction reveals a luminous tail phase for four of
these SL-SNe. 
A high resolution, optical and 
near infrared spectrum from {\em xshooter} provides detection of a broad He~{\sc I}
  $\lambda$10830 emission line in the spectrum (+50d) of SN2012il, 
 revealing that at least some SL-SNe Ic are not completely helium free. 
At first sight, the tail luminosity decline rates that we measure are consistent with
  the radioactive decay of \co, and would require 1-4~\M\/ of \ni\/ to
produce the luminosity. 
These 
\ni\/ masses 
cannot be made consistent with the short diffusion times at
 peak, and indeed are insufficient to power the peak luminosity.
 We instead favour energy deposition by newborn magnetars as
  the power source for these objects.  A semi-analytical diffusion
  model with energy input from the spin-down of a magnetar reproduces
  the extensive lightcurve data well.  The model predictions of 
ejecta velocities and temperatures which are required are in reasonable
 agreement with those determined from our 
observations. 
We derive magnetar energies of $0.4\lesssim
  E$($10^{51}$erg) $\lesssim6.9$ and ejecta masses of $2.3\lesssim M_{\rm
    ej}$(\M) $\lesssim 8.6$.  The sample of five SL-SNe Ic presented
  here, combined with SN 2010gx - the best sampled SL-SNe Ic so far -
  point toward an explosion driven by a magnetar as a viable
  explanation for all SL-SNe Ic.

\end{abstract}



\section{Introduction}

The discovery of unusually luminous optical transients by modern
supernova (SN) surveys has dramatically
expanded the observational and physical parameter space of known SN
types. The  Texas Supernova Search was a pioneer in this area, with
one of the first searches of the local Universe without a galaxy bias
\citep{qu05}. This has been followed by deeper, wider surveys from 
the Palomar Transient Factory \citep[PTF,][]{rau09}, the Panoramic Survey Telescope \& Rapid Response
System \citep[Pan-STARRS,][]{PS1_system}, the Catalina Real-time Transient Survey \citep[CRTS,][]{dr09},
and the La Silla QUEST survey \citep{ha12}, all of which have found unusually
luminous transients which tend to be hosted in intrinsically faint
galaxies.  These Super-Luminous SNe (SL-SNe) show absolute magnitudes at
maximum light of $M_{\rm AB} <-21$ mag and total radiated energies of  order
$10^{51}$ erg. They are factors of 5 to 100  brighter than Type Ia SNe or normal
core-collapse events.  \citet{2012Sci...337..927G} proposed that  
 three distinct physical groups of SL-SNe have emerged. The first group which was
 recognised was the  luminous Type IIn SNe
epitomised by SN 2006gy \citep{sm07,smc07,of07,ag09} which show signs of
strong circumstellar interaction. The second group includes  
Type Ic SNe that have broad, bright lightcurves and the decay rates
imply that they could be due to
pair-instability  explosions powered by large ejecta masses of \ni\/
(3--5 \M). To date only one object of this type (SN 2007bi) has been published and
studied in detail \citep[][]{gy09,yo10}. The third group was labelled
by Gal-Yam (2012) as 
 Super-Luminous Type I Supernovae, or ``SLSN-I''  and the two earliest
examples  are SCP-06F6 and SN 2005ap 
\citep{ba09,qu07} which are characterized at early times 
by a blue continuum and a distinctive ``W''-shaped spectral feature 
at $\sim$4200~\AA.  In this paper we will call this type Super-Luminous Supernovae
of Type Ic (or  SL-SNe Ic), simply because they are Type Ic in the established
supernova nomenclature and are extremely luminous. 

The existence of this class of SL-SNe Ic was firmly established when
secure redshifts were determined with  the identification of narrow
Mg~{\sc ii} $\lambda\lambda$2796,2803 absorption from the host 
galaxies in the optical spectra of $z >0.2$ transients by
\citet{qu11}. The spectra of four PTF transients, SCP-06F6 and
SN 2005ap were then all linked together with these redshifts
establishing a common type \citep{qu11}. 
Subsequently, the identification of host galaxy emission lines such
as [O~{\sc ii}] $\lambda$3727, [O~{\sc iii}]
$\lambda\lambda$1959,5007, H$\alpha$ and H$\beta$ has confirmed the
redshift derived from the Mg~{\sc ii}  absorption in many SL-SNe Ic, such as in
the case of SN 2010gx  \citep{qu10,ma10,pa10}. These distances imply an incredible luminosity
with $u-$ and $g-$band absolute magnitudes reaching about $-22$.
This luminosity allowed these SNe to be easily identified in the
Pan-STARRS Medium Deep fields at $z\sim1$ \citep{ch11}
 and beyond \citep{2012ApJ...755L..29B}. 

 The typical spectroscopic signature of the class of SL-SNe Ic is a
 blue continuum with broad absorption lines of intermediate mass
 elements such as C, O, Si,  Mg  with velocities
10000~$<$~$v$~$<$~20000~\kms. 
 No clear
 evidence of H or He has been found so far in the spectra of SL-SNe
 Ic, whereas Fe, Mg and Si lines are typically prominent  after
 maximum. The study of the well sampled SN 2010gx \citep{pa10} showed
 an unexpected transformation from a SL-SN Ic
 spectrum to that of a normal Type Ic SN. A similar transformation was
 also observed in the late-time spectrum of PTF09cnd \citep{qu11},
 which evolved to become consistent with a slowly evolving Type Ic SN.

 The SL-SNe Ic discovered to date appear to be associated with faint
 and metal poor galaxies at redshifts ranging between $0.23-1.19$
 \citep[typically $M_{g}>-17$
 mag,][]{qu11,2011ApJ...727...15N,ch11,che12}, although the highest
 redshift SL-SNe Ic (PS1-12bam, $z=1.566$) is in a galaxy which is more luminous and more
 massive than the lower redshift counterparts
 \citep{2012ApJ...755L..29B}.  An estimate of the metallicity of the faint
 host galaxy of SN 2010gx, from the detection of the auroral [O~{\sc III}]
 $\lambda$4363 line indicates a very low metallicity of 0.05~Z$_{\odot}$
 \citep{che12}. Research on SL-SNe Ic is progressing rapidly, with thirteen
 of these intriguing transients now identified since the discovery of
 SCP-06F6.  To power the enormous peak luminosity of SL-SNe Ic
 with radioactive decay would require several solar masses of
 \ni\/. However, this is inconsistent with the width of the lightcurves as
 shown by \citet{ch11}. The lightcurves cannot be reproduced with a
 physical model that has an ejecta mass significantly greater than the
 \ni\/ mass needed to power the peak. In the case of SN 2010gx, 
\citet{che12} showed that the tail phase faded to levels which 
would imply an upper limit of around 0.4\M\ of \ni\/.
Among the scenarios proposed to
explain the remarkable peak luminosity are the spin down of a
 rapidly rotating young magnetar \citep{kb10,wo10}, that provides an additional reservoir of energy for the SN \citep{og71,us92,wh00,th04}; 
 the interaction of
 the SN ejecta with a massive (3-5 \M) C/O-rich circumstellar medium
 \citep[CSM,][]{bl10} or with a dense wind \citep{ci11,gb12}; or collisions
 between high velocity shells ejected by a pulsational pair
 instability, which would give rise to successive bright optical
 transients \citep{wo07}.  One of these SNe (SN 2006oz) was discovered 
29 days before peak luminosity and showed a flat plateau in the
rest-frame NUV before its rise to maximum, indicating that finding
these objects early could give constraints on the explosion channel
\citep{le12}. 
In most cases to date however, the lightcurves and energy released at $>$100d is unexplored territory. 
\citet[][]{che12} quantified host of SN2010gx, but difference imaging showed no detection of 
SN flux at 250-300 days to deep limits.  \citet{qu11} detected flux at the position of PTF09cnd at +138d after peak 
in a B-band image, but it's not clear if this flux is from the host or the SN.
In this paper we present the detailed follow-up of five SL-SNe Ic at
$0.100<z<0.245$, namely \rks\/, \xk\/, \fo\/, \paj\/ and
\css\/, and attempt to follow them for as long as
possible to garner further evidence to probe the physical mechanism 
powering these intriguing events. A detailed analysis of their hosts
will be part of a future paper (Chen et el. in prep).

The paper is organized as follows: in Sect.~\ref{sec:sample} we introduce the SNe, and report distances and reddening values. Photometric data, light and colour curves as well as the absolute light curves in the rest-frame are presented in Sect.~\ref{sec:obs}. Sect.~\ref{sec:bol} is devoted to the analysis of bolometric and pseudo-bolometric light curves and the evaluation of possible ejected \ni\/ masses, while in Sect.~\ref{sec:sp}~
we describe and analyse the spectra. Finally, a discussion about the origin of these transients is presented in Sect.~\ref{sec:dis}, followed by a short summary in Sect.~\ref{sec:end}.

\begin{deluxetable*}{lccccc}
\tablewidth{0pt}
\tablecaption{Main properties of the SN sample.\label{table:sample}}
\tablehead{\colhead{}&\colhead{\paj}& \colhead{\xk}& \colhead{\rks} & \colhead{\css}& \colhead{\fo} }
\startdata
Alternative &  PSO J249.4461+06.20815  &  PS1-11xk, PTF11idj,& & CSS111230:& PS1-12fo \\
names&  & CSS110406:&&143658+163057&CSS120121: \\
& &135058+261642 &&& 094613+195028 \\
$\alpha$ (J2000.0)& $16^h37^m47^s.08$ & $13^h50^m57^s.78$ & $01^h39^m45^s.49$& $14^h36^m57^s.64$&$09^h46^m12^s.91$ \\
$\delta$ (J2000.0)& $06\degr12'29''.35$ &  $26\degr16'42''.40$ & $29\degr55'26''.87$ & $16\degr30'57''.17$&$19\degr50'28''.70$\\
$z$ & 0.100 & 0.143& 0.190 &0.245 & 0.175 \\
Peak $g$ (mag)& $-20.42$& $-21.42$&$-20.76$ &$-21.73$ &$-21.56$\\
E(B-V) (mag) & 0.09 & 0.01 & 0.04 & 0.02 & 0.02\\
L$_{\rm  griz}$ peak (x $10^{43}$ erg s$^{-1}$) & 2.09 & 4.47 & 3.24 & 6.45 & $\gtrsim$ 4.47 \\
Light curve peak (MJD) & $55326.4 \pm 4.0 $ & $55686.5 \pm 2.0$ & $55932.7 \pm 2.0$& $55925.5 \pm 3.0$ & $55941.4 \pm 3.0$\\
Host $r$ (mag)&$-16.50$ &$-18.42$ &  $-19.02$&$-16.52$&$-18.18$
\enddata
\end{deluxetable*}

\section{Discovery and target sample}\label{sec:sample}

\subsection{Pan-STARRS1 data : discovery and recovery of the transients}

The strong tendency for these SL-SNe to be hosted in faint galaxies
appears not to be a bias, which suggests a straightforward way of
finding them in large volume, wide-field searches. With the
Pan-STARRS1 survey, we have been running the ``Faint Galaxy Supernova
Survey'' (FGSS) which is aimed at finding transients in faint galaxies
originally  identified in the Sloan Digital Sky Survey catalogue
\citep{2010ATel.2773....1V}. 

The Pan-STARRS1 optical system uses a 1.8m diameter aspheric primary
mirror, a strongly aspheric 0.9m secondary and 3-lens corrector and
has 8m focal length \citep{PS1_optics}.  The telescope illuminates a
diameter of 3.3 degrees and the ``GigaPixel Camera" \citep{PS1_GPCA}
comprises a total of 60 $4800\times4800$ pixel detectors, with
10~$\mu$m pixels that subtend 0.258~arcsec \citep[for more details
see][]{ps1_photladder}. 
The PS1 filter
system is described in \citet{2012ApJ...750...99T}, and is similar to
but not identical to the SDSS $griz$ \citep{yo00} filter
system. However it is close enough that catalogued cross-matching
between the surveys can identify high amplitude transients. In this
paper we will convert all the PS1 filter magnitudes \grizy\/ to the
SDSS AB magnitude system as the bulk of the follow-up data were taken
in SDSS-like filters (see Section 3 for more details).

The PS1 telescope is operated by the PS1 Science Consortium (PS1SC) to
undertake several surveys, with the two major ones being the ``Medium
Deep Field'' survey
\citep[e.g][]{bot10,2012ApJ...745...42T,2012Natur.485..217G,2012ApJ...755L..29B} 
which was optimised for transients
(allocated around 25\% of the total telescope time) and
the wide-area $3\pi$ Survey, allocated around 56\% of the available
observing time.  As described in \citet{ps1_photladder}, the  goal of the $3\pi$
Survey is to observe the portion of the sky North of $-30$ deg declination, with a
total of 20 exposures per year across all five filters for each field
center. The $3\pi$ 
survey plan is to observe each field center 4 times in each of
\grizy\/ during a 12 month period, although this can be interrupted by bad
weather. 
As described by \citet{ps1_photladder},
the 4 epochs in a calendar year are typically split into two pairs called
Transient Time Interval (TTI) pairs which are single observations
separated by 20-30 minutes to allow for the discovery of moving 
objects. 
The temporal distribution of the two sets of TTI pairs is not
a well defined and straightforward schedule. The blue bands (\gps,
\rps, \ips) are scheduled close to opposition to enhance asteroid
discovery with \gps\/ and \rps\/ being constrained in dark time as far as
possible. The \zps\/ and \yps\/ filters are scheduled  far 
opposition  to optimise for parallax measurements of faint
red objects. Although a large area of sky is observed each night
(typically 6000 square degrees), the moving object and parallax constraints mean the
$3\pi$ survey is not optimised for finding young, extragalactic
transients in a way that the Palomar Transient Factory and La
Silla-QUEST projects are. The exposure times at each epoch
(i.e. in $each$ of the TTI exposures) are 43s, 40s, 45s, 30s and 30s
in \grizy.  These reach typical 
 $5\sigma$ depths of roughly 22.0, 21.6, 21.7, 21.4. 19.3 
as estimated from point sources with uncertainties of 0.2$^{m}$. 
\citep[in the PS1 AB magnitude system described by][]{2012ApJ...750...99T}. 

The PS1 images are processed by the Pan-STARRS1 Image Processing
Pipeline (IPP), which performs automatic bias
subtraction, flat fielding, astrometry \citep{ma08} and
photometry \citep{ma07}.  
These photometrically and astrometrically calibrated catalogues
produced in MHPCC are made available to the PS1SC on a 
nightly basis and are immediately ingested into a MySQL database at Queen's
University Belfast. 
We apply a tested  rejection algorithm and cross match the 
PS1 objects with SDSS objects in the DR7 catalogue\footnote{http://www.sdss.org/dr7/} \citep{ab09}. 
We apply the following selection filters to the PS1 data (all
criteria must be simultaneously fulfilled)

\begin{itemize}
\item PS1 source must have $15 < $\gps$< 20$ or  $15 < $\rps$< 20$ or $15 < $\ips$< 20$ or $15 < $\zps$< 20$
\item  SDSS counterpart must have $18 < r_{\rm SDSS} < 23$ 
\item Distance between PS1 source and SDSS source $< 3$ arcsec    
\item PS1 mag must be 1.5 mag brighter than SDSS  (in the
  corresponding filter)  
\item The PS1 object must be present in both TTI pairs and the
  astrometric recurrences $<$ 0.3$\arcsec$. Objects with multiple detections
  must have RMS scatter $<$ 0.1$\arcsec$    
\item The PS1 object must not be in the galactic plane ($|b|
  >5^{\circ}$). 
\end{itemize}

All objects are then displayed through a Django-based interface to a
set of interactive webpages, and human eyeballing and checking takes
place.  We use the star-galaxy separation in SDSS to guide us in what
may be variable stars (i.e. stellar sources in SDSS which increase
their luminosity) or extragalactic transients (i.e. QSOs, AGNs and
SNe). While the cadence of the PS1 observations is not ideal
for detections of young SNe we have found many SNe in intrinsically
faint galaxies. As of January 2013, spectroscopically confirmed objects include
34 QSOs or AGNs and 41 SNe.
Several of these have been confirmed as SL-SNe. \citet{pa10} presented the
data of SN 2010gx recovered in $3\pi$ images, and in many cases the same
object is detected by the Catalina Real-Time Transient Survey (CSS/MSS) and
PTF. As we are not doing difference imaging and only comparing to
objects in the SDSS footprint, there are some cases where objects are
reported by PTF or CRTS and interesting pre-discovery epochs are
available in PS1. As $3\pi$ difference imaging is not being carried
out routinely, we often use the PTF and CRTS announcements to inform a
retrospective search. In this paper, we present five SL-SNe Ic which
were either detected through the PS1 Faint Galaxy Supernova Survey, or
were announced in the public domain. A sixth object
(PTF12dam$\equiv$PS1-12arh) is discussed in a companion paper
(Nicholl et al. in prep.).  In all five cases follow-up imaging and
spectroscopy was carried out as discussed below.

For all the SNe listed here (and throughout this paper) we adopt a
standard cosmology with $H_0 = 72$ \kms\/, $\Omega_{\rm M}=0.27$ and
$\Omega_{\rm \lambda}=0.73$.  There is no detection of Na\,{\sc i} 
interstellar medium (ISM) features from the host galaxies, nor do we have any
evidence of significant extinction in the hosts from the SNe
themselves. This suggests that the absorption in host is low and 
we assume that extinction from the host galaxies is negligible. Although we do detect Mg\,{\sc ii} ISM lines
from the hosts in some cases,  
there is no clear correlation with these line strengths and line of
sight extinction.  In all cases only the Milky Way foreground
extinction was adopted.

\subsection{\paj\/}

\paj\/ was first discovered by PTF on 2010 May 15.5 and
announced on 2010 July 15 \citep{qu10}. The spectra taken by PTF
on May 21.0 UT and June 11.0 UT were reported as a blue continuum with faint features typical
of SL-SNe Ic. Another spectrum obtained by PTF on July 7.0 UT was
similar to PTF09cnd at 3 weeks past peak brightness. Initiated by
\citet{qu10},  UV observations 
were obtained with Swift+UVOT in July 2010, and we analyse those
independently later in Sect.\ref{sec:obs}. 
\paj\/ lies outside the SDSS DR9\footnote{http://www.sdss3.org/dr9/} area \citep{ah12}, hence was not 
discovered by our PS1 FGSS software. However after the
announcement \citep{qu10},  we recovered it in PS1 images taken (in band \rps) on 
2010 May 18 and on 4 other epochs around peak magnitude (in bands 
\gps\rps\ips), listed in Table~\ref{table:10hgi}. 

We detect a faint host
galaxy in deep $griz$-band images taken with the  William Herschel
Telescope on 2012 May 26 and the Telescopio Nazionale Galileo on 2012 May 28. 
At a magnitude $r=22.01\pm0.07$, this is too faint to affect
the measurements of the SN flux up to +90 days.  There are no
host galaxy emission lines detected in our spectra, hence the redshift
is determined through cross-correlation of the spectra of \paj\/
with other SNe at confirmed redshift indicating a redshift of
$z=0.100\pm0.014$, corresponding to a luminosity distance of $d_{\rm L}\sim448$
Mpc. 
The Galactic reddening toward the SN line of sight is
$E(B-V)=0.09$ mag \citep{sc98}.

\subsection{\xk\/}


\xk\/ was discovered in the CRTS (CSS110406:135058+261642) and PTF
surveys (PTF11dij) with the earliest detections on 2011
April 06 and 2011 March 30 respectively \citep{dr11,qu11c}. We also
independently detected this transient in the PS1 FGSS (PS1-11xk) on images
taken on 2011 April 15 \citep{sma11}.  However earlier PS1 data show
that we can determine the epoch of explosion to around one day, at
least as far as the sensitivity of the images allow. On MJD 55649.55
(2011 March 29.55 UT)  a PS1 image (\rps\/ = 40s) shows no detection of the transient to
$r \simeq 21.17$ mag. 
\citet{qu11c} report the PTF detection on 2011 March 30 (MJD=55650) 
the night after the PS1 non-detection at $g\simeq21$. PS1 detections
then occurred 1 and 3 days after this on 55651.6 and 55653.6 (in \ips\/
and \rps) respectively. The photometry is given in Table~\ref{table:11xk}. 
This is the best constraint on the explosion epoch of an SL-SNe to
date, allowing the rise time and light curve shape to be confidently measured. 
The object brightened rapidly, by $\sim3$ mag in
$g-$band in the first 15 days and 1.7 mag in the subsequent 20 days.
It was classified as an SL-SNe Ic by both \citet{dr11} and \citet{qu11c}; their spectra
obtained on May 8.0 and 11.0 UT, respectively, showed a blue continuum with faint features,
similar to PTF09cnd about 1 week past maximum light \citep{qu11c}.

We found a nearby source 
in the SDSS DR9 catalog ($g=21.10\pm0.08$, $r=20.71\pm0.08$), which is the host galaxy as confirmed by deep $griz$ images taken with the William Herschel Telescope on the 26th of May 2012 at a magnitude $g=21.18\pm0.05$, $r=20.72\pm0.04$ . The host emission lines set the SN at $z=0.143$, equivalent to a luminosity distance of $d_{\rm L}=660$ Mpc. The Galactic reddening toward the position of the SN is $E(B-V)=0.01$ mag \citep{sc98}.

\subsection{\rks\/}
\rks\/ was first detected by PTF on 2011 December 21.0 UT
\citep{qu11b}. Spectra acquired on December 27.0 UT and 31.0 UT showed
a blue continuum with broad features similar to PTF09cnd at maximum
light, confirming it as a SL-SN Ic. A non-detection in the $r$-band on
Dec. 11 UT prior to discovery is also reported, setting a limit of
$>$20.6 mag at this epoch. 
\citet{qu11b} detailed a
brightening of 0.8 mag in $r$-band in the first 6 days after the
discovery. Prompt observations with Swift revealed an ultraviolet (UV)
source at the optical position of the SN, but no source was detected
in X-rays at the same epochs. There are no useful early data from
PS1 for this object. 

The host galaxy 
is listed in the SDSS DR9 catalogue with $g=21.59\pm0.11$ mag and $r=20.88\pm0.10$ mag.
Confirmation of these host magnitudes as achieved with our 
deep $gr$-band images taken with the William Herschel Telescope 
on the 22nd of September 2012 at a magnitude $g=21.67\pm0.07$ and $r=20.83\pm0.05$.
The emission lines of the host
and narrow absorptions consistent with Mg~{\sc ii}
$\lambda\lambda$2796,2803 doublet locate the transient at $z=0.19$,
corresponding to a luminosity distance of $d_{\rm L}=904$ Mpc.  The Galactic
reddening toward the position of the SNe is $E(B-V)=0.04$ mag
\citep{sc98}.

\subsection{\css}
\css\/ was first detected by CRTS (CSS111230:143658+163057) on 2011 December 30.5 UT \citep{dr12}. 
The spectra taken by \citet{pr12} on 2012 January 2.5 UT and 17.5 UT reveal a blue continuum with absorption feature typical of a luminous Type Ic SN. 

 The closest galaxy is 
 $\sim23\arcsec$ S/W of the object position and is hence too far to be the
 host. There is no obvious host coincident with the position of this SN in SDSS DR9. 
 We detect a faint host galaxy in deep $gri$-band
 images taken with the William Herschel Telescope on 20 July 2012. At
 a magnitude $r=23.94\pm0.20$, this is too faint to affect the
 measurements of the SN flux out to +120 days.  The redshift of the
 object has been determined to be z=0.245 from narrow \Ha\/ and
 [O~{\sc iii}], equivalent to a luminosity distance of $d_{\rm L}=1204$
 Mpc. The foreground reddening is $E(B-V)=0.02$ mag from \citet{sc98}.

\subsection{\fo\/}

\fo\/ was first detected in the PS1 FGSS on 2012 January 19.9 UT
\citep{sma12} and also independently discovered by CRTS on the 21st of January 2012
\citep[CSS120121:094613+195028][]{dr12}. On January 29 UT we obtained a spectrum of the SN, which
resembled SN 2010gx 4d after maximum light. The merged PS1 and
CRTS data suggest a rise time of more than 2 weeks, different from
that of SN 2010gx \citep{pa10} but similar to PS1-11ky \citep{ch11}.
An initial analysis of observations from Swift revealed a marginal
detection in the u, b, v and uvm2 filters, with no detection in the
uvw1 and uvw2 filters, or in X-rays \citep{ma12}. However our
re-analysis of the Swift data reveals a detection above the 3$\sigma$
level in the uvw1 and uvw2 filters (see Tab.~\ref{table:swift}). No
radio continuum emission from the SN was detected by the EVLA
\citep{ch12}.

In our astrometrically calibrated images the SN coordinates are
$\alpha=09^h46^m12^s.91$ $\pm 0^s.05$, $\delta=+19\degr50'28''.70$
$\pm 0''.05$ (J2000). This is within 0.12" of the faint galaxy SDSS
J094612.91+195028.6 ($g=22.13\pm0.08$, $r=21.46\pm0.07$). The emission
lines of \Ha, \Hb\/ and [O~{\sc iii}] from the host give a redshift of
$z=0.175$, corresponding to a luminosity distance of $d_{\rm L}=825$ Mpc.
The Galactic reddening  towards \fo\/ given by \citet{sc98} is
$E(B-V)=0.02$ mag, three times lower than the value reported by
\citet{ma12}.

\begin{figure*}
\includegraphics[width=18cm]{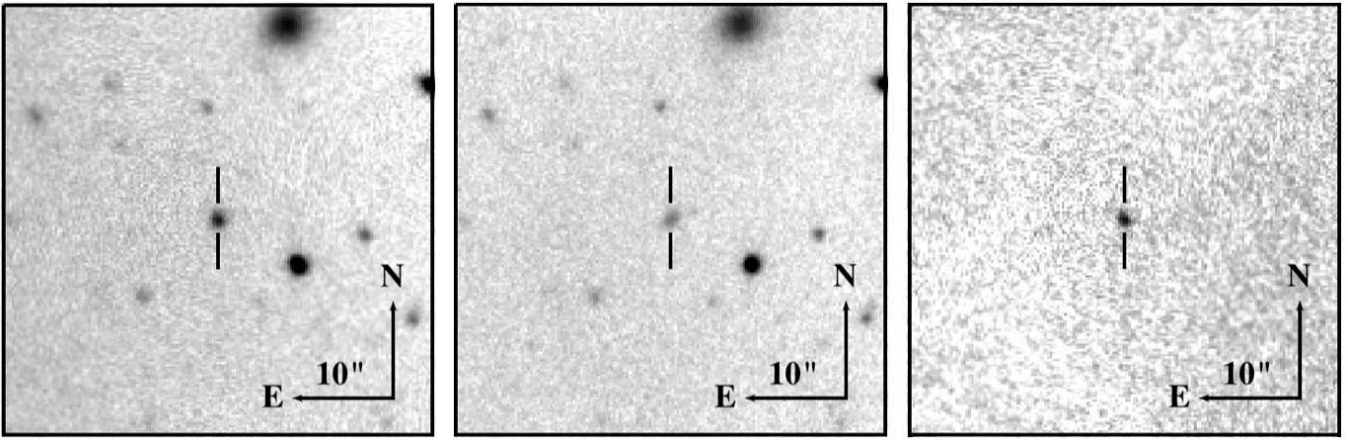}
\caption{From left to right: \paj\/+host galaxy on MJD 55615.2, host galaxy on MJD 56075.0 used as template image and the final subtracted image showing the SN.} 
\label{fig:sub}
\end{figure*} 

\begin{figure*}
\includegraphics[width=18cm]{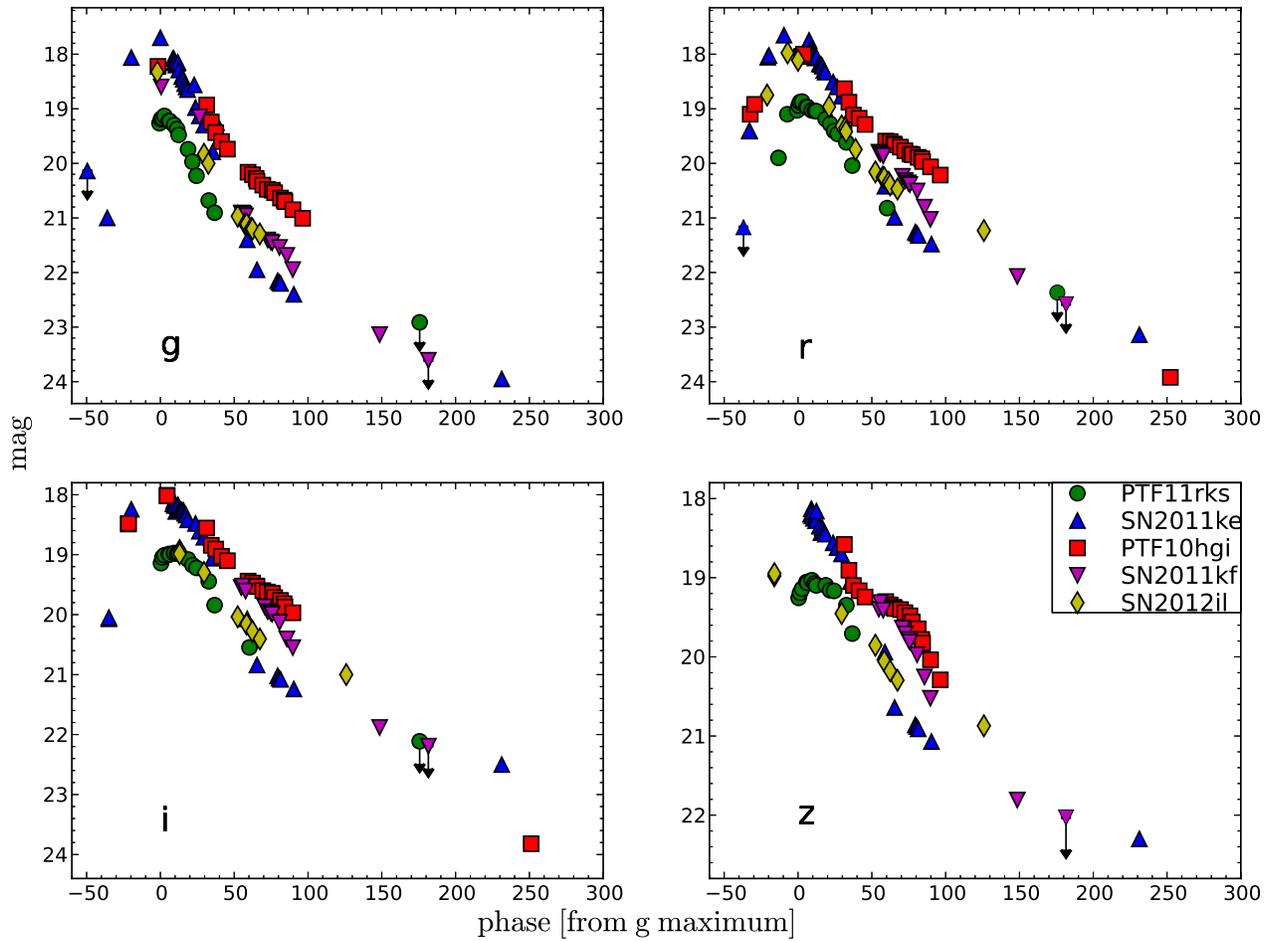}
\caption{Observed {\it g r i z} light curves of \rks\/ (green circles), \xk\/ (blue triangles), \paj\/ (red squares) , \css\/ (purple upside-down triangles) and \fo\/ (gold diamonds). The phase is from the respective maximum in the $g$ band. Detections from the ATels are shown here and reported in Appendix~\ref{sec:tab}, in Tabs.~\ref{table:10hgi},~\ref{table:11xk},~\ref{table:11rks},~\ref{table:css} and~\ref{table:12fo}.} 
\label{fig:lc}
\end{figure*}

\section{Follow-up Imaging and photometry}\label{sec:obs}

Optical and near infrared (NIR) photometric monitoring of the five SNe
was carried out using the telescopes and instruments listed in Appendix~\ref{sec:tab}.
The main sources of our photometric follow-up
were the SDSS-like $griz$ filters in the cameras at the Liverpool
Telescope (RATCam), William Herschel Telescope (ACAM), and the Faulkes
North Telescope (MEROPE). Further data in $BV$ and $JHK$ filters were
taken for some of the SNe with the EKAR 1.8m Telescope (AFOSC), the
ESO NTT (EFOSC2) and the Nordic Optical Telescope (NOTCam).
Swift+UVOT observations 
have been taken  for four out of five SNe in the UV filters uvw2, uvm2
and uvw1 (and for three SNe in the Swift $u$ filter) and we
analysed  these publicly available data independently.  Aside from \xk\/, ground  based
SDSS-like {\it u} observations were sparse, and for two SNe of our sample nonexistent. 

Observations were reduced using standard procedures in the 
IRAF\footnote{Image Reduction and Analysis Facility, distributed by
  the National Optical Astronomy Observatories, which are operated by
  the Association of Universities for Research in Astronomy, Inc,
  under contract to the National Science Foundation.}
environment. The magnitudes of the SNe, obtained through a point spread function 
(PSF) fitting, were measured on the final images
after overscan correction, bias subtraction, flat field correction and
trimming.  
When necessary we applied a template subtraction technique on later epochs
\citep[through the HOTPANTS\footnote{http://www.astro.washington.edu/users/becker/hotpants.html} package based on the algorithm presented in][]{al}.
The instruments used to obtain reference images
were the William Herschel Telescope and the Telescopio
Nazionale Galileo.
The same images were used to measure the host magnitudes
and listed in Appendix~\ref{sec:tab} (Tabs.~\ref{table:10hgi},~\ref{table:11xk},~\ref{table:11rks},~\ref{table:css}~\&~\ref{table:12fo})
and labelled as {\em ``Host"}.
When we did not have template images, we used SDSS images as template
to remove the flux of the host.
The magnitudes of SDSS stars in the fields of the
transients were used to calibrate the observed light curves
(Fig.~\ref{fig:lc}). All Sloan magnitudes - as well as the NTT $U$ and $R$ magnitudes - were converted
to the SDSS AB magnitude system and colour corrections were applied.
PS1 magnitudes were also converted to SDSS magnitudes following the prescription in \citet{2012ApJ...750...99T}. 
$B$ and $V$ magnitudes are reported in the Vega system.
The \paj\/ field was not covered by SDSS, so the
average magnitudes of local sequence stars were determined on
photometric nights, and subsequently used to calibrate the zero points
for the non-photometric nights. Magnitudes of the local sequence stars
are reported in Appendix~\ref{sec:sspaj} (Tab.~\ref{table:ss10hgi})
along with their rms (in parentheses).

For the Swift $u$ band data, we determined magnitudes in the UVOT instrumental system \citep{poo08} and
subsequently converted to Sloan $u$ by applying a shift of $\Delta
u\approx0.2$ mag. The shift has been computed for each SN from a
comparison of the magnitudes of the reference stars in the SNe fields
in the UVOT and Sloan photometric systems. The only exception is
\paj\/, where due to the absence of ground-based $u$ images, we
applied the average shift of the other SNe. The UV magnitudes
are reported in Appendix~\ref{sec:tab} (Tab.~\ref{table:swift}).

NIR observations are not shown in Fig.~\ref{fig:lc} as these
were only obtained for \rks. The $JHK$ photometry was calibrated to
the 2MASS system (Vega based), using the same local sequence stars as for the
optical calibration. Thus the values reported are Vega magnitudes.

\subsection{Light curves}\label{sec:lc}

\subsubsection{\paj}
Pre-peak observations are available only in the $r-$band, suggesting a rise time comparable to \xk\/. \paj\/ shows a bell-like shaped light curve around peak. The post maximum lightcurve shows a constant decline in all the bands until 40d. After 40d, the decline rate of \paj\/ changes to have a slope similar to the decays shown by the other SNe. The change in the $i-$band slope is not as evident, while the $z-$band light curve is also dissimilar to the other bands. The magnitudes beyond 90d are evaluated using the template subtraction with 646-648d epochs as template images.

\subsubsection{\xk}
\xk\/ was detected during the rise phase, and we continued to observe the SN until it disappeared behind the Sun in late August 2011. 
The non-detection of the transient the day before the discovery gives us the best constraint on the explosion epoch of any SL-SN to
date, allowing the rise time and light curve shape to be confidently measured.
The light curve is bell shaped around peak in the observed-frame $g-$band, and more similar to the light curve of SCP-06F6 \citep{ba09}. The post maximum light decrease is slower at redder wavelengths, as in the previous object. It follows a constant slope until 50d, when the slope changes to a slower decline. \xk\/ then continued to fade at the same rate until the last available photometric point at $\sim200$d post maximum. The reference template (339d) was used to retrieve the magnitudes after 51d.

\subsubsection{\rks}
The transient was discovered just before the $g-$band peak. The pre-discovery limit of December 11 \citep{qu11b} indicates a rise time on the order of 20 d, followed by a slower decline post-maximum. The $r-$band light curve shows an asymmetric peak as for SNe 2005ap and 2010gx \citep{qu07,pa10}, in contrast to the rounded peaks of the light curves of PS1-10awh and PS1-10ky \citep{ch11}. The SN fades by $\sim2.1$ mag over the first 30d in rest frame $g$ band, with a slower decline in the redder bands. The decrease is faster than that of the other SL-SNe Ic, although a more rapid decline at redder wavelengths is common in SNe \citep[see Fig.~1 in][]{pa10}. After 50 days the SN faded below our detection limit, even in deep imaging. A small, but non-negligible, flux contribution from the host has been found after 28d and was removed using template subtraction with the reference images at 218d. While for the $i$ and $z$ bands we used the SDSS images as template.

\begin{figure*}
\includegraphics[width=18cm]{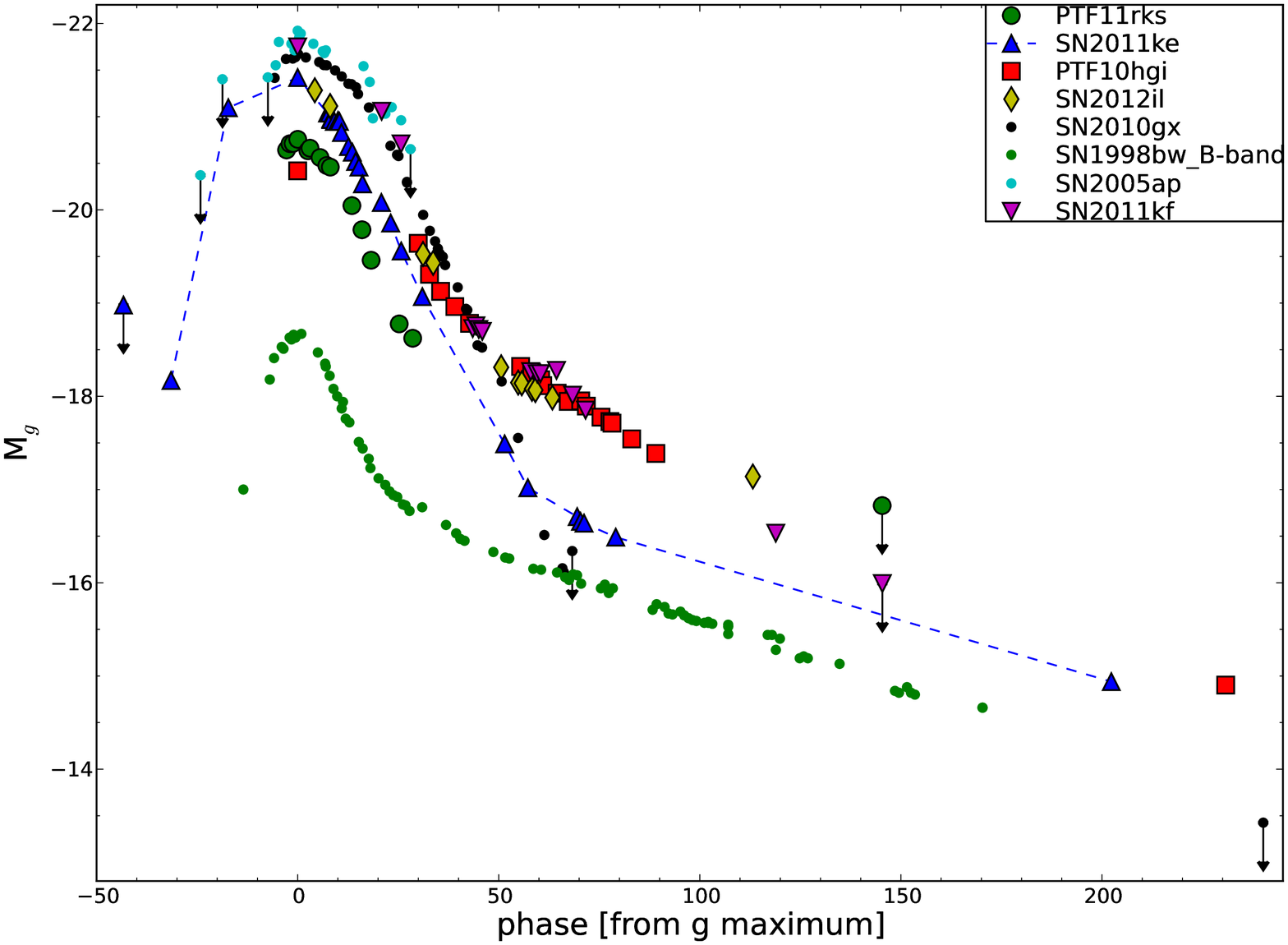}
\caption{$g-$band absolute light curves of \rks\/, \xk\/, \paj\/, \css\/ and \fo\/ and a number of super-luminous events as well as the stripped envelope SN 1998bw. The light curves for each SN have been derived by correcting the observed broadband photometry for time dilation, distance modulus, foreground extinction, and differences in effective rest-frame bandpass ($K$-correction). The last \paj\/ and \fo\/ points were converted from the $r$ mag applying a colour correction derived from \xk\/ and \css\/ at similar epochs.} 
\label{fig:absg}
\end{figure*}

\subsubsection{\css}

The light curve of \css\/ is the least well sampled as our monitoring
started some 20 days after the ATel discovery announcement. We assume
that the reported point of \cite{dr12} is at peak which is supported
by the spectral and colour evolution of the SL-SN during its
subsequent evolution.  During the first 50d, the decline in the $g-$band resembles
those of the other SL-SNe Ic. But between 50-150d the decline rate
changes markedly and the fading is slower. 
The reference template (164d) was used to retrieve the magnitudes after 71d.
Because of the proximity of the template epoch and last SN epoch, we also used 
SDSS images as secondary template. The values retrieved with the two different templates
were in agreement, strengthening simultaneously the lack of the SN at 145d and the 
detection of the faint host in the deep images of 164d.

\subsubsection{\fo}
\fo\/ was discovered before it  peaked in the $g-$ and
$r-$bands. The first two epochs available are in 
the PS1 \zps band and the CSS unfiltered system  \cite{dr12}. 
We can not set a robust constraint on the rise time for \fo,
but it is likely at least two weeks. The shape of the light curve around peak
in $r-$band is possibly asymmetric as in \rks\/ and SN 2010gx, 
although we are somewhat constrained in this statement due to the
uncertainty of the peak epoch.  As for
\xk\/, we see a clear change in the decline rate after 50 days, when
\fo\/ has a slower decline (shown in Fig.~\ref{fig:lc}. 
This change in decline to a slower fading rate is illustrated with the
latest detection in all three filters $riz$ at $\sim$113d after peak.
The reference template (327d) was used to retrieve the magnitudes at 113d and the $g$
magnitudes after 58d.

\subsection{Absolute Magnitudes}

In calculating absolute magnitudes (and subsequently bolometric 
magnitudes),  we have assumed negligible internal host galaxy reddening for all the objects,
and applied only foreground reddening, with the values 
reported in Sect.~\ref{sec:sample}. No Na ID absorption features due to gas in the hosts were
observed. However, we can not exclude  possible dust
extinction from the hosts, therefore the absolute magnitudes reported
here are technically lower limits. Given that the hosts are all dwarf 
galaxies, and the transients have quite blue spectra around peak it
appears that any correction would be small. We computed 
$k$-corrections for each SN using the spectral sequence we have
gathered. For photometric epochs for which no spectra were available, 
we determined a spectral energy distribution (SED) using the multi-color photometric
measurements available. This SED was then used as a spectrum template
to compute the $K$-corrections. 
Comparisons between the two methods ($K$-correction directly from
spectra, or with the photometric colours) showed no significant
differences.  We also determined $K$-corrections for SN2010gx 
using the spectral method  and using photometric colours. Again we
found consistency  between the two methods.  After applying 
foreground reddening corrections and $K$-corrections 
we estimated absolute rest-frame peak
magnitudes (cfr. Tab.~\ref{table:sample}).

In Fig.~\ref{fig:absg}, we compare the rest frame $g-$band absolute
light curves (in the AB system) of the SNe studied here with those of
other low-$z$ super-luminous events and the  well studied Type Ic SN 1998bw
\citep{ga98,mck99,so00,pat01}. The epochs of the maxima were computed
with low-order polynomial fits and by comparison of the light curves
and their colour evolution with those of other SL-SNe, 
and are listed in Tab.~\ref{table:sample}.
The absolute
peak magnitudes of \rks\/ and \paj\/ are fainter than the bulk of
SL-SNe, although they are still $\sim2$ mag brighter than SN
1998bw. Interestingly, the two faintest SL-SNe Ic display different
decline rates to each other, \rks\/ is similar to SN 2010gx whereas \paj\/ decreases
at a slower rate.  The other three objects have peak magnitudes
comparable with that of SN 2010gx (M$_g\approx-21.67$)  and show a similar decline. The
decline slope changes in four of the five objects after 50 days in the
rest frame (while for the other, \xk\/, our data do not constrain it).
The light curves then settle on a tail resembling the
decay of \co. This is apparent in  Fig.~\ref{fig:absg}  as the tails of the
SL-SNe are similar to that of SN 1998bw which is known to be powered
by \ni. The light curves follow the \co\/ decay within errors of 10\%,
the biggest discrepancy is for \css\/ which falls more rapidly between
100-200
days. 
SN 2007bi \citep{gy09,yo10} also
followed the \co\/ decay at late times, but with a tail that is
$\sim2$ mag brighter than the SL-SNe Ic, and with a much slower
overall evolution. We also note that the light curves of our objects
flatten at slightly different epochs, with the tail for \xk\/
commencing $\sim10$ days after the last point in the lightcurve for SN
2010gx.

\subsection{Colour Evolution}

\begin{figure}
\includegraphics[width=\columnwidth]{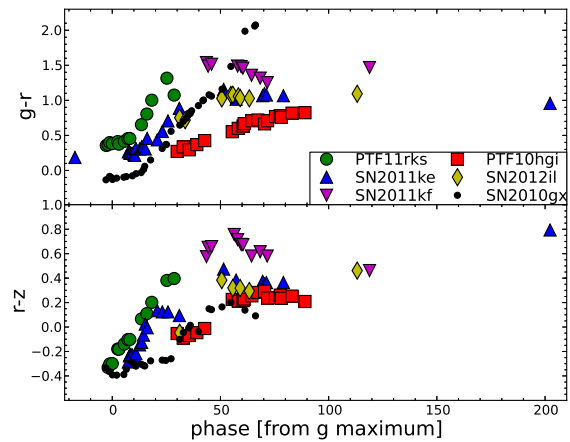}
\caption{Comparison of the dereddened and $K-$corrected colour evolution. \rks, \xk, \fo, \paj\/ and \css\/ are shown together with the well sampled SN 2010gx.} 
\label{fig:col}
\end{figure}

We computed rest-frame colour curves, after accounting for the
reddening and redshift effects of time-dilation and $K-$correction.
The colour curves are useful probes of the temperature evolution of
the SNe. We also calculated rest-frame colour evolution of SN 2010gx,
the only other SL-SN Ic with a good coverage in SDSS filters at a
similar redshift.  In Fig.~\ref{fig:col} the SL-SNe 
show a constant colour close to $g-r=0$ from the pre-peak phase to
$\sim15$d. This evolution is similar to the colour evolution of the
higher redshift PS1-10ky in the observed bands $i_{\rm P1} - z_{\rm P1}$  
\citet{ch11}.

The constant colour until 15 d implies that the SED does not evolve
over these epochs. Up to maximum light, the spectra of these SL-SNe
appear to be blue, with  the  only strong features being the O~{\sc ii} lines 
in this range covered by the $gr$ filters
\citep{pa10,qu11,ch11,le12}. 
Hence this could be
due to an approximately  constant temperature. This is also  illustrated in Fig. 8 of
\citet{ch11} for the higher redshift PS1-10ky. Close to peak the O~{\sc ii} lines disappear, leaving
the spectra featureless for $\sim10$ days, while after peak the
temperature begins to decrease (see Sect.~\ref{sec:velT}). A monotonic
temperature decline between 14000~K and 12000~K (blackbody peak
$2100$~\AA~$\lesssim\lambda\lesssim2500$~\AA\/) in objects with
featureless spectra does not strongly affect the colour evolution for
$\lambda\gtrsim\lambda_{\rm peak}$, as the slopes of blackbodies at
these two temperatures are quite similar. To detect differences in
temperature between a 12000~K and a 14000~K blackbody requires colour
curves which sample rest wavelengths below 3800~\AA\/, such as the
$g-z$ colour of PS1-10ky. And indeed this object did show an increase in $g-z$.

\begin{figure}
\includegraphics[width=\columnwidth]{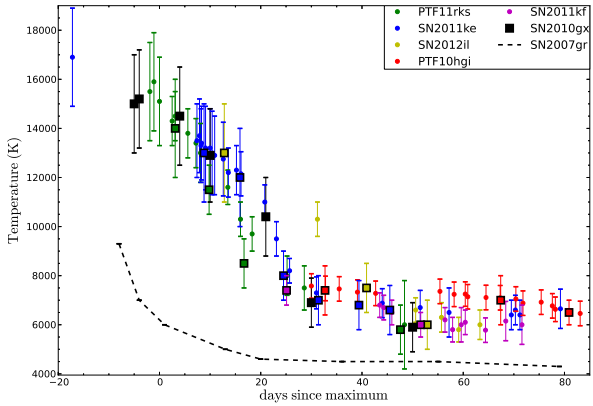}
\caption{Evolution of the continuum temperature of \rks\/ (green), \xk\/ (blue), \fo\/ (gold), \paj\/ (red) and \css\/ (purple) are reported with those of SN 2010gx (black).  The dashed black line shows the temperature evolution in the normal Type Ic SN 2007gr. Dots denote measurements from photometry, whereas squares are spectroscopic measurements. } 
\label{fig:T}
\end{figure}

After this early period of constant colour, the $g-r$ colour
increases, reaching another phase of almost constant value at
$\sim40$d, perhaps indicating a decrease in the cooling rate. There are some
exceptions to this behaviour, as seen for \rks\/ and \paj\/. The $g-r$
colour of \rks\/ increases earlier and with a steeper slope than the
other SNe, but unfortunately our data stop before the possible second
period of constant colour. In contrast, the $g-r$ colour for \paj\/
increases much more slowly, and only reaches the possible late
constant phase at $\sim80$d.


\begin{figure*}
\includegraphics[width=18cm]{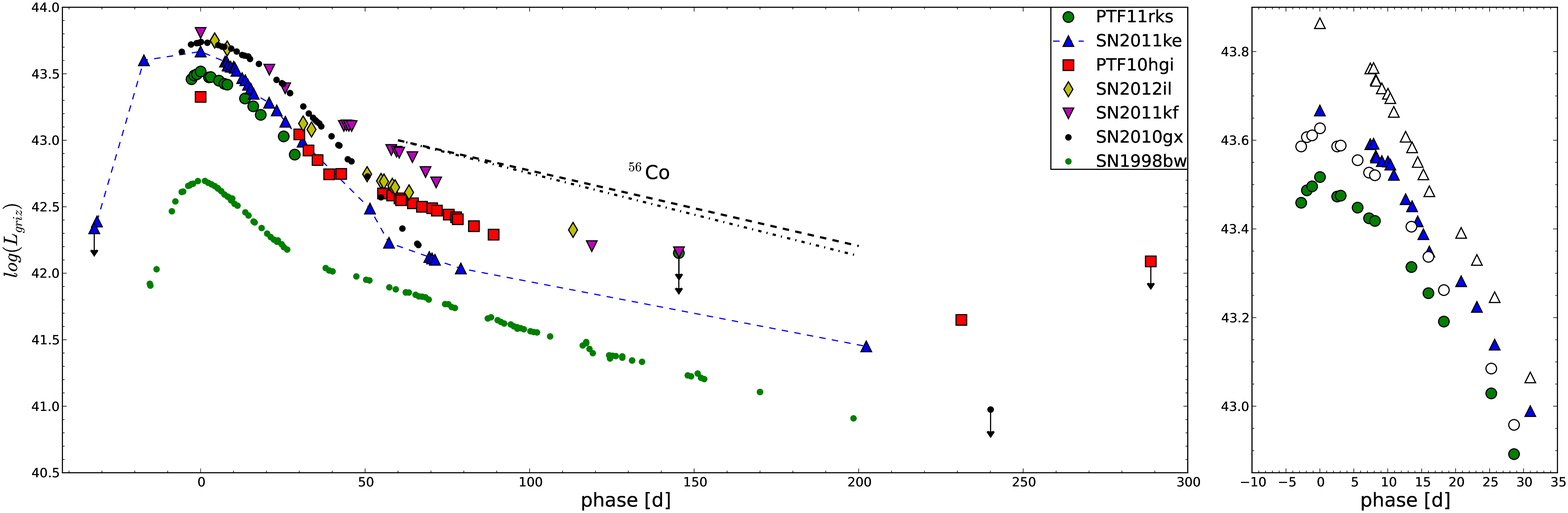}
\caption{Left: {\it griz} bolometric lightcurves of \rks\/, \xk\/,
  \paj\/, \css\/, \fo\/, SN2010gx and the Type Ic SN 1998bw
  ($BVRI$ bolometric lightcurve). These bolometric lightcurves are
  computed after correcting the observed broadband photometry for time
  dilation and applying $K$-corrections. The dashed line is the slope
  of \co\/ to \fe\/ decay, 
the dot-dashed line is the decay slope with a 10\% error. 
Right: Comparison between the  {\it griz} bolometric light curve
(filled symbols) and the $UVgriz$ bolometric lightcurve of \rks\/ and
\xk\/ which include the measured UV flux from Swift photometry.} 
\label{fig:bol}
\end{figure*}

The $r-z$ colours of the sample show a roughly constant increase from
peak to $\sim50-60$d, when the colour evolution
appears to flatten.  The two exceptions are again 
\rks\/ and \paj\/; the former increases in $r-z$ more rapidly than the
other objects, whereas the latter does not become as red, and
experiences a clear decrease in $r-z$ after 80d. The $r-z$ colour
evolution of SN 2010gx is similar to that of \xk\/ and \paj.

\subsection{Temperature evolution}\label{sec:T}

In Fig.~\ref{fig:T} the
evolution of the temperature is plotted. This is derived from a
blackbody fit to the continuum of our spectra (see Sect.~\ref{sec:sp}), and compared to those
of SN~2010gx and SN~2007gr.  
We also fit colour temperatures at rest-frame with a blackbody
and the measurements are in good agreement with those from spectra.

Only \rks\/ has a good temperature coverage around peak, whereas
our spectroscopic data are not well
sampled at that phase. While we can not clearly confirm the apparent
constant temperature seen in SN 2010gx until $\sim10$d, the epochs
either side at $\pm10$d are suggestive of a roughly constant temperature
phase.  After $\sim10$d a clear decline in temperature is seen, with a
rate of decline of $\sim$2500 K over 10d. This decline continues until
the SN reaches a constant temperature of $\sim6000$ K prior to, and
during, the pseudo nebular phase.

\section{Bolometric luminosity}\label{sec:bol}
 
Simultaneous UV--optical--NIR photometry at all epochs is required to
obtain a direct measurement of the bolometric luminosity. This is
typically difficult to attain at all epochs during a SN lightcurve, and
we do not have complete wavelength coverage for the five SL-SNe. Nevertheless,
valid corrections can be applied to the observed photometric bands to
compute the bolometric flux. 

The effective temperatures of the photospheres of SL-SNe Ic during
their first 30-50 days after explosion are between $T_{\rm bb} \sim
13000 - 19000$\,K \citep[see Table 10 and][]{pa10,ch12}. This means
that their fluxes peak in the UV ($\lambda<3000$~\AA) during this
period while our $griz$ bands typically cover from rest-frame
3800~\AA\/ redwards.  Thus a significant fraction of the flux is not
covered by the optical $griz$ imaging.  At around 20d after peak, the
effective temperatures tend to drop below 10000 K, hence the SEDs peak
between 3000\,\AA\ and 4000\,\AA. Although the peak of the SED moves
redward, a significant amount of the bolometric flux is radiated in
the UV even during these late stages.   In the following, we will use the term
``$griz$-bolometric lightcurve'' to refer to a bolometric lightcurve
determined using only the specified filters (in this example, $griz$) with
the flux set to zero outside the observed bands. 


Initially, the broad band magnitudes in $griz$ were converted into
fluxes at the effective filter wavelengths, then were corrected for
the adopted extinctions (cfr. Sect.~\ref{sec:sample}).  A SED was
then computed over the wavelengths covered and the flux under the SED
was integrated assuming there was zero flux beyond the integration
limits.  Fluxes were then converted to luminosities using the
distances previously adopted. We initially determined the points on
the $griz$-bolometric lightcurves at epochs when $griz$ were available
simultaneously (or very close in time). For epochs with coverage in
less than the four filters we were able to estimate the
$griz$-bolometric lightcuves.  Magnitudes from the missing bands were
generally estimated by interpolating the light curves using low-order
polynomials between the nearest points in time.  For some points this
interpolation was not possible and we used one of two methods. 
The first was an extrapolation assuming
constant colours from neighbouring epochs, the second was using 
colours from the other SL-SNe at similar epochs. For example, we
used the latter method for the last point on the lightcurve for \paj.  
The {\it  griz}-bolometric light curves estimated using this technique are
plotted in the left panel of Fig.~\ref{fig:bol}.

\begin{figure}
\includegraphics[width=\columnwidth]{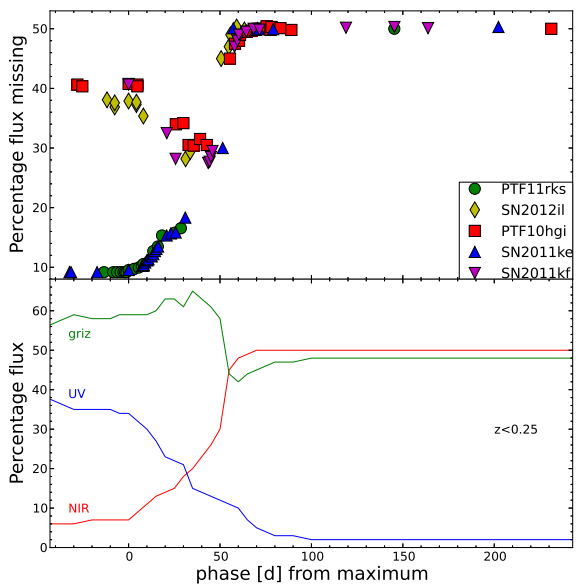}
\caption{Top: Percentage of flux missed, assuming a blackbody fit, in NIR by \xk\/ and \rks\/, and in UV+NIR by \paj\/, \css\/ and \fo\/  (the regions not covered by our photometry for each object). Bottom: Average percentage of the bolometric flux in UV (blue), optical (green) and NIR (red) for a representative
SL-SN Ic at $z<0.25$.} 
\label{fig:fluxm}
\end{figure}

Useful Swift photometry for UV flux measurements exists for \rks\/ and
\xk\/ (see Table\,\ref{table:swift}). Which allow us to compare the
{\it griz}-bolometric light curves and the $UVgriz$-bolometric
lightcurves (pseudo-bolometric hereafter), where the $UV$ component is determined from the $uvw2$,
$uvm2$ and $uvw1$ filters covering 1800-3000\AA. 
The difference between these two bolometric light curves, with and without the 
radiated energy below 3500~\AA\/ is shown in the right panel of
Fig.~\ref{fig:bol}.  

Inclusion of the Swift photometry results in  
maximum luminosities for \rks\/ and \xk\/   of 
L$\approx4.27\times10^{43}$ erg s$^{-1}$ and
L$\approx7.08\times10^{43}$ erg s$^{-1}$ respectively. 
Hence the  {\it griz}-bolometric
fluxes, are a factor 1.5 lower than when including the 
1800-3000\AA\ range covered by $uvw1-uvw2$ filters. 
While one could fit a black-body curve to the observed $griz$ SEDs (or
the spectra) and integrate under the curve to determine the emitted total
flux across all wavelengths, this would not account for the strong line absorption shown in the rest-frame
UV spectra  \citep[for example, clearly seen in the high-z objects of][]{ch11}. Hence from here on we will use $griz$-bolometric
lightcurves for consistency on all objects, but we should bear in mind
the additional contribution from the restframe UV that we have
quantified in the right-hand panel of Fig.~\ref{fig:bol}.  
The maximum luminosities reached by our computed 
{\it griz}-bolometric light curves are $\mathrm
{L_{\rm \rks}}\approx3.24\times10^{43}$ erg s$^{-1}$, $\mathrm
{L_{\rm \xk}}\approx4.47\times10^{43}$ erg s$^{-1}$, $\mathrm
{L_{\rm \css}}\approx6.45\times10^{43}$ erg s$^{-1}$, $\mathrm
{L_{\rm \fo}}\gtrsim4.47\times10^{43}$ erg s$^{-1}$ and $\mathrm
{L_{\rm \paj}}\approx2.09\times10^{43}$ erg s$^{-1}$. As expected,
these  are lower than those reported by \citet{ch11} for the
$z\simeq0.9$ objects  PS1-10ky and
PS1-10awh due to the lack of restrframe UV coverage for our 
low redshift sample.

\begin{figure*}
\includegraphics[width=18cm]{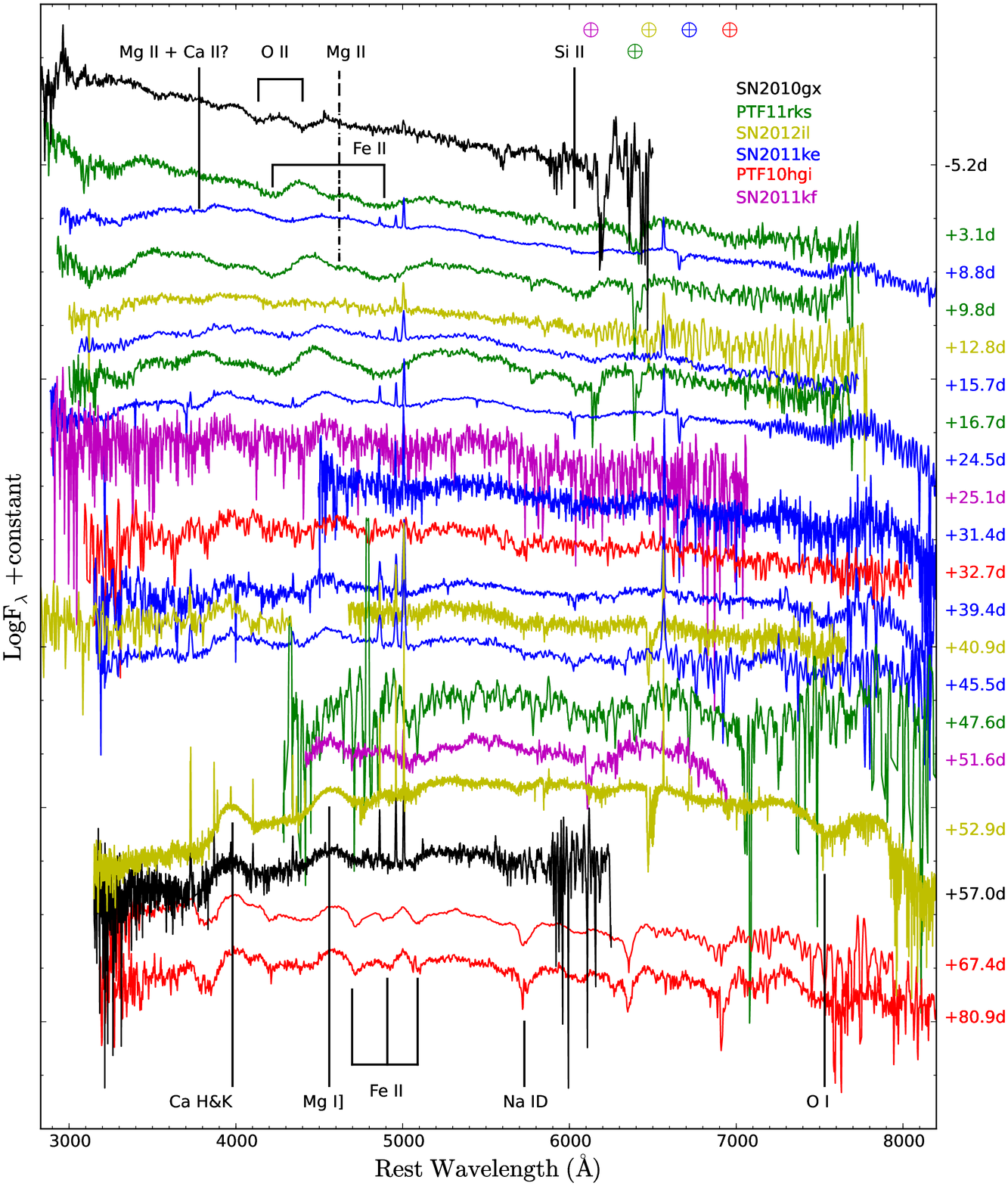}
\caption{Spectra of \rks\/ are in green. \xk\/ in blue, \fo\/ in gold, \paj\/ in red, \css\/ in magenta and SN 2010gx \citep{pa10} in black. The phase of each spectrum relative to light curve peak in the rest frame is shown on the right. The spectra are corrected for Galactic extinction and reported in the rest frames. The $\oplus$ symbols mark the positions of the strongest telluric absorptions. The most prominent features are labelled.} 
\label{fig:spevol}
\end{figure*}

The comparison in Fig.~\ref{fig:bol} further quantifies the  large 
bolometric luminosities of these SL-SNe Ic - as discussed by 
\citet{pa10}, \citet{qu11}, \citet{ch11} and \citet{le12}. There is
clearly some  diversity in the lightcurve peaks and widths. The 
low redshift of these objects makes it possible to follow the
evolution
beyond 100d after peak for the first time. Only one other object
(SN 2010gx)  has been investigated in this phase \citep{pa10,che12} and no detection was found at greater than
100 days.  Quantifying the host contribution and using image subtraction
to recover the SL-SN flux in these late phases is essential (as
discussed in Sect.\,\ref{sec:obs}).  It transpires that 
SL-SNe Ic show a large diversity in this
phase, quite different to the relatively homogenous behaviour around
peak. 
After 50d post maximum, all four of the  SL-SNe Ic for which we have data  (\xk\/,
\fo\/, \css\/ and \paj) show an abrupt change in the slope of the
$griz$-bolometric lightcurve. The slope flattens and is 
quite  similar to that of the decay of \co\/ to \fe. 
 \xk\/ and \paj\/ appear to 
decline even slower than the \co\/ slope.  

Additionally, at these later phases we know from detailed coverage of
CCSNe that a significant amount of radiation will be emitted in the
near-infrared when the photospheric temperature drops below 10000\,K.
This flux can be, mostly, captured by $JHK$ photometric
observations. As we lack this complete wavelength coverage for our
SL-SNe, we employed an SED method to determine the correction.  We
use the photospheric temperatures (derived in Sec.~\ref{sec:T}) to
derive simple blackbody SEDs and integrate the flux redwards of the rest-frame
$z$-band.  The flux missed in the NIR by our $griz$-bolometric
measurements typically increases with time, and reaches roughly 
50\% after $\sim60d$ post maximum. We plot the rest-frame NIR contributions as
an average of all the SL-SNe 
presented here, for a representative redshift of $z<0.25$, in the 
the bottom panel of Fig.~\ref{fig:fluxm}. 
The optical component refers to the $griz$ bands (green line), 
and the NIR contribution beyond the $z$-band is denoted with the red line. 
We also used this method to have a secondary estimate of the UV
contribution (blue line). In this case we integrated the flux under the black body
spectra below the $g$-band. 
We compared the UV flux contribution  with those evaluated from the
two SNe  which have Swift UV photometry and find the two are consistent 
within the errors. Fig.~\ref{fig:fluxm} summarises the flux
contributions from the different wavelength regimes and allows
the $griz$-bolometric lightcurves to be corrected when required.

If these tail phase luminosities were powered
by \co\/ then it would require that there is full $\gamma$-ray
trapping in the ejecta. This is not typically seen in Type Ic SNe. 
For example, the $BVRI$ bolometric lightcurve of SN 1998bw is shown 
for comparison which decays faster than the nominal \co\/ half-life. 
\citet{so00} showed that if one assumes a fixed energy  source 
(i.e. some mass of \co), then the trapping efficiency decreases with 
time ($\propto t^{-2}$). Hence at the epochs of these SL-SNe
(100-200 days)  only around $\sim45\%$ of the $\gamma$-rays would be
trapped if they had similar ejecta mass and density profiles to other
Type Ic SNe.  This seems to be in
contradiction to the measured slopes which appear either to follow
the \co\/ decay timescale or even be slightly shallower.  Despite this issue, we shall initially assume (for 
illustrative purposes) that the tail phases are actually powered
solely by radioactive decay. This allows a corresponding mass of \ni\/ to be
determined. Later in this paper we shall show that the tail phase
luminosity may be powered by magnetar energy injection rather than 
\co\/ decay. 
Although four of the SL-SNe do show a flattening in their luminosity, 
it appears SN 2010gx does not, at least not at the detectability level
of \cite{che12}. The data we have for \rks\/ does not allow a
conclusion. 

We initially make the assumption that $\gamma$-rays from \co\/ decay are fully
thermalized during the full durations of the lightcurves we  measure. 
We know that for typical SN Ic ejecta this is not the case, but it
allows us to derive  illustrative masses for \co\/ powering. 
The \ni\/ masses can thus be estimated using the formula
\begin{equation}\label{eq}
\textstyle M(^{56}{\rm Ni})_{\rm SN} = 7.87\times10^{-44} L_{\rm t}e^{ \left[\frac{(t-t_{0})/(1+z)-6.10}{111.26}\right]} M_{\odot},
\end{equation}
\citep[e.g. as employed by][]{h03a}, where t$_{\mathrm{0}}$ is the explosion epoch, 6.1~d is
the half-life of $^{56}$Ni and 111.26~d is the \textit{e}-folding time
of the $^{56}$Co decay, which release 1.71 MeV and 3.57 MeV,
respectively as $\gamma$-rays \citep{wo89,ca97}.

This method gives \ni\/ masses of M(\ni)$_{\mathrm{\xk}}\sim0.5$ \M\/,
M(\ni)$_{\mathrm{\fo}}\sim1.2$ \M\/, M(\ni)$_{\mathrm{\paj}}\sim1.1$
\M\/, while for \css\/ we retrieved an approximate
M(\ni)$_{\mathrm{\css}}\lesssim1.9$ \M\/. 
The measured decline for \css\/ was slightly steeper than the
fully trapped \ni\/ tail. 
We also estimated an
upper limit M(\ni)$_{\mathrm{\rks}}\sim1.3$ \M\/, for \rks\/  based on the last
epoch in which the SN was detected. These values should be considered
as lower limits because of our limited rest frame wavelength
coverage. At this phase, the contribution from the NIR plays an 
important role in the bolometric luminosity of SNe, as described
above. Indeed, the SED of SN 1987A \citep[data
from][]{ha88,bo89}, 
also suggests that as much as 50\% of the total flux for our
transients could be outside  the $griz$ bolometric lightcurves (at these epochs SN 1987A has already reached a constant
temperature in the tail phase). Thus, to obtain a truer idea of the
\ni\/ mass required to power these tail phases,  the measured
luminosities
should be increased by roughly a factor two. In summary, we find that {\em if the luminosity in the tail phase is
powered by \co\/} then the ejected \ni\/ values required to power the
lightcurves lie between
1.0~\M\/ and 2.4~\M\/, with an upper limit of 3.8 \M\/ for
\css\/.  These \ni\/ masses cannot power the peak of the lightcurves,
as shown by \citet{pa10,qu11,ch11,le12}. As we discuss below, we
consider that the luminosity we detect in this 
phase is not necessarily due to radioactive \co\/ decay energy
injection.  

\section{Spectroscopy}\label{sec:sp}

\begin{deluxetable*}{lccccc}
\tablewidth{0pt}
\tablecaption{Journal of spectroscopic observations. \label{table:sp}}
\tablehead{\colhead{Date}& \colhead{MJD}& \colhead{Phase\tablenotemark{a}} & \colhead{Range}& \colhead{Resolution\tablenotemark{b}}& \colhead{Instrumental} \\
\colhead{}& \colhead{}& \colhead{(days)} & \colhead{ (\AA)}& \colhead{(\AA)}& \colhead{configuration} }
\startdata
\cutinhead{\rks}
09/01/12&55936.36&3.1  &3400-9200& 13 & NOT+ALFOSC+Gm4\\
17/01/12&55944.34&9.8  &3500-9200& 13 & NOT+ALFOSC+Gm4\\
26/01/12&55952.57&16.7 &3500-9200& 13 & NOT+ALFOSC+Gm4\\
03/03/12&55989.38&47.6 &4300-8300&  5 & WHT+ISIS+R158R\\
\cutinhead{\xk}
15/05/11&55696.52&8.8   &3300-10000& 14& TNG+DOLORES+LRB,LRR 	\\
22/05/11&55704.46&15.7  &3400-9200& 14 & CAHA+CAFOS+b200  \\
01/06/11&55714.48&24.5  &3300-10000& 14& TNG+DOLORES+LRB,LRR   \\
09/06/11& 55722.45&31.4&5100-9700 &5& WHT+ISIS+R158R   \\
18/06/11&55731.49&39.4  &3300-10000& 14& TNG+DOLORES+LRB,LRR \\
25/06/11&55738.49&45.5  &3600-8800& 16& NTT+EFOSC2+Gm13 \\
\cutinhead{\fo}
30/01/12&55956.53&12.8   &3400-9200& 13 & NOT+ALFOSC+Gm4 	\\
03/03/12&55989.51&40.9  &5000-7800&  5 & WHT+ISIS+R300B,R158R  \\
17/03/12&56003.56&52.9  &3000-23000&2& VLT+XSHOOTER    \\
\cutinhead{\paj}
20/07/10&55398.41&32.7  &3400-9200& 14 & CAHA+CAFOS+b200   \\
28/08/10&55436.53&67.4  &3200-8000& 16 &GEMINI+GMOS+R150    \\
11/09/10&55451.37&80.9  &3200-8300& 10 &WHT+ISIS+R300B,R158R \\
\cutinhead{\css}
30/01/12&55956.74&25.1  &3400-9200& 13 & NOT+ALFOSC+Gm4   \\  	
03/03/12&55989.75&51.6   &4700-8300&  10 & WHT+ISIS+R158R  
\enddata
\tablenotetext{a}{phases with respect to the g-band maxima and corrected for time dilation}
\tablenotetext{b}{FWHM of night sky emission lines}
\tablecomments{The telescope abbreviations are the same used in the Appendix, in Tabs.~\ref{table:10hgi},~\ref{table:11xk},~\ref{table:11rks},~\ref{table:css} \&\/~\ref{table:12fo} plus TNG~=~3.6m Telescopio Nazionale Galileo + DOLORES; CAHA = 2.2m Telescope at Calar Alto Observatory + CAFOS; VLT = 8.2m ESO Very Large Telescope + XSHOOTER; GEMINI~=~8.2m Gemini Telescope North + GMOS.}
\end{deluxetable*}

\begin{figure*}
\includegraphics[width=18cm]{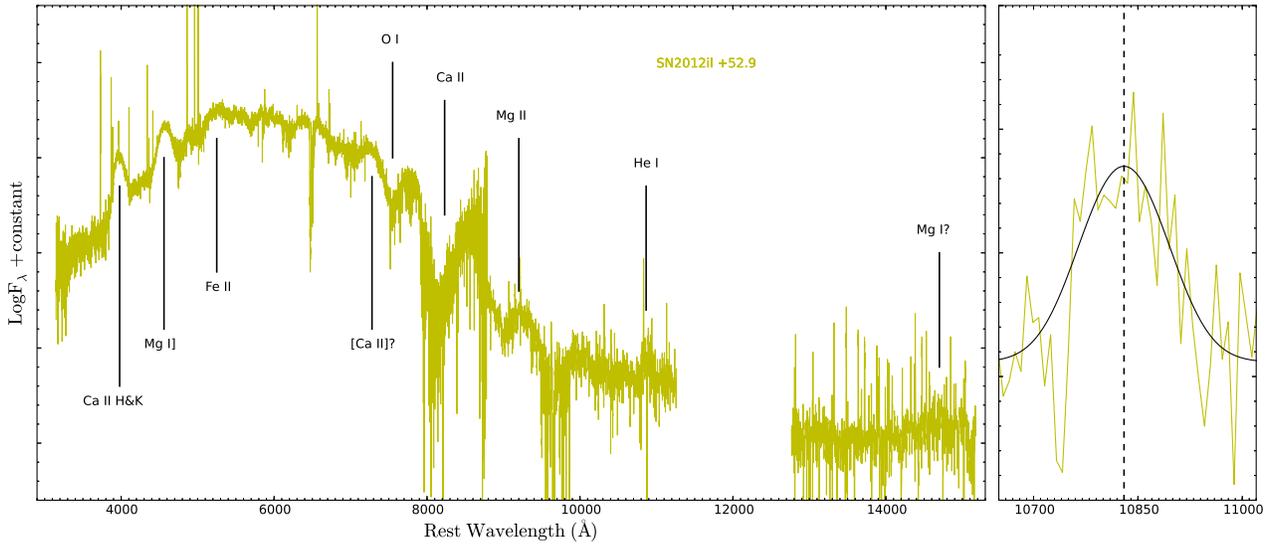}
\caption{Spectrum of \fo\/ at $\sim$53d post maximum. The spectrum is corrected for Galactic extinction and reported in its rest frame. The most prominent features are labelled. In the right panel a zoom of the He~{\sc i} line (spectrum binned by a factor 10) is shown with the gaussian fit of the line. The dashed vertical line marks the expected position of He~{\sc i} $\lambda$10830.} 
\label{fig:tot12fo}
\end{figure*}

All spectra were reduced (including trimming, overscan, bias
correction and flat-fielding) using standard routines within
IRAF. Optimal extraction of the spectra was adopted to improve the
final 
signal-to-noise (S/N) ratio. Wavelength calibration was performed
using spectra of comparison lamps acquired with the same
configurations as the SN observations. Atmospheric extinction
correction was based on tabulated extinction coefficients for each
telescope site. Flux calibration was performed using
spectro-photometric standard stars observed on the same nights with
the same set-up as the
SNe. The flux calibration was checked by comparison with the photometry, integrating the spectral flux transmitted by standard {\it griz} filters and adjusted by a multiplicative factor when necessary. The resulting
flux calibration is accurate to within 0.1 mag.

The collected spectra are shown in Fig.~\ref{fig:spevol} together with
spectra of SN 2010gx \citep{pa10} for comparison. A version with all
the spectra convolved to the same resolution and binned to the same
pixel scale is shown in Appendix~\ref{sec:spbin}.  In our
spectroscopic sample we do not have pre-peak spectral coverage, which
typically shows C~{\sc ii}, Si~{\sc iii} and Mg~{\sc ii} at UV
wavelengths ($<3000$\AA) and O~{\sc ii} in the optical region
\citep{qu11}. The only lines which we may expect to be visible in our
wavelength regions (namely O~{\sc ii}) have already disappeared by
$\sim3-12$ days. At those epochs the SL-SNe spectra are featureless
with the notable exception of \rks\/, which shows weak broad
absorption profiles of heavy elements such as Fe~{\sc ii}, Mg~{\sc ii}
and Si~{\sc ii} between 3000~\AA\/ and 6500~\AA\/. A weak Ca~{\sc ii}
H\&K absorption line is barely detected in \rks\/, as is the case for
\xk\/ and \fo. Redwards, Mg~{\sc ii} $\lambda$4481 and the Fe~{\sc ii}
multiplet $\lambda\lambda$4924, 5018, 5169 are visible in
\rks\/. Other Fe~{\sc ii} lines are barely visible in the region
around 4500~\AA\/, while a shallow absorption due to Si~{\sc ii}
$\lambda$6347 is also present. None of these lines are clearly
detectable in \xk\/ and \fo\/ at a similar phase.

Two weeks after maximum, \xk\/ shows Mg~{\sc ii}, Fe~{\sc ii} and
Si~{\sc ii} lines together with a clearer Ca~{\sc ii} feature, albeit
still shallower than for \rks.  The spectra obtained around 30d for
\css\/ and \paj\/ have a low signal to noise (S/N) ratio of
$\sim10$ which makes line analysis problematic. However, the
comparison with the other spectra indicates
broad absorption profiles from Mg~{\sc ii} and Fe~{\sc ii}. The only
line visible in the red part of the spectrum ($>7000$\AA) is O~{\sc i}
$\lambda$7775 in \xk\/.

\begin{figure}
\includegraphics[width=\columnwidth]{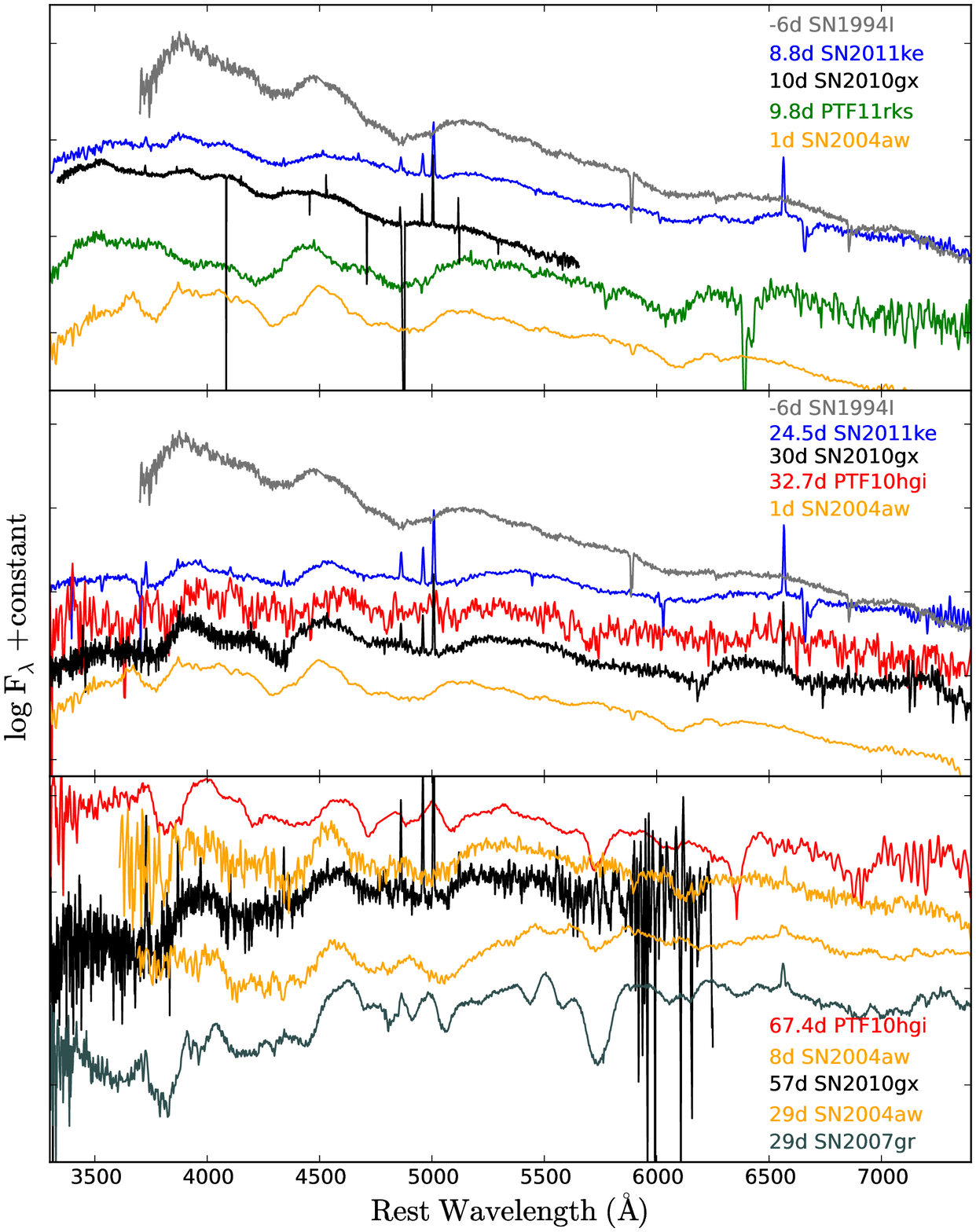}
\caption{Top: comparison of early-time spectra of \xk\/ and \rks\/ with that of SN 2010gx \citep{pa10} and those of Type Ic SNe 2004aw \citep{ta06} and 1994I \citep{ba96}. At $\sim$10d post maximum \rks\/ is already more developed than \xk\/ and SN 2010gx. Middle: comparison at about $\sim30$d post peak of the same objects as for the previous panel, but showing \rks\/ instead of \paj\/. Bottom: comparison of the spectrum of  \paj\/ at $\sim67$d with those of SN 2010gx at the pseudo nebular phase, SN 2004aw at maximum light during the photospheric period, and Type Ic SN 2007gr \citep{va08,hu09}. The spectrum of \paj\/ is more similar to those of SNe 2004aw and 2007gr at 29d rather than those of SN 2010gx at 57d and SN 2004aw at 8d \citep[as noted in][]{pa10}.}
\label{fig:comp}
\end{figure}

From $\sim30$d after maximum onwards there is no strong evidence of
new broad features emerging blueward of 8000\AA, with the exception of
the rise of Mg~{\sc i}] $\lambda$4571. The absorption lines of Mg and
Fe become shallower, resulting in the emission components becoming
more prominent 
from $\sim$50d onwards. The Fe~{\sc ii} emission line at
$\sim5200$~\AA\/ is broader than Mg~{\sc i}] suggesting  it may be a
blend. 
The only exception to this general trend is \paj. In the last two
available spectra \paj\/ shows Ca~{\sc ii} H\&K (possibly blended with
Mg~{\sc ii}), three distinct absorptions related to the individual
components of the Fe~{\sc ii} multiplet ($\lambda\lambda$4924, 5018,
5169) and Na~{\sc i}D $\lambda\lambda$5890,5896.
Although the spectral evolution is relatively homogeneous, two of the
sample have noticeable differences. 
In the first two weeks of evolution of \rks\/ the  absorption line strengths
look stronger, with  $\overline{\mathrm{EW}}_{\mathrm{\rks}}\sim4\times\overline{\mathrm{EW}}_{\mathrm{SL-SNe    Ic}}$. 
The later phase line formation in  \paj\/ also looks different, with
narrower absorption lines more similar to velocities seen in Type Ic
SNe (see the comparison with SN 2007gr in Fig.\ref{fig:comp})
  Weak, narrow emission lines (\Ha, \Hb\/ and
[O~{\sc iii}] $\lambda$4959, $\lambda$5007) from the host galaxies are
also visible in all spectra, except those of \paj. These will allow 
metallicity and star formation rate measurements in the host, which is
underway in a companion paper (Chen et al., in prep). 

Fig.~\ref{fig:tot12fo} shows the complete spectrum of \fo\/ taken at
$\sim53$d with VLT+XShooter, which has the widest wavelength coverage
for a SL-SNe obtained to date. As previously mentioned, bluewards of
8000~\AA\/, Ca~{\sc ii}~H\&K, Mg~{\sc i}], shallow Fe~{\sc ii} and
O~{\sc i} are present. A strong Ca~{\sc ii} NIR triplet
$\lambda\lambda$8498, 8542, 8662 is visible, with the bulk of the
absorption component at $v=12000\pm2500$ \kms\/. This is higher than
the other absorption components (e.g. O~{\sc i} $v\sim9500$ \kms) and
comparable with the velocities of the broad emission components of
Ca~{\sc ii} H\&K, Mg~{\sc i}].  At about 7300\AA\/ a weak emission
line is evolving and we tentatively identify it as [Ca~{\sc ii}]
$\lambda\lambda$7291,7324 emission feature.  
The presence of [Ca~{\sc ii}] and the emission lines around 4500~\AA\/
imply that the SNe is evolving, slowly, toward the nebular phase. This
seems to coincide with the change in
slope of the light curve.  In the NIR the S/N$\sim$5 in the continuum
is less than in 
the optical (S/N$\sim$30), although it is still adequate to identify
the strongest lines. We identify Mg~{\sc ii} around 9200~\AA\/ and
Mg~{\sc i} $\lambda$15024, although the last identification is less
certain due to the low S/N and the proximity of a strong telluric
feature.  Most significantly, we identify He~{\sc i}
$\lambda$10830, the first sign of the presence of He in this group of SL-SNe
Ic.
We note that the only other ions which  have transitions at this
wavelength
are Ar~{\sc ii}, Fe~{\sc i}, Si~{\sc I} and C~{\sc I}. These are
expected to be intrinsically weaker than He~{\sc i}. 
There are no strong sky lines at the observed postion of this line $\sim12725$\AA\/.
Although we do not see other He~{\sc i} lines in the spectrum, this is
not unexpected.  In the Case B recombination in 
the  temperature regime $T<10000$ K and electron density $10^2\leq n_e\leq 10^6$ -
the $\lambda$10830 line is expected to be stronger than $\lambda$5876
line (the 
strongest in the optical region) by a factor 2 to 10.

The spectral evolution of the SL-SNe sample in this paper, provides additional information to that reported in \citet{pa10}. In the top panel of Fig.~\ref{fig:comp} the comparison of the early time spectra with those of SN 2010gx \citep{pa10}, and the Type Ic SNe 1994I \citep{ba96} and 
2004aw \citep{ta06} highlights a difference in the line evolution between \rks\/ and other SL-SNe. The spectrum at 9.8d is similar to that of a Type Ic close to maximum light showing a faster transition to a Type Ic SNe than the others. In contrast, other SL-SNe such as \xk\/ match normal Type Ic SNe only after $\sim$30d (middle panel Fig.~\ref{fig:comp}), resembling the spectral transition of SN 2010gx at this epoch.
This makes the temporal evolution of \rks\/ very similar to those of canonical SNe Ic.
Again the low S/N of the spectrum of \paj\/ (32.7d) precludes a precise analysis, although there are hints of a peculiar line evolution at wavelengths redder than 5500~\AA.
In the bottom panel, SN 2010gx resembles SN 2004aw at $\sim10$d after
peak whereas \paj\/ matches normal Type Ic SNe at $\sim$30d, with a
good match to SN 2007gr. From this comparison it appears that \rks\/
and \paj\/ evolve to into Type Ic SN on timescales of about 20 days quicker than other SL-SNe. The ``fainter" luminosity (M$>-21$) of these two objects and their faster evolution to Type Ic SNe may provide another clue to understand the evolutionary path of these transients. It appears from these objects that the lower the peak luminosity, the faster the evolution of the photospheric spectra, with a time delay of 10 days to the Type Ic phase instead of the usual 30 days shown by most SL-SNe Ic. 

\subsection{Expansion velocity}\label{sec:velT}

\begin{figure}
\includegraphics[width=\columnwidth]{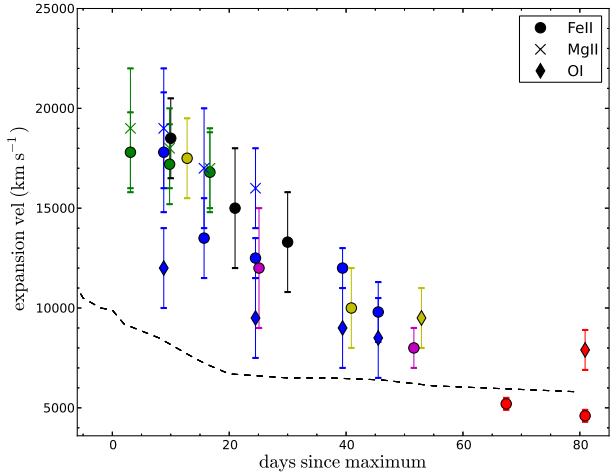}
\caption{Expansion, photospheric velocity as measured from the different lines. Phases are relative to light curve peak. Measurements of \rks\/ (green), \xk\/ (blue), \fo\/ (gold), \paj\/ (red) and \css\/ (purple) are reported with those of SN 2010gx (black). The dashed black line shows the Fe~{\sc II} evolution in the normal Type Ic SN 2007gr.} 
\label{fig:vel}
\end{figure}

The expansion velocities measured for \rks\/, \xk\/, \fo, \paj\/ and
\css\/ are reported in Tab.~\ref{table:vel}, and are compared with
those of SN 2010gx and the standard Type Ic SN 2007gr \citep{hu09} in
Fig.~\ref{fig:vel}. In the photospheric phase these were derived from
the minima of the P-Cygni profiles and their errors were established
from the scatter between several independent measurements. During the
 phase in which the objects appear to be transitioning to the nebular
 phase (i.e. beyond about 50 days)  the velocities were computed as the FWHM of the
emission lines (we will call this the pseudo-nebular phase).
These are not tracers of the photospheric velocity and are reported for completeness; they will be discussed further in Sect.~\ref{sec:model}. 
We used Fe~{\sc ii} $\lambda5169$, Mg~{\sc ii}
$\lambda4481$ and O~{\sc i} $\lambda7775$ to measure velocities during
the photospheric phase, and Ca~{\sc ii}~H\&K and Mg~{\sc i}]
$\lambda4571$ during the pseudo-nebular phase.  The Fe~{\sc ii}
velocity 
evolution monotonically declines for all transients, ranging from
$\sim18000$ \kms\/ around peak to $\sim8000$ \kms\/ at 50d. After this
epoch we have only \paj\/ spectra, and they still show clear
absorption components of Fe~{\sc ii}. The decline in velocity is
faster and it reaches $\sim4600$ \kms\/ at the last epoch of 80 days,
which is quite similar to the Fe~{\sc ii} velocity of SN 2007gr. Mg~{\sc ii} in the photospheric
phase is always seen at higher velocity than Fe~{\sc ii}, and
decreases linearly by $\sim1000$ \kms\/ every 10 days. The O~{\sc i}
velocity is comparable with that of Fe~{\sc ii}, with the exception of
\xk\/ where it is slower than all other ions (albeit with large
uncertainties due to the low S/N). Ca~{\sc ii} and Mg~{\sc i}] appear
after $\sim$40d from maximum light and show the same intensity and
velocity of $\sim11000$ \kms\/ in the entire sample. 

The \paj\/ spectra after 60d show Ca~{\sc ii} absorption at late
epochs, similar to normal Type Ic SNe.  Fig.~\ref{fig:vel} shows
evidence for decreasing line velocity with time, especially from 10d
post peak as seen for SN 2010gx \citep[previously reported as a
comparison in][]{ch11}. Our analysis shows a clear sign of change in
the rate of photospheric expansion ($30-40$d), with a decline which
resembles that typically observed during the photospheric phase of a
CCSN explosion. 

\begin{deluxetable*}{lcccccc||cc}
\tablewidth{0pt}
\tablecaption{Observed blackbody temperature, expansion photospheric velocities from Fe~{\sc ii} $\lambda5169$, Mg~{\sc ii} $\lambda4481$ and O~{\sc i} $\lambda7775$ in our SL-SNe sample on the left. Expansion velocities from Ca~{\sc ii} H\&K and Mg~{\sc i}] $\lambda4571$ on the right.\label{table:vel}}
\tablehead{\colhead{Date}& \colhead{MJD}& \colhead{phase\tablenotemark{a}} & \colhead{T}& \colhead{v (Fe II)}& \colhead{v (Mg II)}& \colhead{v (O I)} & \colhead{v (Ca II)} & \colhead{v (Mg I])}\\
 \colhead{}& \colhead{}& \colhead{(days)} & \colhead{(K)}& \colhead{(\kms)}& \colhead{(\kms)}& \colhead{(\kms)} & \colhead{(\kms)} & \colhead{(\kms)}}
\startdata
\cutinhead{\rks\/}
09/01/12&55936.36&3.1  &$14000\pm2000$ & $17800\pm2000$ & $19000\pm3000$&  & &\\
17/01/12&55944.34&9.8  &$11500\pm1000$& $17200\pm2000$ & $18000\pm2000$&  & &\\
26/01/12&55952.57&16.7 &$8500\pm1000$&  $16800\pm2000$ &  $17000\pm2000$&   &  & \\
03/03/12&55989.38&47.6  &$5800\pm1000$&   &&      & &\\
\cutinhead{\xk\/}
15/05/11&55696.52&8.8   &$13000\pm2000$& $17800\pm3000$& $19000\pm3000$  &  $12000\pm2000$ & & 	\\
22/05/11&55704.46&15.7  &$12000\pm2000$& $13500\pm2000$& $17000\pm3000$ &   &  & \\
01/06/11&55714.48&24.5  &$8000\pm1000$& $12500\pm1000$ &$16000\pm2000$ &$9500\pm2000$ &   &    \\
09/06/11& 55722.45&31.4& $7000\pm1000$& & & &   &      \\
18/06/11&55731.49&39.4  &$6800\pm1000$& $12000\pm1000$& &$9000\pm2000$&  &  $10000\pm1500$ \\
25/06/11&55738.49&45.5  &$6600\pm1000$& $9800\pm1500$ & &$8500\pm2000$ & $10000\pm1500$&  $10000\pm1500$\\
\cutinhead{\fo\/}
30/01/12&55956.53&12.8   &$13000\pm2000$& $17500\pm2000$& &   &   &	\\
03/03/12&55989.51&40.9  &$7500\pm1000$& $10000\pm2000$&   &  &  & \\
17/03/12&56003.56&52.9  &$6000\pm1000$& &   & $9500\pm1500$ & $12000\pm2500$& $12000\pm2500$ \\
\cutinhead{\paj\/}
20/07/10&55398.41&32.7  &$7400\pm1000$&  &  &   & & \\
28/08/10&55436.53&67.4  &$7000\pm1000$& $5200\pm300$ &    & &  $8300\pm800$&  $9800\pm1000$\\
11/09/10&55451.37&80.9  &$6500\pm500$& $4600\pm300$ & &   $7900\pm1000$ &  $8000\pm800$ & $9500\pm1000$ \\
\cutinhead{\css\/}
30/01/12&55956.74&25.1  &$7400\pm1000$&  $12000\pm3000$&   &  &&\\  	
03/03/12&55989.75&51.6   &$6000\pm1000$& $8000\pm1000$ & & &&  
\enddata
\tablenotetext{a}{corrected for time dilation}
\end{deluxetable*}

\section{On the nature of SL-SNe Ic}\label{sec:dis}

The SN sample presented in this paper provides new data for understanding the nature
of super-luminous events. The similarity within
this family is well established: high luminosity, similar spectral
evolution and an origin in faint host galaxies.
The overall spectral evolution is indeed similar to that of SNe Ib/c, although SL-SNe
spectroscopically evolve on a much longer timescale. 
However, the observed parameters of SL-SNe present several problems in interpreting the explosion.
The enormous luminosity at peak cannot be powered by radioactive \ni\/
\citep{pa10,qu11,ch11,le12}
which is the canonical energy source
for SNe Ic emission.  However, we showed in Section~\ref{sec:bol} that the
tail phase luminosity in some objects declines in a similar fashion as
to what would be expected from  \co\/ decay.

In the following subsections we use this most
extensive dataset to constrain plausible models for the origin of
these SNe, particularly using the bolometric luminosity from -30 to
200d and the temperature and velocity information.
In this paper we focus on quantitative modeling of
the magnetar spin down scenario (Sect.~\ref{sec:model}). Several other models
have been proposed, which we briefly review
in Sect.~\ref{sec:alt}.
The bolometric light curves used in this Section are corrected for
flux missed in {\em both}  UV and NIR as described in
Section\,\ref{sec:bol}. 

\subsection{Alternative models}\label{sec:alt}
Alternative models have been proposed: the spin down of a rapidly
rotating young magnetar \citep{kb10,wo10,2012MNRAS.426L..76D}; interaction of the SN
ejecta with a massive (3-5 \M) C/O-rich circumstellar medium
\citep[CSM,][]{bl10} ; shock breakout from dense mass loss \citep{ci11}; or
pulsational pair instability in which collisions between high velocity
shells are the source of multiple, bright optical transients
\citep{wo07}.

The first scenario considered is the
pulsational pair-instability model, where the luminosity is powered by
the collision of shells of material ejected at different times by
the pulsations \citep[see][for the implementation of this on SL-SNe Ic]{cw12}.  The outbursts are expected to be energetic, reaching
very high peak luminosities and creating hot (T$_{\rm eff} \approx 25000$ K),
optically thick photospheres \citep{wo07}. 
PS1 has occasionally observed the explosion sites of the five SL-SNe presented
here, and SN 2010gx on 3-5 occasions (per event) in one of \grizy\/ filters, 
reaching typical AB magnitudes of 22, 21.6, 21.7, 21.4, 19.3 
(see Sect.\ref{sec:sample}). 
This would
correspond to absolute magnitudes of roughly $-16$ to $-19$. 
We found no detection of any previous outburst to these magnitude
limits in the 1-2 year periods before explosion. Typically there were
5-10 epochs of images from PS1. This is not a constant 
monitoring period, and it does not rule out that pre-explosion 
outbursts occur, but we have found no evidence for them.

The second scenario is that of circumstellar interaction proposed by
\citet{ci11} in which the interaction between the ejecta and the CSM
converts the ejecta's kinetic energy into radiation.  This requires 
the diffusion radius of the SN to be about the radius
of the stellar wind expelled by the SN progenitor.  Based on this 
model of \citet{ci11},  \citet{ch11} determined the dense wind and
ejecta parameters for the observed $z\simeq1$ SL-SNe. This requires 
a mass-loss rate of 6 \M\/ in the  last year before the SN explosion, 
with an outer radius of around 40000 \R, and ejecta mass of 10 \M
\citep[consistent with the estimates reported in][]{mo12}. 
The rise times of our low-z SNe are similar to that of PS1-10awh
\citep[from][]{ch11} - around 10-30 days which would lead to similar
estimates of masses.  The detection of He~{\sc i} in \fo\/ at $\sim53$d 
is perhaps some hint that the dense CSM wind could be plausible, since
it is difficult to find any known massive star example in the Local
Universe which has a dense and extended  wind comprising 6 \M\/ of 
C+O material only. Although the S/N in the NIR is low, the He~{\sc i}
line is relatively broad, indicating that it arises from ejecta. Although most SL-SNe have similar rise times which 
are feasible in the diffusion model of \citet{ci11}, the light curves
of SN 2010gx \citep{pa10}, \rks\/ and \fo\/ are not symmetric, which 
one would expect in this dense wind scenario. 
However, \citet{gb12} showed as the rise and fall times can be different.
They also showed that the tail phase can be explained as diffusion from the inner layers which can slow the decline.
We have not investigated in this scenario in depth, but a combination of this model plus a \ni\/ tail would be
unlikely because of the necessity of full $\gamma$-ray trapping (see Sec.~\ref{sec:bol}).

\begin{figure*}
\vspace{0pt}
\includegraphics[width=\columnwidth]{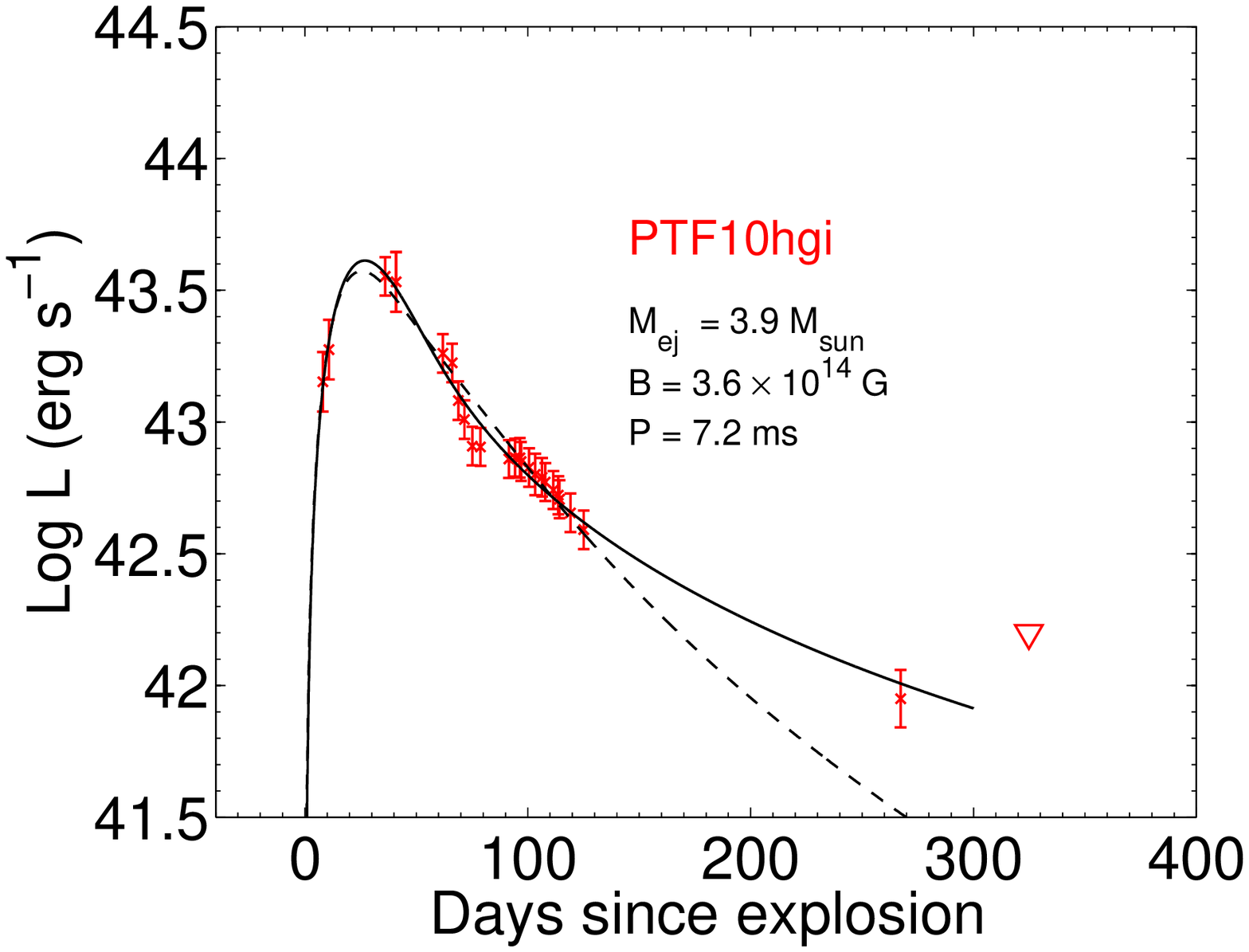}
\includegraphics[width=\columnwidth]{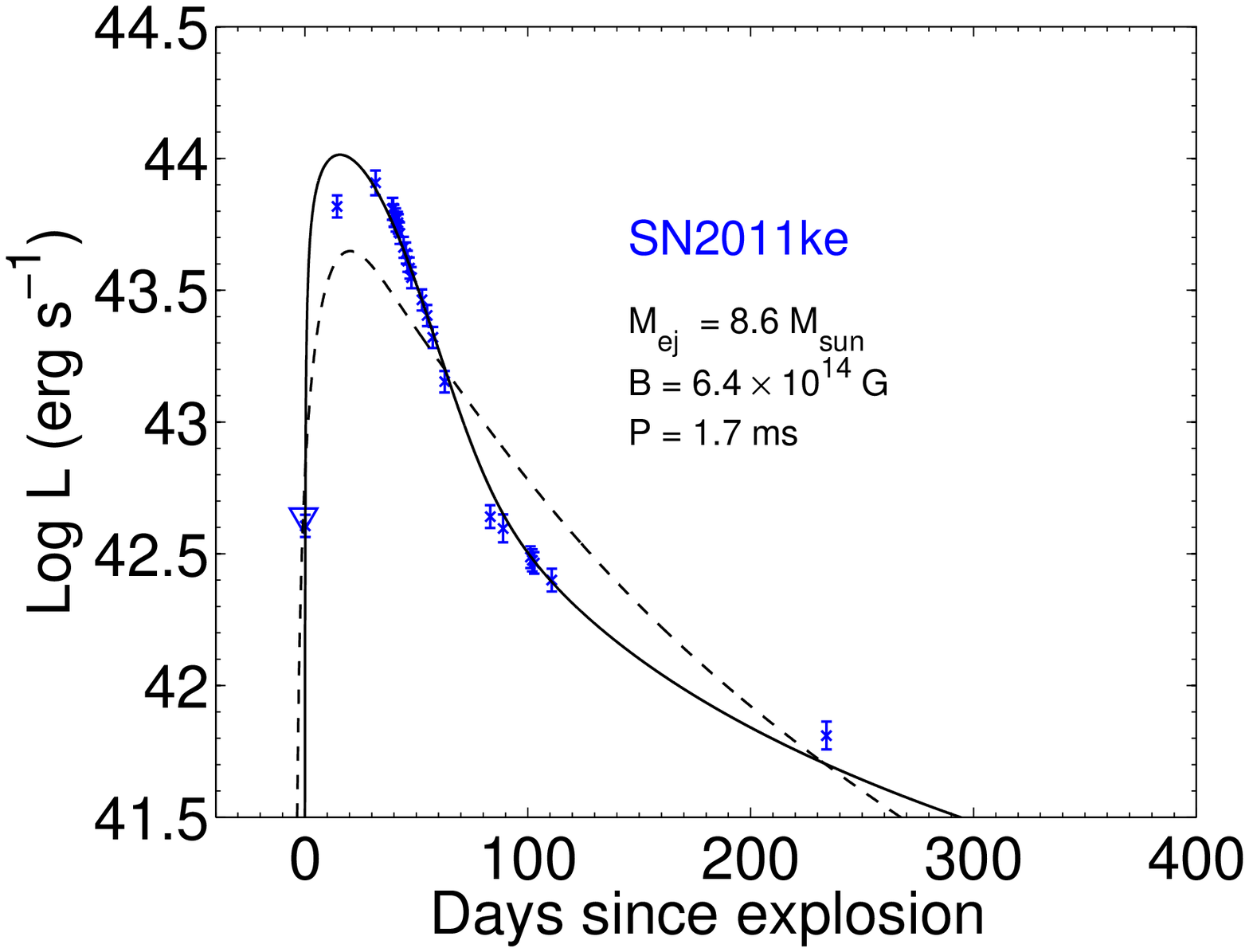}\\
\includegraphics[width=\columnwidth]{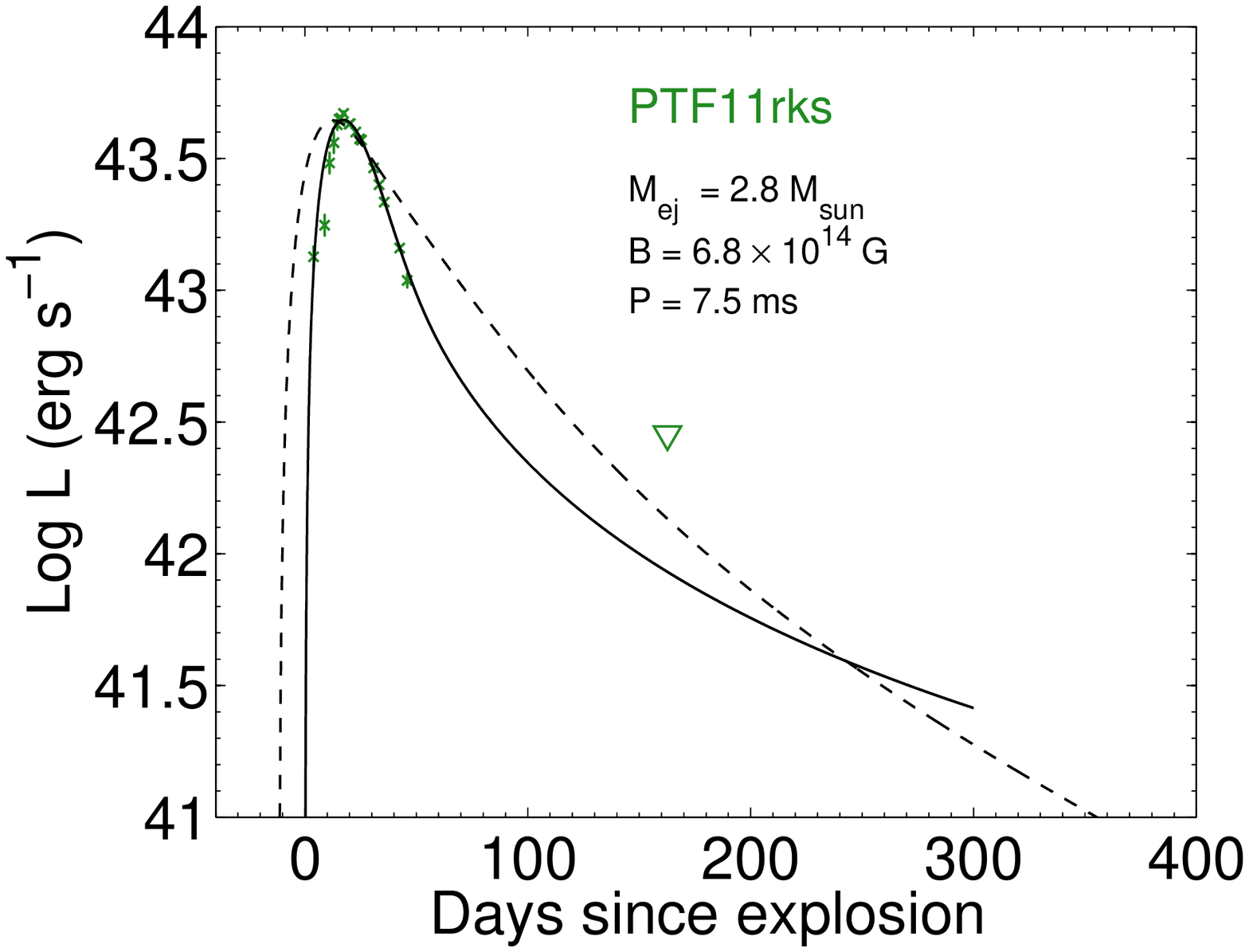}
\includegraphics[width=\columnwidth]{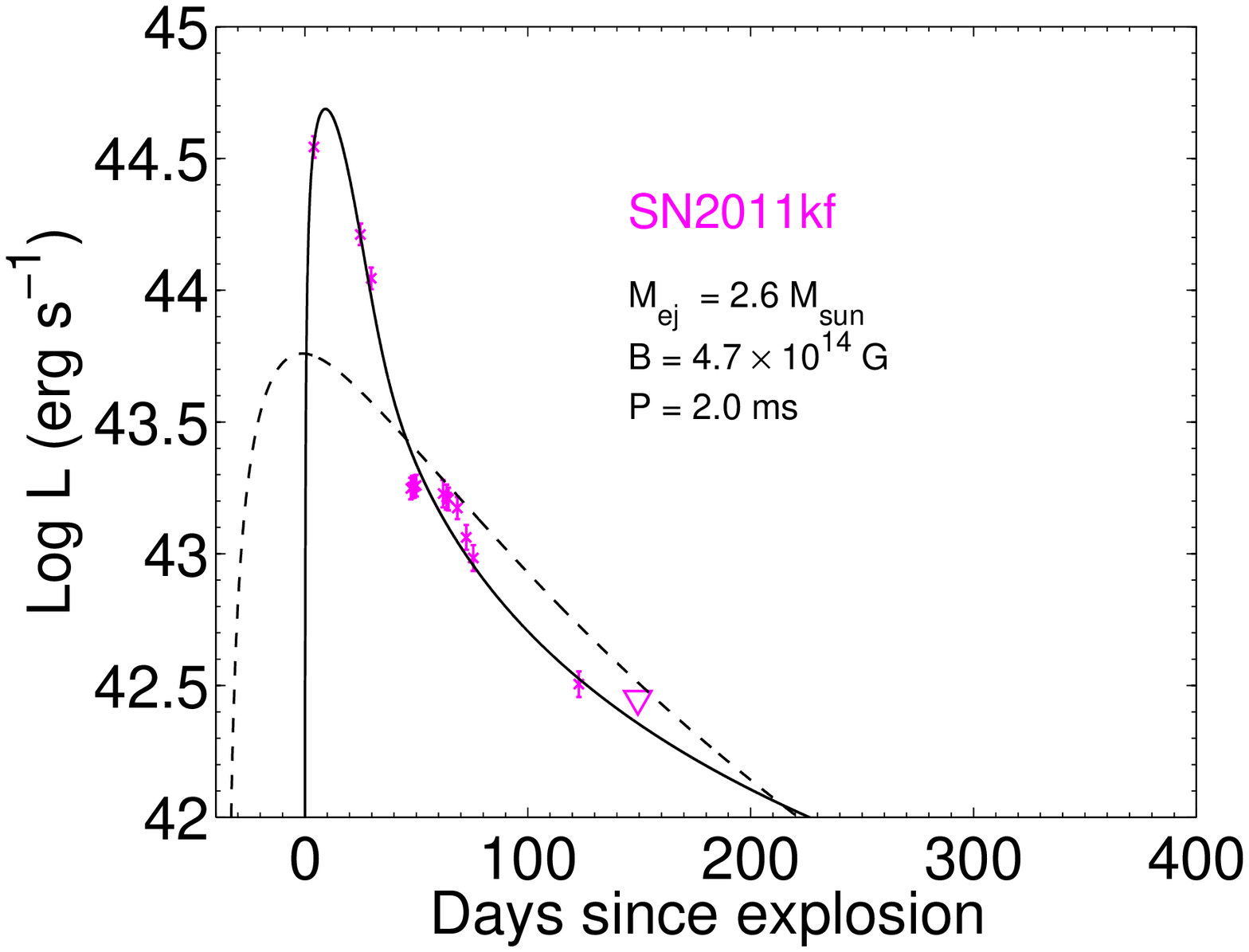}\\
\includegraphics[width=\columnwidth]{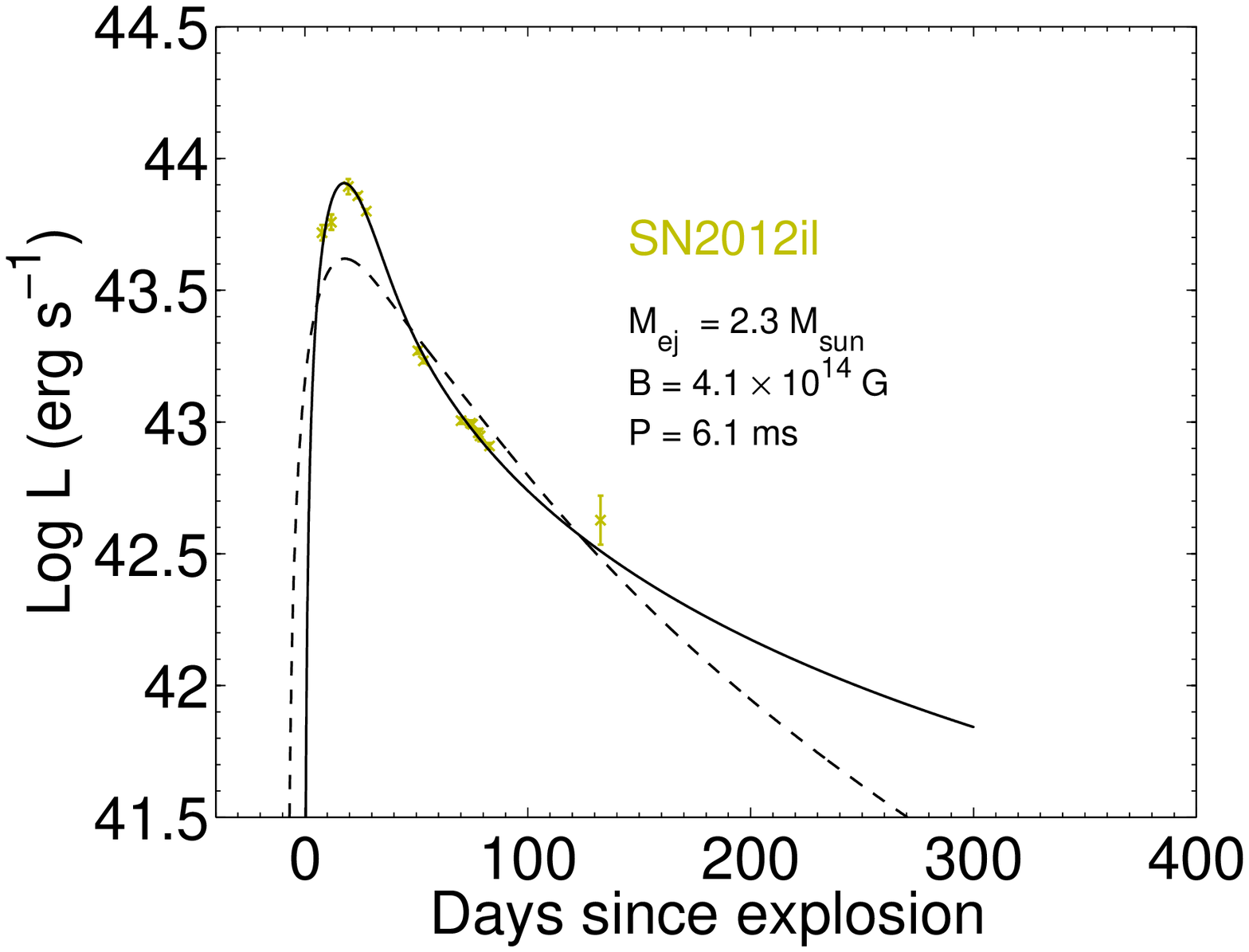}
\includegraphics[width=\columnwidth]{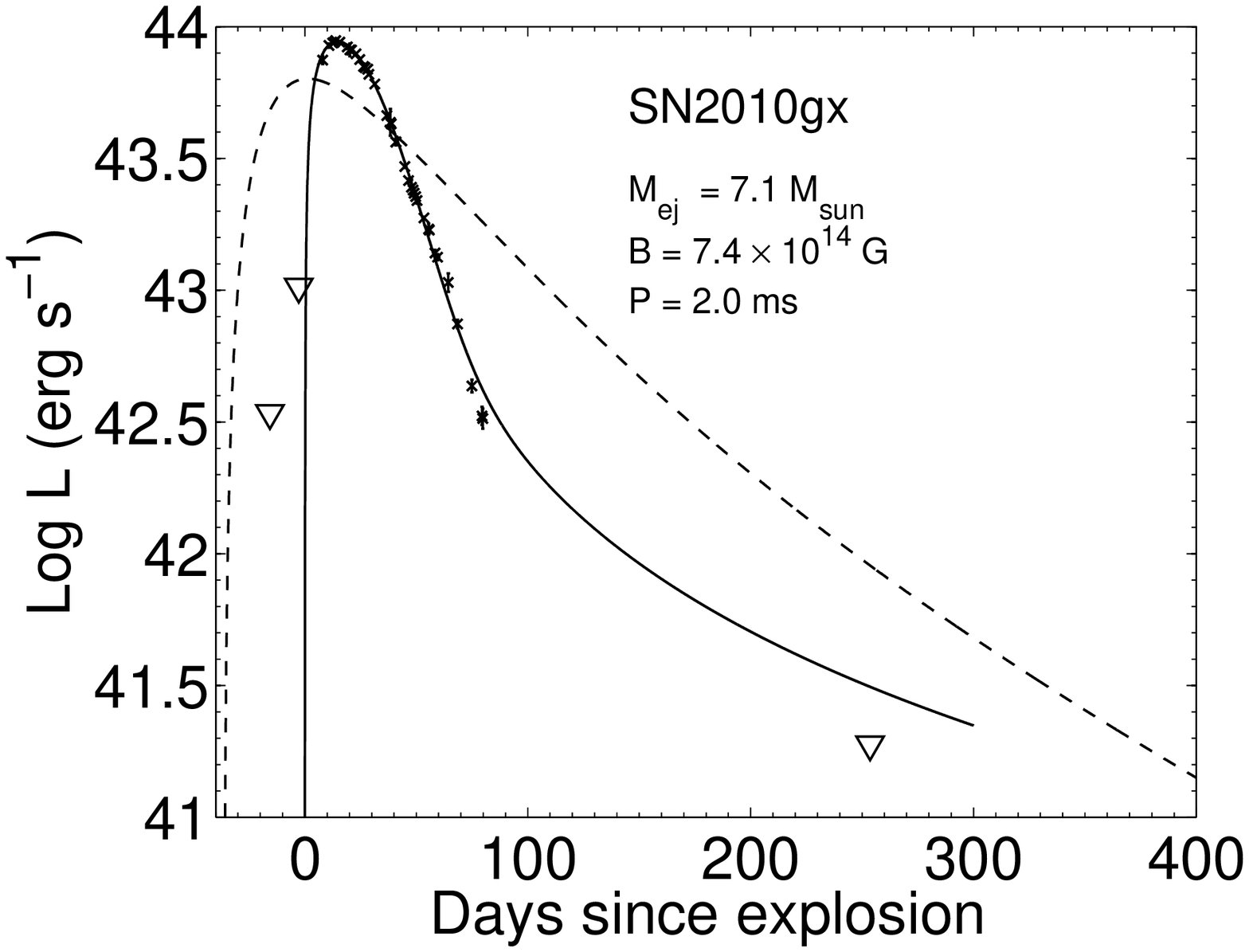}
\caption{The bolometric light curves of \paj\/, \xk, \rks, \css, \fo\/ and SN 2010gx and the diffusion semi-analytical model that best fit the light curve (black solid line). Limits are shown as empty upside down triangles. Best fit of the \ni\/ model (black dashed line) for each SN is also reported. The $x$-axes refer to magnetar models.}
\label{fig:fits}
\end{figure*}







A variant of the previous scenario was proposed by \citet{bl10}
claiming high luminosities from a radiative shock in a massive
C-O shell. The shells have radii and density profiles that are similar
to the dense wind of \citet{ci11}, with a density gradient of
$\rho(r)\propto r^{-1.8}$ and radii of the order $10^{5}$ \R. Considering a total
ejecta mass $M_{\rm ej}\lesssim1$~\M, which collides with a shell of mass
5~\M, these models have initial energies ($2-4\times10^{51}$ erg)
higher than those retrieved by the expansion velocities during the
pseudo nebular phase in our set of spectra ($\sim1\times10^{51}$ erg).
Moreover, the light curves are too shallow and they are not able to
reproduce the SN 2010gx decline after 30d post maximum light.
We find no unequivocal signs of interaction in the spectra of the
objects. 

In Sects.~\ref{sec:lc} and \ref{sec:bol}  we have presented detections
of the SNe at later times than published to date, due to our focus on
low redshift candidates. We detect a flattening of the lightcurve and
a  tail phase in four out of five transients and this slope appears to
be consistent with \co\/ decay. 
In summary these interaction scenarios can 
reproduce the peak energy and diffusion time. However the ejecta/shell
velocity should be lower (by about a factor of 2) than those observed. 
And one still needs another power source for the late time luminosity
that we  now detect in these SL-SNe Ic.

As shown previously by \cite{cw09,pa10,qu11,ch11,cw12a,le12}, the lightcurve peaks
cannot be fit with a physically plausible \ni\/ diffusion model like
normal SNe Ic, and our similarly shaped lighcurves result in the same
conclusion.  If the tail phase was actually due to \co\/ powering,
then approximately 1-4~\M\/ of \ni\/  would be required.  But this
would not be enough to power the peak luminosity solely through
radioactive heating. 
In Fig.~\ref{fig:fits} we also show the best fitting \ni-powered models, under the 
assumptions that the \ni\/ mass must be $<$50\% ejecta mass, and that the ejecta 
velocities are less than 15000 \kms. The first assumption is based on 
the implausibility of a pure \ni\/ ejecta; \citet{um08} found 
typical \ni\/ masses which are at most 20\% of the ejecta, while if 
the ejecta was comprised of more  \ni\/ than this, we would expect to see spectra dominated by Fe-group 
rather than intermediate mass elements. The 
velocity constraint is motivated by the observed velocities in our sample.
Thus we derived kinetic energies of $6.2\lesssim
  E$($10^{51}$erg) $\lesssim15.0$, ejecta masses of $5.9\lesssim M_{\rm
    ej}$(\M) $\lesssim 13.0$ and \ni\/ masses of $2.9\lesssim M_{\rm
    ^{56}{\rm Ni}}$(\M) $\lesssim 6.0$ (see Appendix~\ref{sec:nifit} for further details).
From the fits appear that no 
physical and consistent solutions for \ni\/ heating can be determined, 
as found by previous authors \citep{pa10,qu11,ch11,le12}. 
One could invoke a combination of CSM interaction to explain the peak
luminosity and then \ni\/ masses of 1-4~\M\/ to account for the tail
phases. But as discussed above, this requires full $\gamma$-ray
trapping
and somewhat fine tuning of the two scenarios to work in unison.  

\begin{figure}
\vspace{0pt}
\includegraphics[width=4.2cm,trim=13mm 9mm 11mm 0mm,clip]{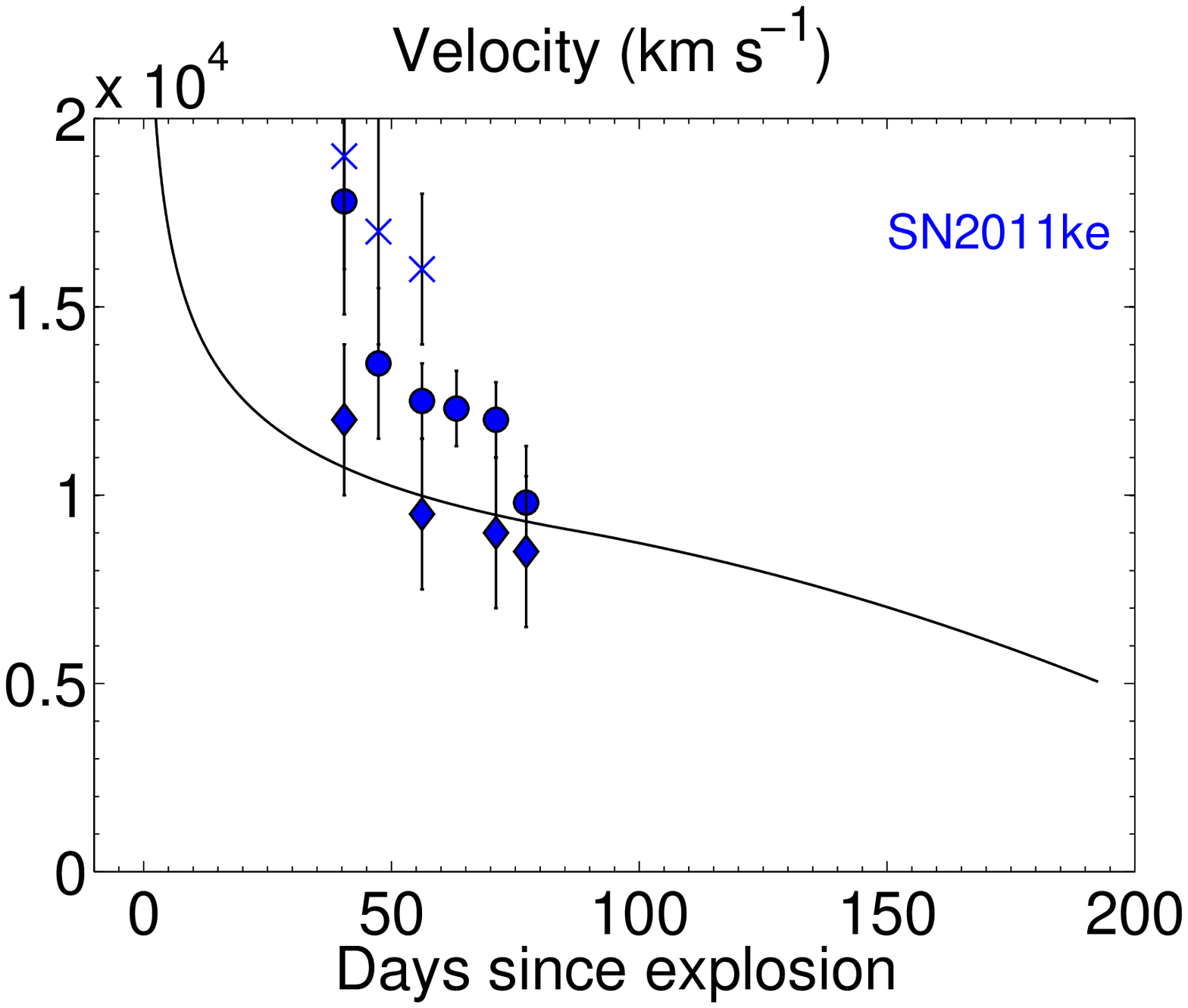}
\includegraphics[width=4.2cm,trim=13mm 9mm 11mm 0mm,clip]{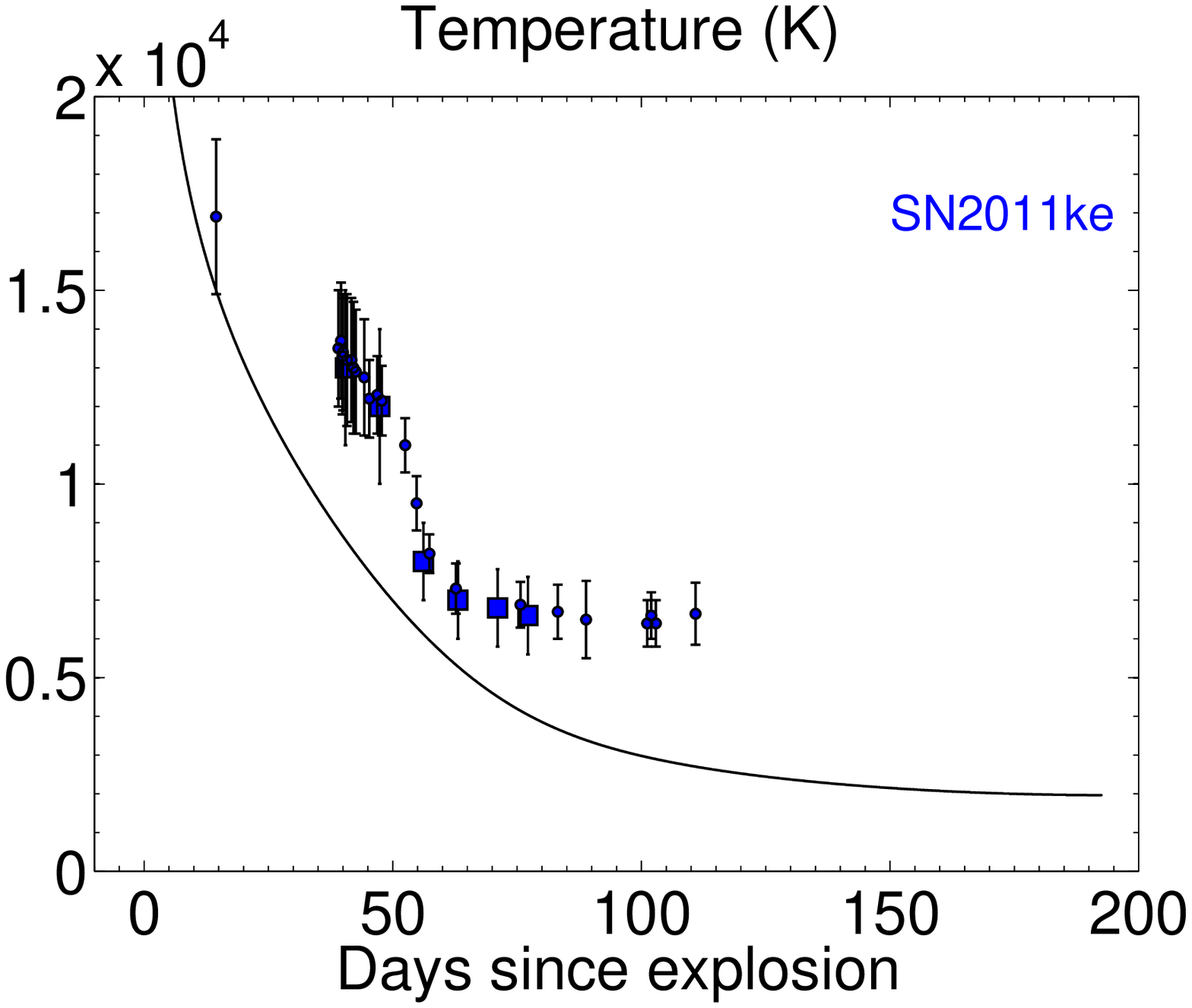}\\
\includegraphics[width=4.2cm,trim=13mm 9mm 11mm 0mm,clip]{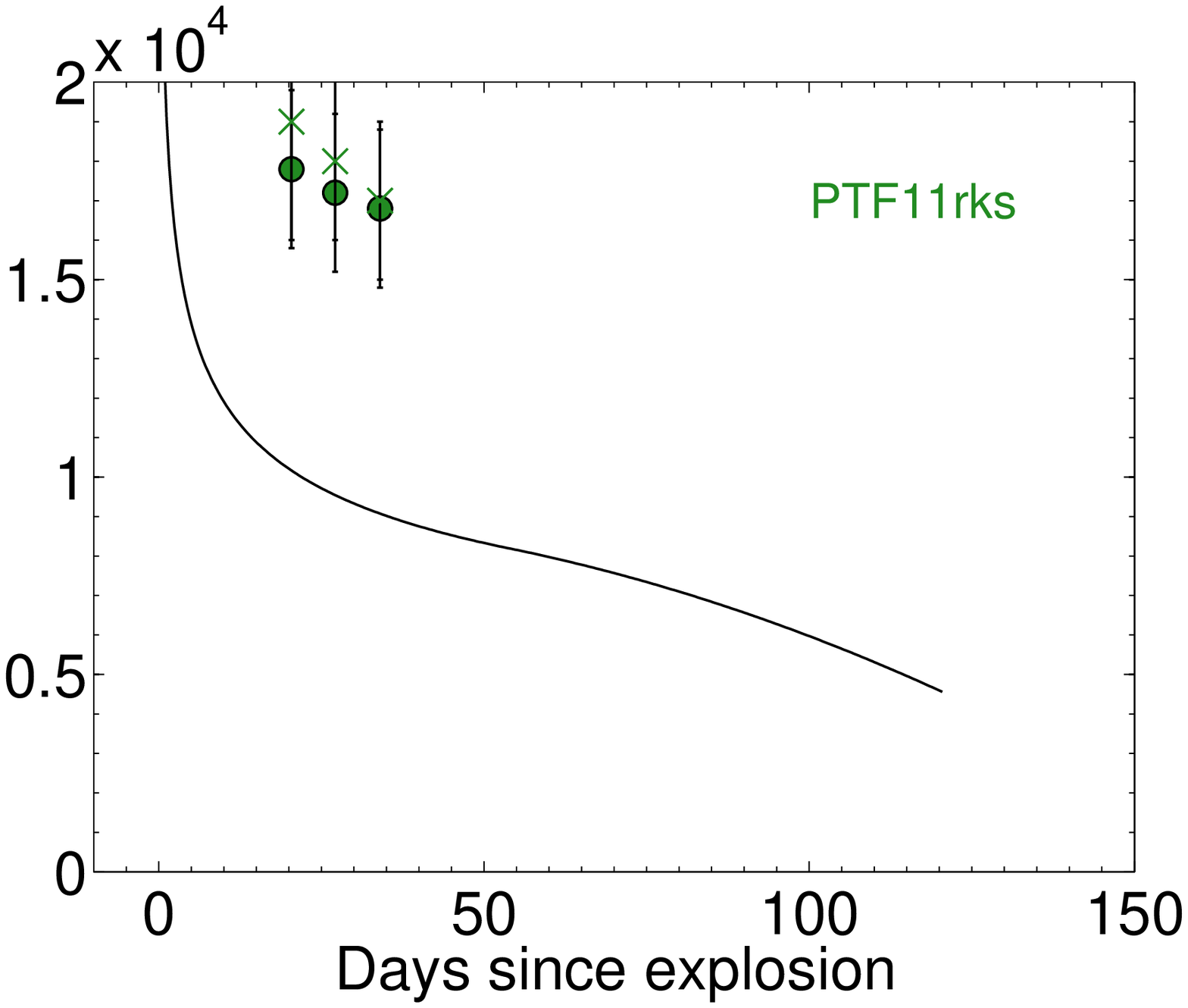}
\includegraphics[width=4.2cm,trim=13mm 9mm 11mm 0mm,clip]{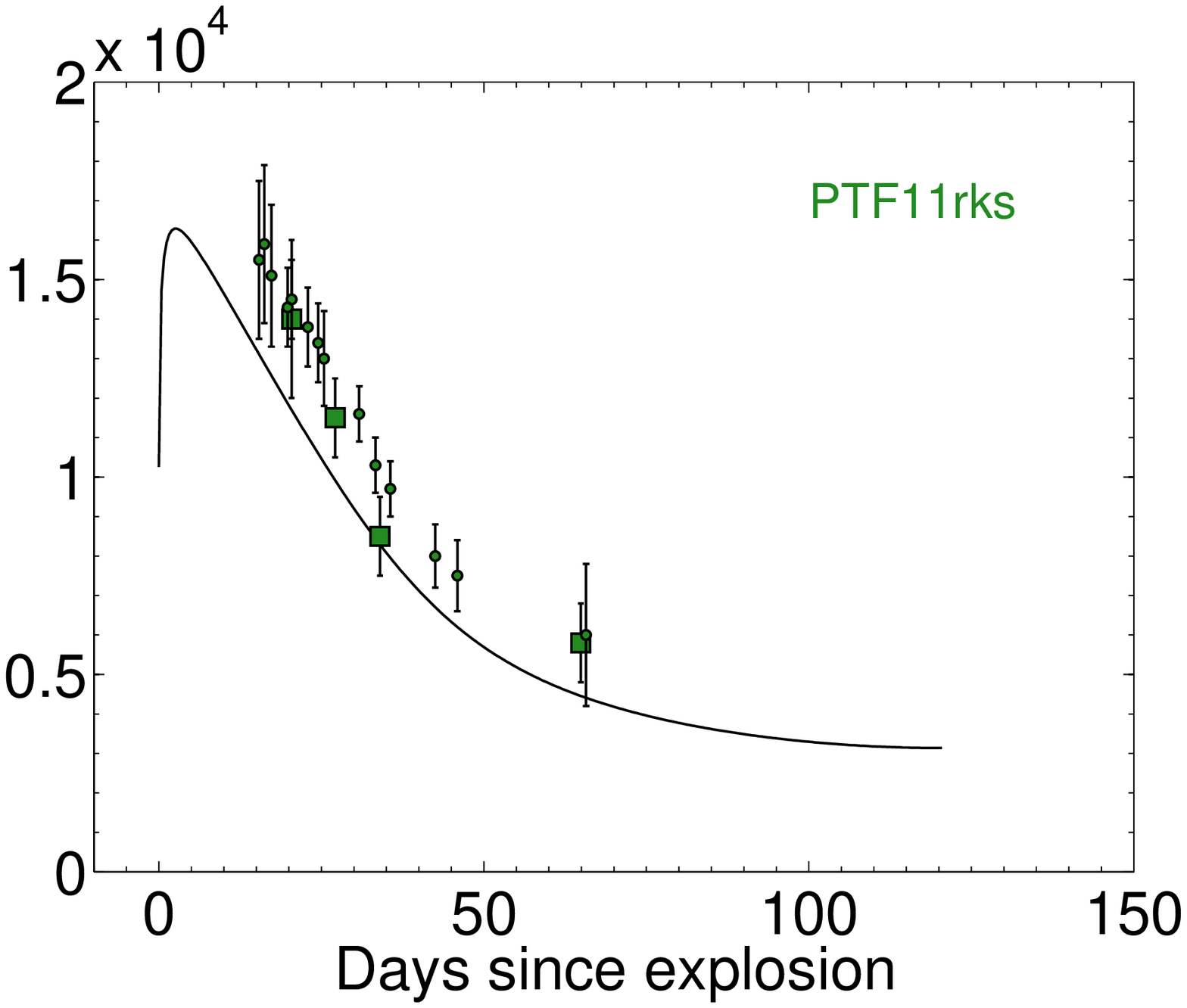}\\
\includegraphics[width=4.2cm,trim=13mm 9mm 11mm 0mm,clip]{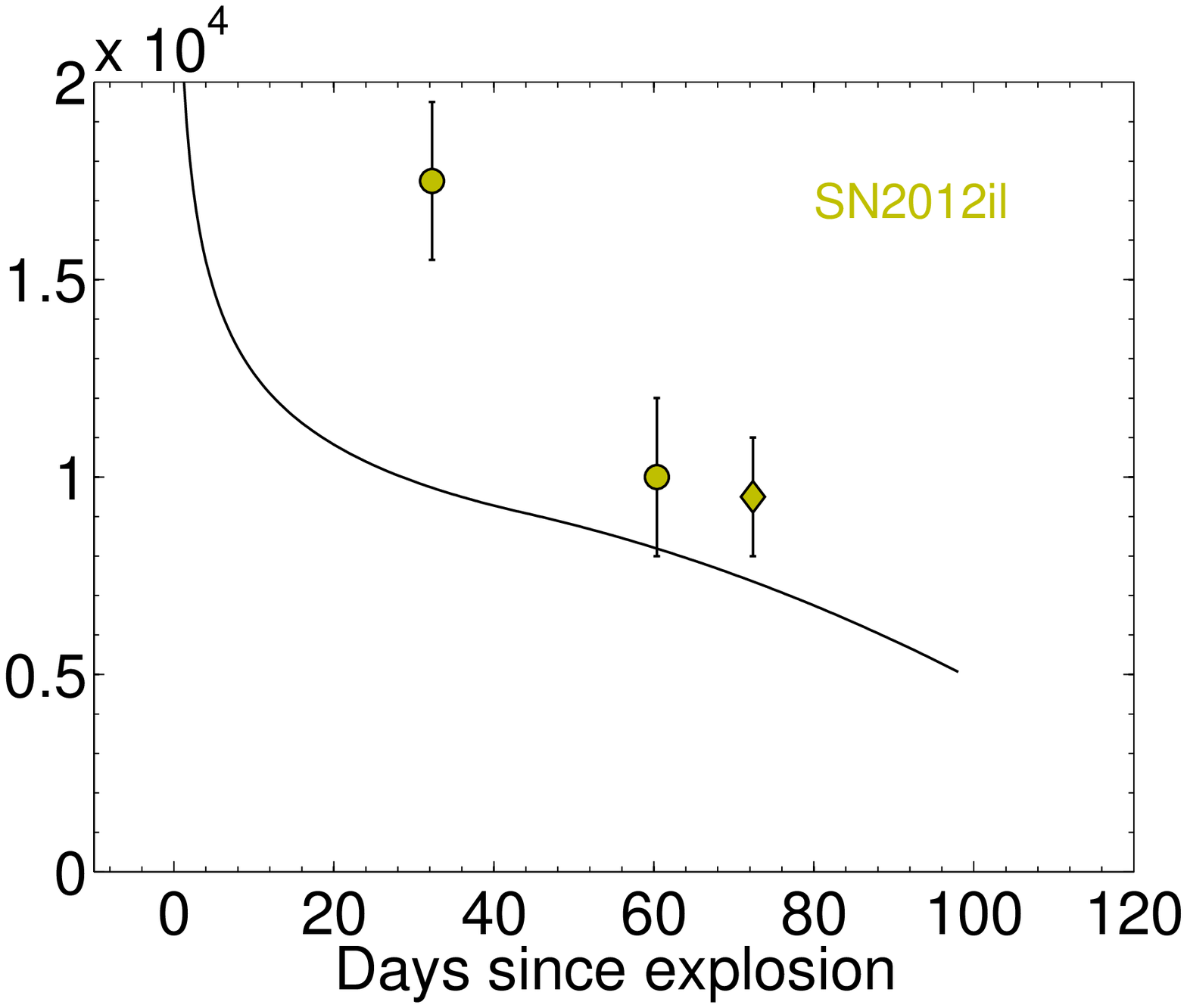}
\includegraphics[width=4.2cm,trim=13mm 9mm 11mm 0mm,clip]{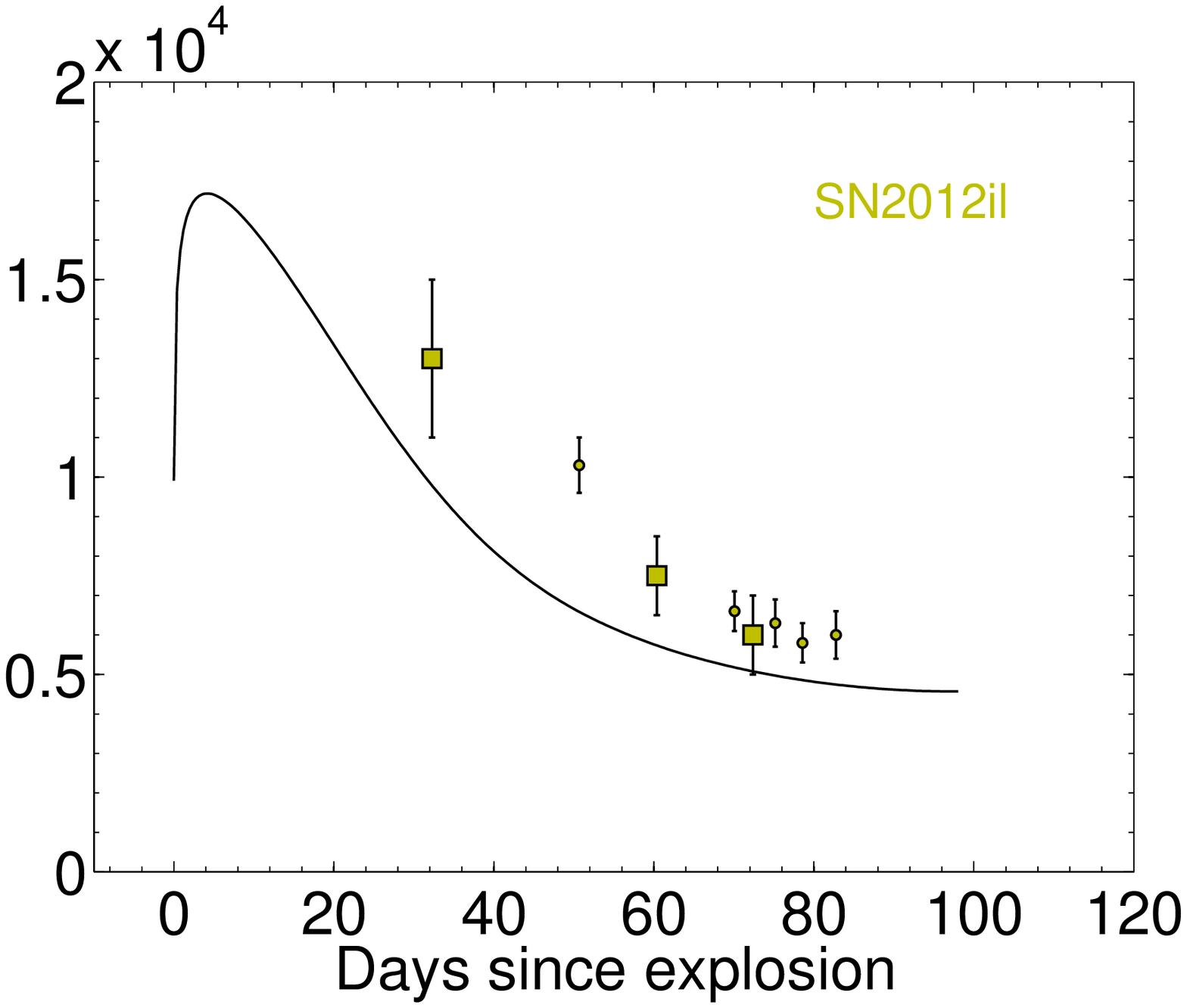}\\
\includegraphics[width=4.2cm,trim=13mm 9mm 11mm 0mm,clip]{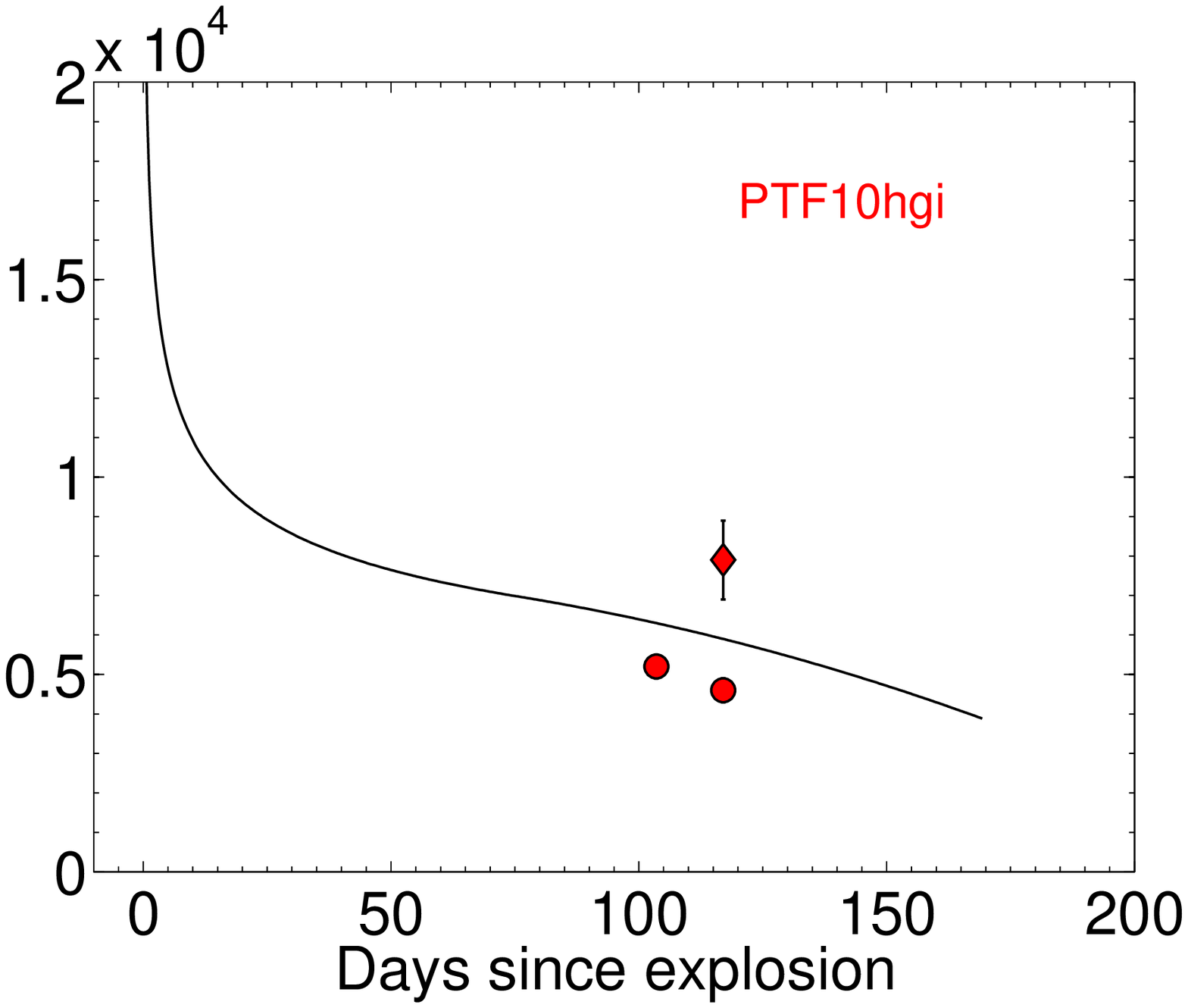}
\includegraphics[width=4.2cm,trim=13mm 9mm 11mm 0mm,clip]{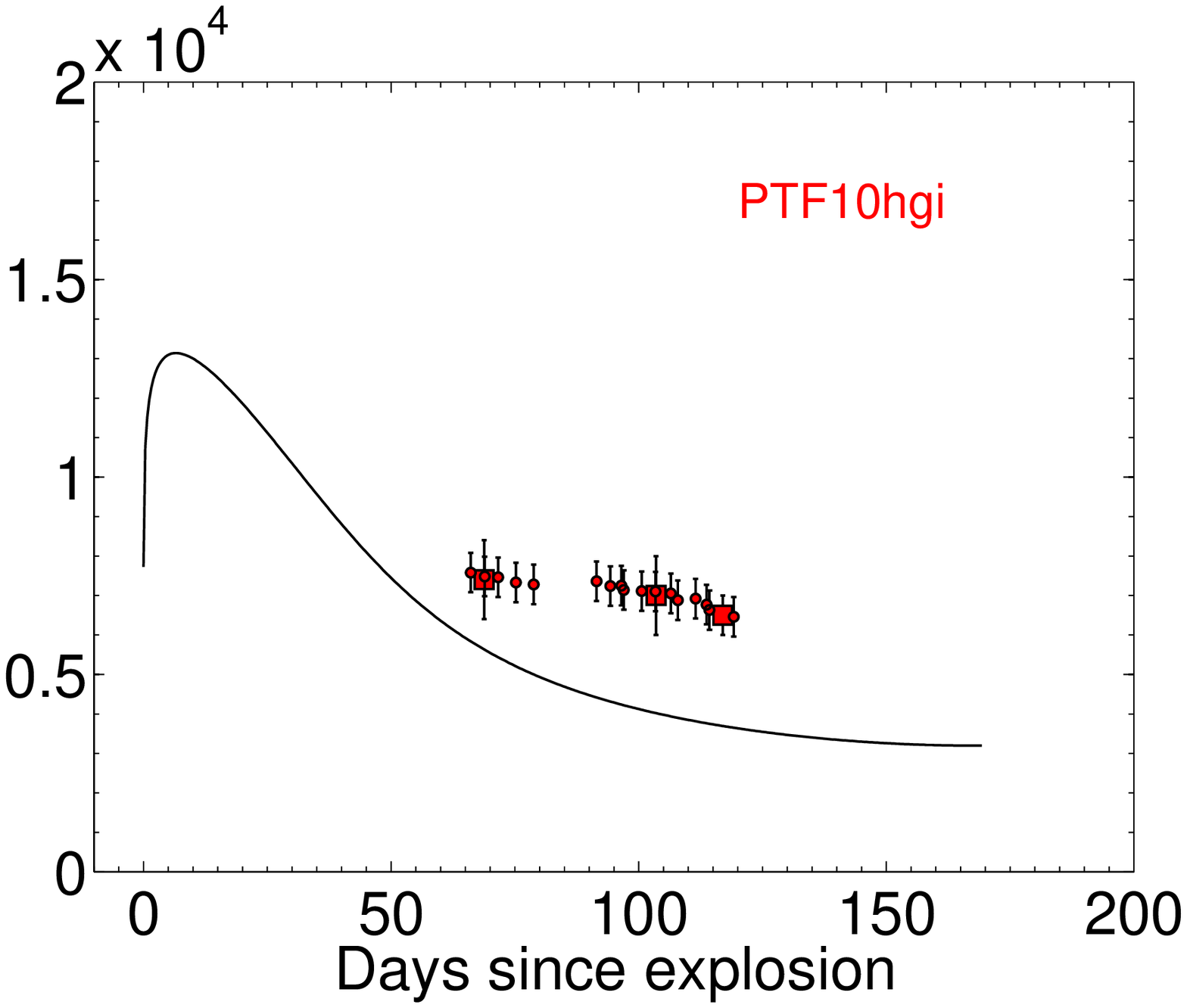}\\
\includegraphics[width=4.2cm,trim=13mm 9mm 11mm 0mm,clip]{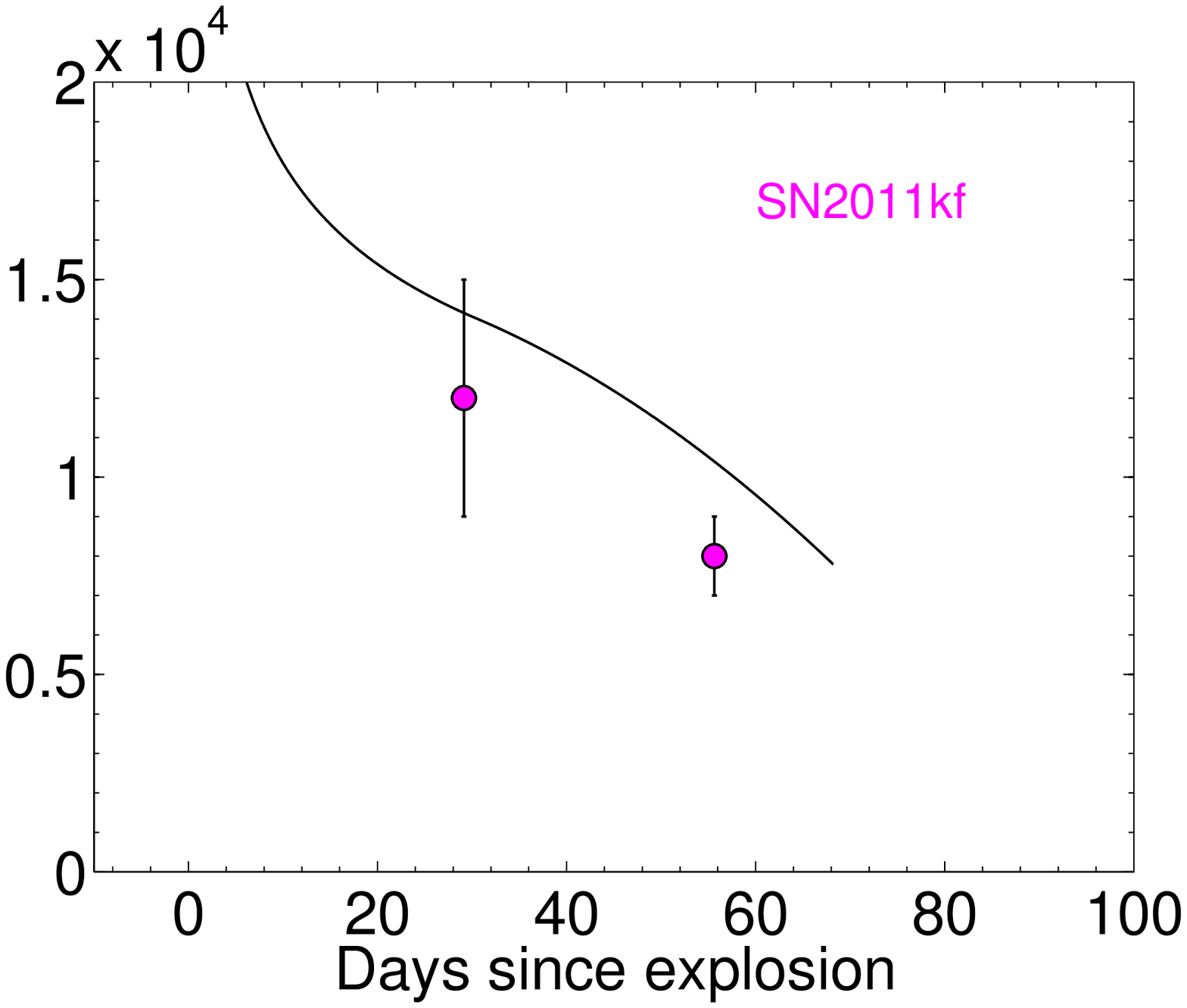}
\includegraphics[width=4.2cm,trim=13mm 9mm 11mm 0mm,clip]{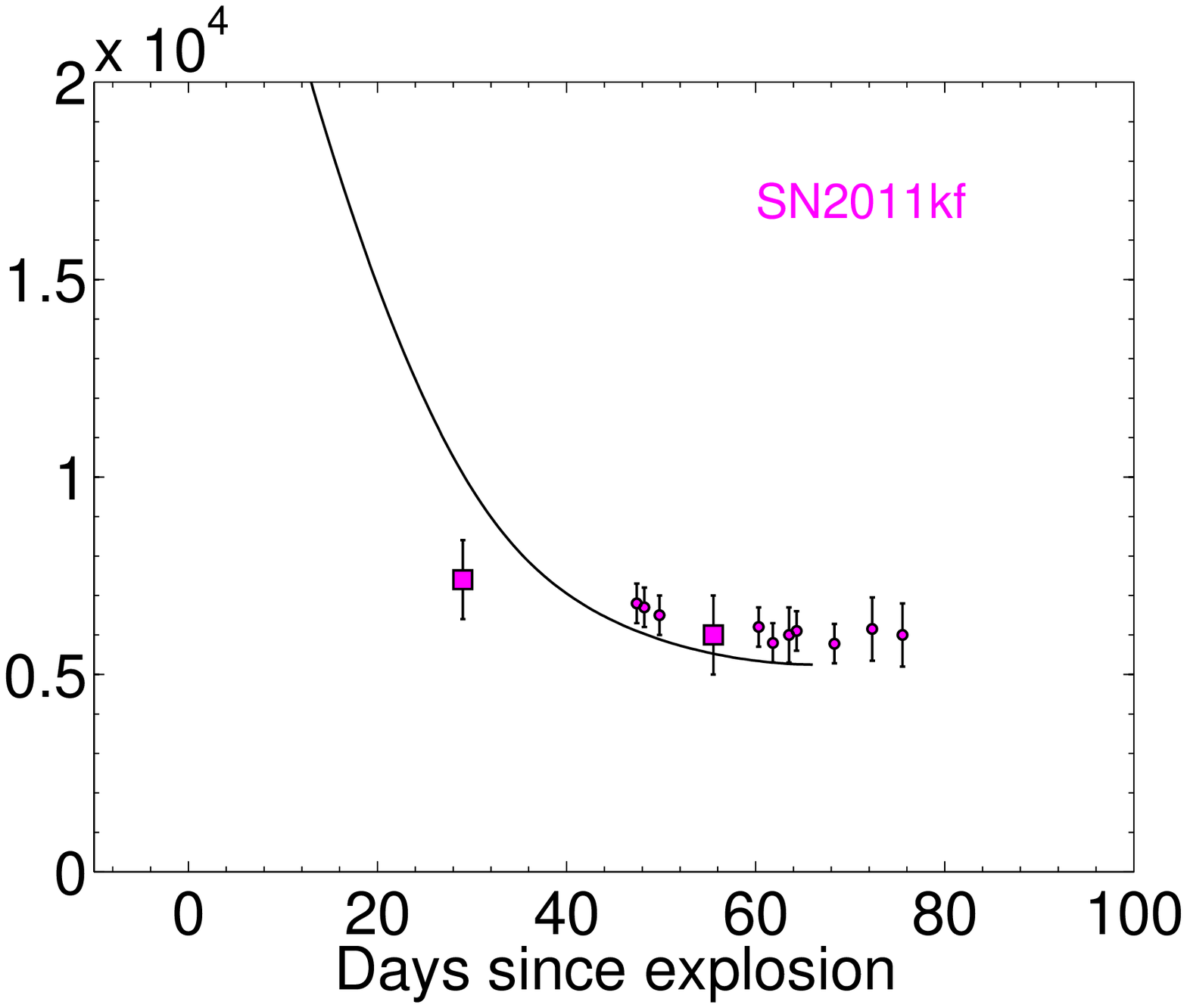}\\
\includegraphics[width=4.2cm,trim=13mm 0mm 11mm 0mm,clip]{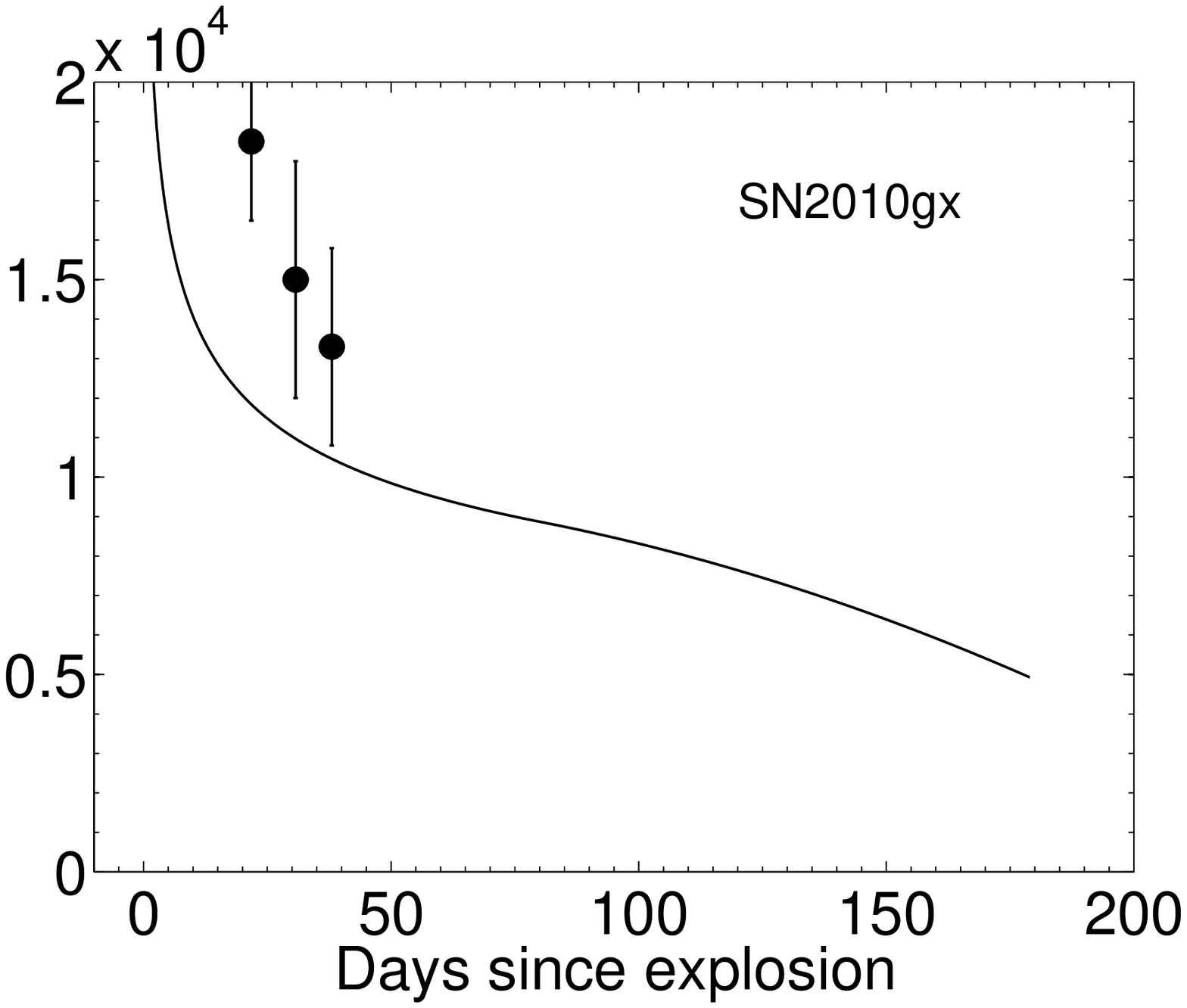}
\includegraphics[width=4.2cm,trim=13mm 0mm 11mm 0mm,clip]{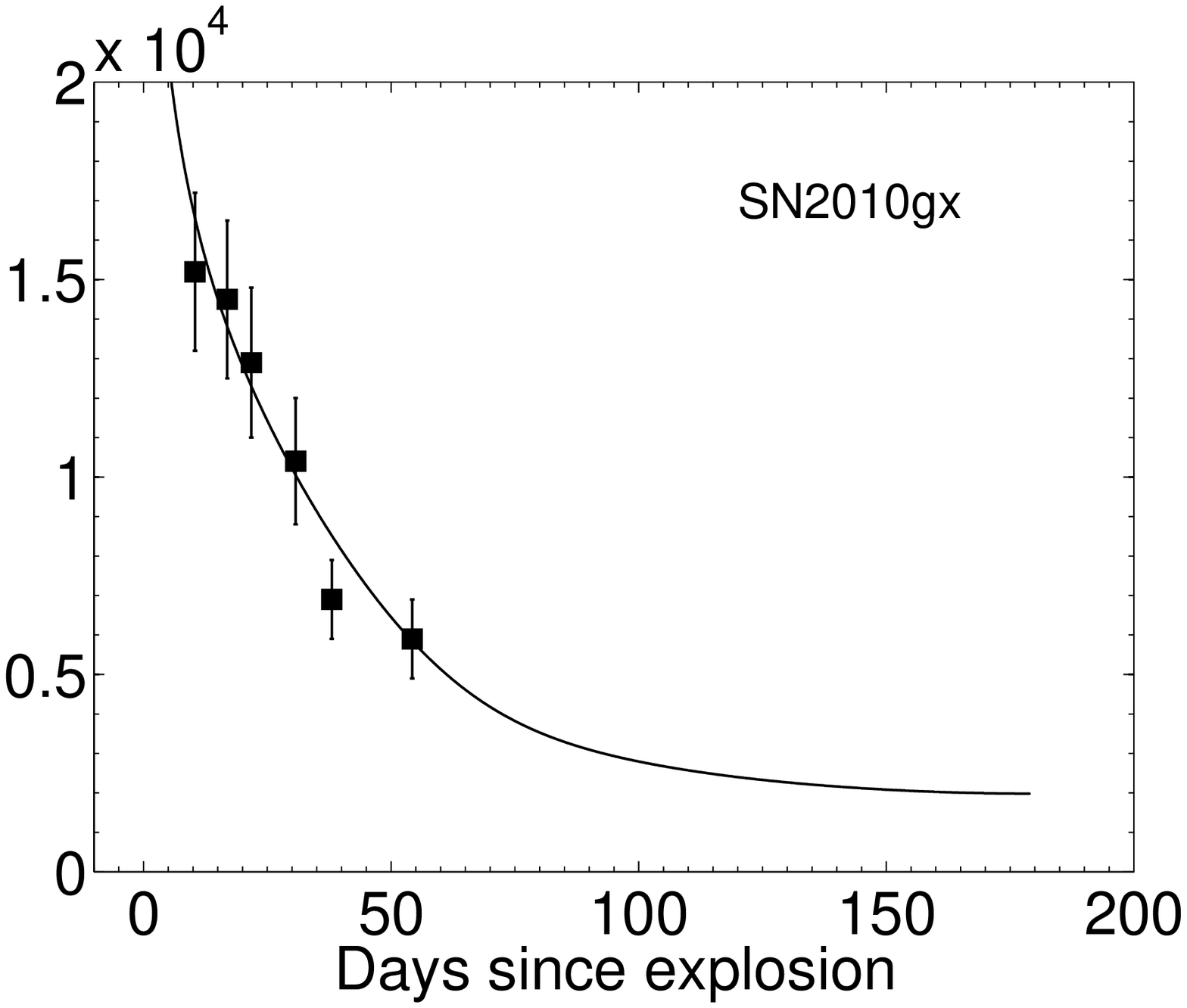}\\
\caption{Left: photospheric velocities of \xk\/, \rks\/, \fo\/, \paj\/, \css\/, SN 2010gx and and their values in the semi-analytical models best fitting the light curve data. Different symbols correspond to different ions as in Fig.~\ref{fig:vel} Right: temperature evolution of \xk\/, \rks\/, \fo\/, \paj\/, \css\/, SN 2010gx and their values in the semi-analytical models best fitting the light curve data.}
\label{fig:fitvelT}
\end{figure}

\subsection{Magnetar model}\label{sec:model}

\citet{kb10,wo10,2012MNRAS.426L..76D} have already proposed 
that a rapidly spinning magnetar can deposit its rotational energy
into a supernova explosion and significantly enhance the luminosity. 
This appears to be an appealing scenario as the model is fairly  
simple, and this additional power source can potentially transform a
canonical Type Ic SN into a SL-SN Ic. To investigate this further and 
quantitatively compare our extensive lightcurves with this model
we have derived a semi-analytical diffusion models. 
We use standard diffusion equations derived by \citet{ar82} and add
magnetar powering \cite[as in][]{kb10} to fit the light curves of our five
objects. A full description can be found in
Appendix~\ref{sec:mag}. 
Assuming full trapping of the magnetar radiation\footnote{
Which is the case if the SED of the magnetar is dominated by X-ray radiation,
as in the Crab pulsar for instance \citep{crab}.}, the ejecta
mass $M_{\rm ej}$, explosion energy $E_{\rm k}$, and the opacity $\kappa$ only
influence the bolometric light curve through their combined effect on
the diffusion time-scale parameter (see Appendix~\ref{sec:mag})

\begin{deluxetable*}{lccccc||ccc}
\tablewidth{0pt}
\tablecaption{Best-fit parameters for magnetar modelling of the bolometric light curves and $\chi^2_{\rm red}$ value on the left. Derived parameters on the right.\label{table:fit}}
\tablehead{\colhead{Object}& \colhead{$\tau_{\rm m}$}& \colhead{$B_{14}$} & \colhead{$P_{\rm ms}$} & \colhead{$t_{\rm 0}$}& \colhead{$\chi^2_{\rm red}$} & \colhead{$E^{\rm mag}$} & \colhead{$M_{\rm ej}$}& \colhead{$V_{\rm core}^{\rm final}$} \\
\colhead{}& \colhead{(day)}& \colhead{} & \colhead{} & \colhead{(MJD)} & \colhead{} & \colhead{($10^{51}$ erg)} & \colhead{(\M)} & \colhead{(\kms)}}
\startdata
\xk\/ & 35.0 & 6.4 & 1.7 & 55650.65&1.8& 6.9 & 8.6 & 12400\\
\css\/      & 15.4  & 4.7 & 2.0 & 55920.65& 5.8& 5.0 & 2.6 & 19600\\
\fo\/     & 18.4  & 4.1 & 6.1 & 55918.56& 3.9& 0.5 & 2.3 & 10500\\
\paj\/    & 26.2  & 3.6 & 7.2 & 55322.78& 0.5& 0.4 & 3.9& 7700\\
\rks\/    & 21.0 & 6.8 & 7.5 & 55912.11 & 5.0& 3.6 & 2.8& 16500\\
SN 2010gx   & 32.4 & 7.4 & 2.0 & 55269.22 & 3.9& 5.0 & 7.1 & 11900
\enddata
\end{deluxetable*}

\begin{equation}
\tau_{\rm m} = 10~{\rm d} \left(\frac{M_{\rm ej}}{1~M_\odot}\right)^{3/4}{ \left(\frac{E_{\rm k}}{10^{51}~{\rm erg}}\right)^{-1/4}\left(\frac{\kappa}{0.1~{\rm cm^{2}g^{-1}} }\right)}^{1/2}~.
\end{equation}

\begin{figure}
\includegraphics[width=\columnwidth]{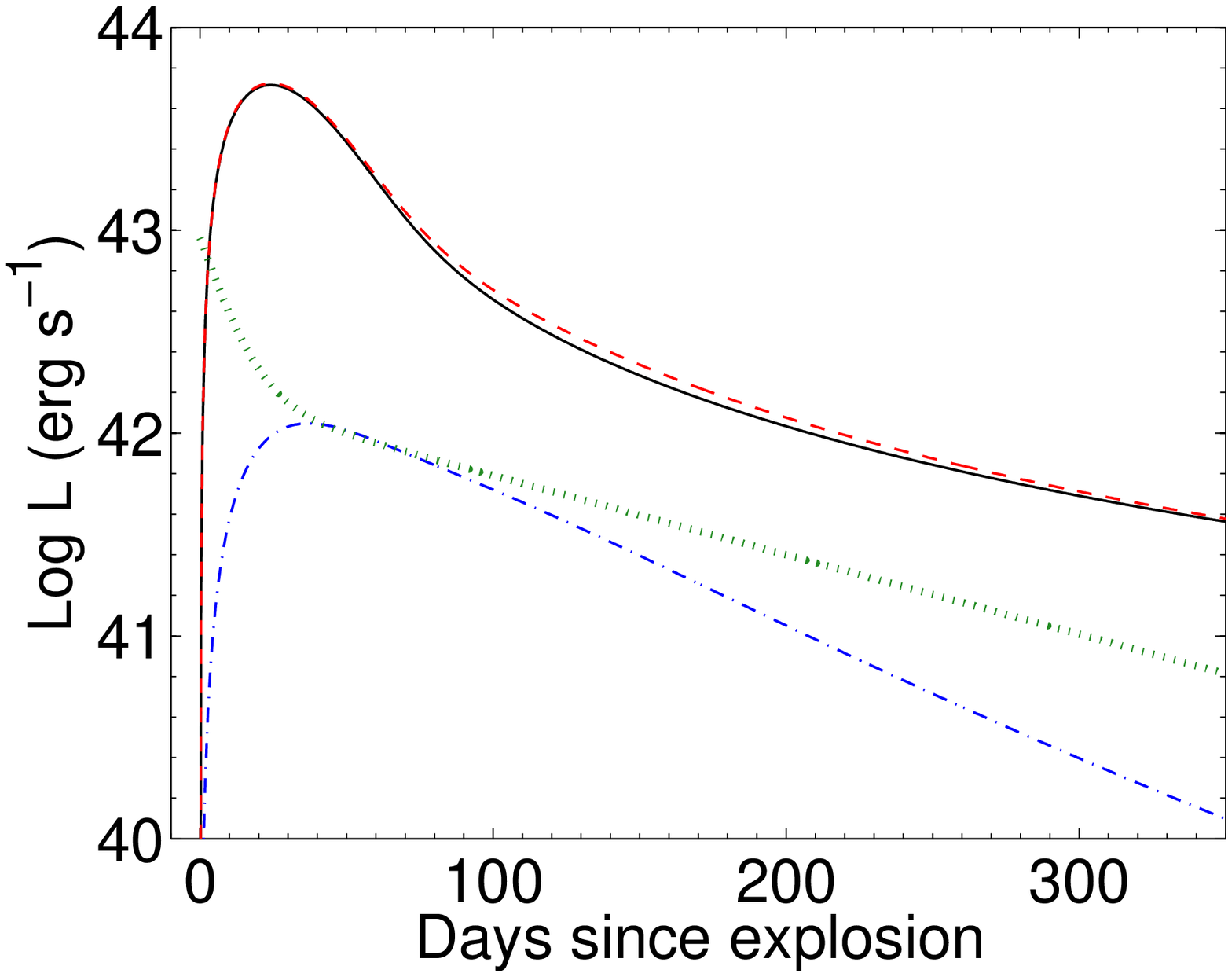}\\
\includegraphics[width=\columnwidth]{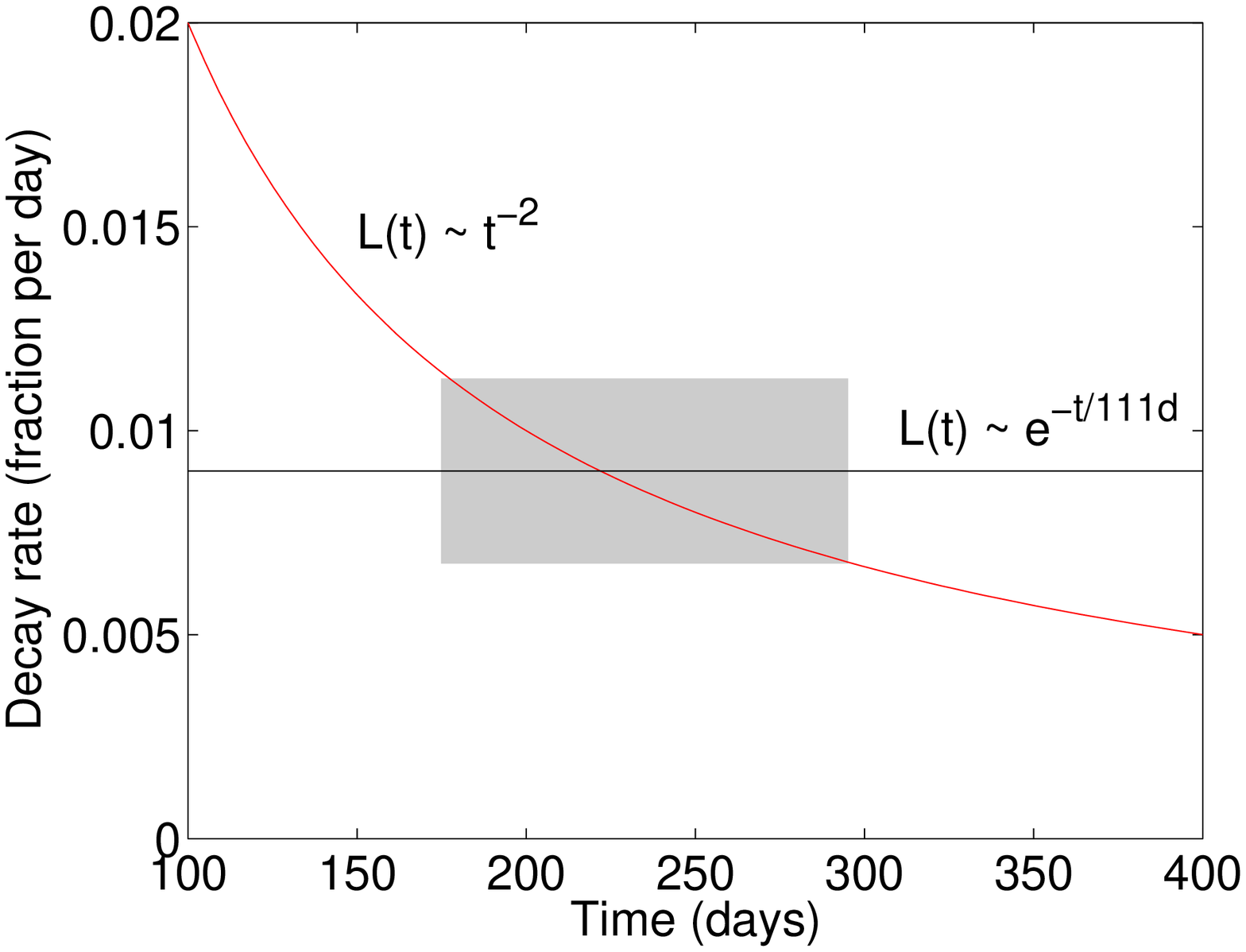}
\caption{Top: Bolometric light curves for $B_{14} = 5$, $P_{\rm ms} = 5$, $M_{\rm ej} = 5~M_\odot$ without \ni\/ contribution (black solid) and with 
M(\ni) = 0.1 \M\/ (red dashed). Also shown is the total energy emitted by M(\ni) = 0.1 \M\/ (green dotted) and the light curve produced by this amount of \ni\/ in a 5 \M\/ ejecta with E$_{\rm k}$~=~$10^{51}$~erg (blue dot-dashed). Bottom: Comparison between magnetar and \co\/ decay rate. The grey box is the region where the two slopes are similar to within 25\%.}
\label{fig:compfit}
\end{figure}

The magnetar luminosity depends on two parameters, 
 the magnetic field strength $B_{14}$  (expressed in terms of
 $10^{14}$G) and the initial spin period $P_{\rm ms}$ (in
 milliseconds). 
Combined with the explosion date $t_0$, we therefore have four free parameters to fit. Tab.~\ref{table:fit} lists the best-fit parameters for each object, and Fig.~\ref{fig:fits} 
shows the fits. As the $\chi^2$ fitting gives good matches to models {\it without} 
\ni\/, we have no need to introduce \ni\/ as an additional free parameter. 
All the models have M(\ni)~=~0~\M\/ and we investigated the sensitivity to the assumed \ni\/ mass by recomputing
the fits including 0.1 \M\/ of \ni\/ in the ejecta (the typical \ni\/ yield in core collapse SNe). We find virtually
the same fit parameters as we do without the nickel. As can be seen from Fig.~\ref{fig:compfit} (top), 
the two magnetar models ($B_{14} = 5$, $P_{\rm ms} = 5$, $M_{\rm ej} = 5~M_\odot$) with (red dashed)
and without \ni\/ (black solid) are similar. 
The late decline rate in the magnetar model ($>$100d) is actually
quite similar to the \co\ decay rate as shown in Fig.~\ref{fig:compfit} (bottom), but fully trapped 
$\gamma$-rays are required. As Fig.~\ref{fig:compfit} shows, this full
trapping is different from the typical light curve of a
radioactivity-powered Type Ib/c SN in which full trapping is not
observed. Recently, \citet{de12} showed that fall-back accretion
can give a similar asymptotic behaviour of the light curve (L$_{\rm t}\propto t^{-5/3}$)
which is a scenario that would need further investigation.

The lightcurves are quite well reproduced with appropriate choices of
parameters, and the tail phase luminosities that we measure can also 
be explained with this model. 
 The diffusion time scale parameters are between 15-35 days, which corresponds to ejecta masses of 2-9 $M_\odot$ for $\kappa = 0.1$ cm$^2$ g$^{-1}$ (see Appendix~\ref{sec:mag} for further details about $\kappa$) and

\begin{equation}\label{eq:apa2}
E_{\rm k}=10^{51} + \frac{1}{2} \left( E^{\rm mag}-E^{\rm rad} \right) {\rm (erg)}
\end{equation}

where $E^{\rm mag}$ is the total energy of the magnetar and $E^{\rm rad}$ is the total radiated energy of the SN.  
We use a factor of 1/2  for an approximation of the average kinetic energy over the magnetar energy input phase, which we show in the Appendix~\ref{sec:test} produces good agreement with more detailed time dependent calculations of $E_{\rm k}$ \footnote{
We also investigated the sensitivity to this assumption by computing masses (as well as photospheric velocities and temperatures) also from
$E_{\rm k} = 10^{51} + 0.4 E^{\rm mag}$
where the 40\% conversion to kinetic energy is a typical value \citep{wo10}.
We found very small differences in the derived quantities.}. 
These ejecta masses are consistent with the ones derived for
radioactivity-powered Type Ib/c ejecta \citep{Ensman1988,
Shigeyama1990_2,va11,dro11,el13}. Furthermore,
Fig.~\ref{fig:fitvelT} shows that the evolution of photospheric
velocities and temperatures match the observed ones reasonably well. 
While the velocity and temperature evolution as estimated from our 
models are crude, this good agreement is an important test for the physical self-consistency of the 
model.
From the lightcurve fits to ejecta mass and kinetic energy, we estimate the ejecta velocities (V$_{\rm core}$, see eq.~\ref{eq:vel})
which we compare to the observed velocities of the emission lines (Ca~H\&K and Mg~I] $\lambda$4571, see Tab.~\ref{table:vel}), finding them reasonably similar. 

From the plot of bolometric light curves shown in Fig.\ref{fig:bol}, it appears 
that there is a gap between the faintest SL-SNe Ic, and the brightest 
normal Type Ic. One possibility is that this is an observational
bias.  As SL-SNe Ic are intrinsically rare  \citep{qu11,qu13}, we find
them at  moderate redshifts simply due to the large survey volume 
required to find them. If intermediate objects were also as rare, then 
we would require wider field searches to encompass more local volume
as they may evade detection at the higher redshifts of known
population of SL-SNe. 
An alternative explanation is that the mechanism powering the 
SL-SNe has a minimum energy. From the semi-analytic magnetar model of 
\citet{kb10} which we have used to match our light curves, we see 
that the peak magnetar luminosity is inversely proportional to the 
square of the magnetic field \citep[eq. 4][]{kb10}. Hence, the 
question arises, could the apparent lower luminosity limit to the SL-SNe 
Ic be caused by some physical upper limit to the magnetic field in a 
magnetar? For our faintest SL-SNe, and using a minimum plausible ejecta 
mass of 1 \M, we determine an upper limit to the magnetar B-field of 
$B<1.4\times10^{15}$ G from Eq. 4 of \citet{kb10}. If we consider that 
\citet{kb10} assume an angle between the B field and the spin axis 
of the magnetar of 45$^{\mathrm o}$, then a factor of two higher than this value could be a plausible maximum. We hence adopt $B<3\times10^{15}$ G.

The most conservative limit we can set is that the magnetic energy in 
the magnetar must be less than the gravitational binding energy of the 
neutron star \citep{ch53}. This implies that

\begin{equation}
B<10^{18} {\rm \left(\frac{M_{NS}}{1.4~M_{\odot}}\right)} {\rm \left(\frac{R_{NS}}{10~km}\right)}^{-2} {\rm G}
\end{equation}

which does not set a particularly stringent upper limit on the B-field.
This limit is consistent with the $B$ values retrieved from all the galactic magnetars studied so far which have
$B\sim10^{14}-10^{15}$ G \citep{wt06}.
However, magnetic fields are known to be a possible source of braking in stars \citep{me11}, while the magnetar models require an extreme 
magnetic field and fast rotational period.
An explanation for both a small rotational period and a large magnetic field could be a large-scale helical dynamo, 
that is possible when the rotation period is comparable to the timescale of the convective motions \citep{du92}.

\section{Conclusion}\label{sec:end}
We have presented extensive photometric and spectroscopic coverage of
five of the lowest redshift  SL-SNe Ic.  For one of them, \xk\/, we
present a lightcurve from -30d to 200d, showing a well constrained
rise time and a clear detection at 200d indicating a flattening of the
luminosity. In four out of six SL-SNe, we show that there is
significant luminosity in this late tail phase, and we illustrate that
these measurements can aid our understanding of 
the power source and the possible progenitors of these ultra luminous transients.

The five SNe, namely \paj, \xk, \rks, \css\/ and \fo\/ have absolute
magnitudes $-21.73\lesssim M_{g}$(mag) $\lesssim -20.42$ similar to
previous SL-SNe as well as a spectral evolution resembling SN 2010gx
\citep{pa10}.  There is some variation in the sample as two of the 
objects are fainter than the rest and spectroscopically evolve
faster. \paj\/ and \rks\/ have peak absolute magnitudes fainter than
$M_{g}>-21$ mag. The spectra of \rks\/ evolves faster than the rest
and  at 10d post maximum it resembles a normal Type Ic at peak. 
In contrast, SN 2010gx and three other SNe presented here typically 
take 30 days to evolve to this phase. The latest spectrum we have  of \paj\/
shows well developed lines of Fe~{\sc ii} at velocities which are
more comparable
with standard Type Ic SNe, significantly slower than the bulk of the
SL-SNe Ic sample.  Our {\em xshooter} spectrum of  \fo\/
+52d is the only NIR spectrum of an SL-SNe Ic and we 
detect a broad He~{\sc i} $\lambda$10830 in emission
 implying that at least some SL-SNe Ic are not completely
He-free.  During the epochs of $\pm10$ d around peak the 
temperatures and velocities are nearly constant and do not show a
clear decrease until after that period.  We find that the decay
timescale in the 100-200d period is similar to that expected from the
radioactive decay of \co\/, but it requires the $\gamma$-rays to be fully
trapped. This is in contrast to the faster decline observed in most Type
Ib/c SNe, where $\gamma$-ray leakage has a significant effect from 50 days
onwards \citep{so00}. Hence it is unlikely that this is due to large 
amounts (1-4~\M\/) of \ni\/ produced in the explosion.

We applied a semi-analytical diffusion model with energy input
from a spinning down magnetar to fit the lightcurves
of our five objects and SN 2010gx, including the diffusion peak and
the tail phase detections.  All lightcurves, including the tail
phases are reproduced with feasible physical values for a magnetar. 
powered SL-SN. 
We require $3.6\lesssim B_{14}$
$\lesssim7.4$, $1.7\lesssim P_{\rm ms}$$\lesssim7.5$ consistent with
$B$ of known galactic magnetars ($B_{14}\sim1-10$) and with
physically plausible periods ($P_{\rm ms}>1$). We derived energies of
$0.4\lesssim E^{\rm mag}$($10^{51}$ erg) $\lesssim6.9$ and ejected masses of
$2.3\lesssim M_{\rm ej}$(\M)$\lesssim 8.6$.

The well sampled data of the five SL-SNe Ic presented here combined
with those of SN 2010gx point toward a SN explosion driven by a
magnetar as a viable explanation for all the SL-SNe Ic. The
lightcurves are reproduced and the model temperatures and velocities 
are in reasonable agreement with the observational data. 
However even if this is a reliable model, it still leaves other open
questions such as 
\begin{itemize}
\item Do H-rich SNe powered by magnetars exist\footnote{Some studies in order to investigate the existence of H-rich SNe powered by magnetar have been done by \citet{bk11}.} ?  Possibly SN 2008es
  \citep{ge09} could be an example of such a luminous type II SN. In
  other words, why do we observe so many more H-free than H-rich SL-SNe?
\item What is the role of metallicity in the progenitor stars
  evolution that will produce a H-free SL-SNe? It appears that they
  are all associated with faint dwarf galaxies and almost certainly
  low metallicity progenitors. 
\item Do realistic spectral calculations of magnetar driven SN reproduce the main observed features at late time? 
\item Where is the peak of the magnetar SED and how does the magnetar radiation deposit and thermalize in the nebular phase?
\end{itemize}

To address these,  further observations are required combined with
modelling. Theoretical modelling of high quality data in the nebular phase
to determine the ejecta masses, composition and the mass of 
\co\/ contributing to the luminosity seems the most likely way to make
progress. 
 Optical spectroscopy at $\sim$300d post
explosion appears to us to be the next step in probing the nature of
these events. At $z\sim0.1$, the typical AB magnitudes of these
sources are $\sim23\pm0.5$, requiring approximately 1 night of 8m
telescope time to gather a spectrum with high enough signal-to-noise
to measure the expected Ic-like emission lines with confidence. 

\acknowledgments
We thank Dan Kasen for providing us with the data of the simulations of \citet{kb10} in order to compare them with the output of our code.
The research leading to these results has received funding from the
European Research Council under the European Union's Seventh Framework
Programme (FP7/2007-2013)/ERC Grant agreement n$^{\rm o}$ [291222] (PI
: S. J. Smartt).
AP SB and MTB are partially supported by the PRIN-INAF 2011 with the project ``Transient Universe: from ESO Large to PESSTO". 
GL is supported by the Swedish Research Council through grant No. 623-2011-7117.
JPUF acknowledges support from the ERC-StG grant EGGS-278202. The Dark Cosmology Centre is funded by theDNRF.
ST acknowledges support by the TRR 33 ``The Dark universe" of the German Research Foundation.
The Pan-STARRS1 Surveys (PS1) have been made possible through contributions of the Institute for Astronomy, the University of Hawaii, the Pan-STARRS Project Office, the Max-Planck Society and its participating institutes, the Max Planck Institute for Astronomy, Heidelberg and the Max Planck Institute for Extraterrestrial Physics, Garching, The Johns Hopkins University, Durham University, the University of Edinburgh, Queen's University Belfast, the Harvard-Smithsonian Center for Astrophysics, the Las Cumbres Observatory Global Telescope Network Incorporated, the National Central University of Taiwan, the Space Telescope Science Institute, the National Aeronautics and Space Administration under Grant No. NNX08AR22G issued through the Planetary Science Division of the NASA Science Mission Directorate, the National Science Foundation under Grant No. AST-1238877, and the University of Maryland.
We greatly appreciate the work of the support astronomers at the Telescopio Nazionale Galileo, the Copernico Telescope, the 2.2m Telescope at Calar Alto, the Liverpool Telescope, the Nordic Optical Telescope, the New Technology Telescope, the William Herschel Telescope, the Faulkes Telescope North and the Gemini North.
Funding for SDSS-III has been provided by the Alfred P. Sloan Foundation, the Participating Institutions, the National Science Foundation, and the U.S. Department of Energy Office of Science. The SDSS-III web site is http://www.sdss3.org/.\\
{\it Facilities:} PS1, WHT, TNG, Gemini, Ekar, NTT, LT, NOT, Calar Alto Observatory, Faulkes Telescope North.

\clearpage

\appendix
\section{Tables}\label{sec:tab}

\begin{deluxetable*}{lcccccccc}
\tablewidth{0pt}
\tablecaption{Observed (non {\it K}-corrected) photometry of \paj\/ plus associated errors in parentheses. Also reported are the host galaxy magnitudes,
measured after the SN has faded. \label{table:10hgi}}
\tablehead{\colhead{Date}& \colhead{MJD}& \colhead{Phase\tablenotemark{a}} & \colhead{u} & \colhead{g}& \colhead{r}& \colhead{i} & \colhead{z} & \colhead{Telescope}}
\startdata
15/05/10 &  55331.61 &-28.0&    &    &  19.10 ( - )   &    &     &  ATel 2740\\
18/05/10 &  55334.57 &-25.3&    &    &  18.90 (0.07)   &      &    &  PS1\\
26/05/10 &  55342.33 &-18.2&   &    &     &    18.49 (0.07)  &    &  PS1\\
26/05/10 &  55342.34 &-18.2&  &    &     &   18.48 (0.06)   &    &  PS1\\
15/06/10  & 55362.40 &0.0&    & 18.23 (0.05)  &      &     &   	& PS1\\
19/06/10  & 55366.45 &3.7&    &   &  18.04 (0.10)    &     &   	& PS1\\
20/06/10  & 55367.75 &4.9&   &    &   18.00 ( - )   &      &      &  ATel 2740\\
21/06/10 &  55368.53 &5.4&    &    &    & 18.01 (0.05)     &    &  PS1\\
21/06/10 &  55368.55 &5.4&    &    &    &   18.02 (0.05)   &    &  PS1\\
13/07/10 &  55390.81 &25.8& 20.08  (0.12)  &     &     &     &     &  UVOT \\
17/07/10 &  55395.39 &30.0&  &  18.93 (0.05)  & 18.63 (0.08) &   18.56 (0.05)  &  18.58 (0.14) &     LT\\
18/07/10 &  55396.00 &30.5&  20.80  (0.15)  &     &   &        &      &  UVOT\\
20/07/10 &  55398.47  &32.8&  &  19.24 (0.08)  & 18.88 (0.09) &   18.84 (0.08) &   18.91 (0.10) &     LT\\
23/07/10 &  55401.50 &35.5& &  19.44 (0.08)  & 19.11 (0.09)  &  18.91 (0.08)  &  19.10 (0.10)   &   LT\\
27/07/10  & 55405.39 &39.1&   &  19.60 (0.05)  & 19.17 (0.06)  &  19.03 (0.03)  &  19.16 (0.10)   &   LT\\
31/07/10  & 55409.40 &42.7& & 19.74 (0.02) &  19.29 (0.03)  &  19.10 (0.04)  &  19.24 (0.10)  &    LT\\
14/08/10  & 55423.39 &55.4&    &  20.16 (0.03) &  19.59 (0.03)  &  19.44 (0.04)  &  19.30 (0.10)  &    LT\\
17/08/10 &  55426.42  &58.2&  &  20.21 (0.04) &  19.60 (0.05)  &  19.47 (0.05)  &  19.34 (0.11)    &  LT\\
19/08/10  & 55428.87  &60.4& &  20.28 (0.17)  & 19.64 (0.07)  &  19.52 (0.05)  &  19.37 (0.09)   &   FTN\\
20/08/10  & 55429.43  &60.9& &  20.33 (0.13)  & 19.66 (0.06)  &  19.53 (0.06)   & 19.39 (0.16)    &  LT\\
24/08/10   &55433.38 &64.5&  &  20.40 (0.07) &  19.70 (0.05)  &  19.60 (0.07)  &  19.40 (0.05)    &  LT\\
27/08/10  & 55436.40 &67.3&   &  20.48 (0.29)  & 19.77 (0.12)  &  19.62 (0.09)  &  19.44 (0.26)     & LT\\
30/08/10  & 55439.82 &70.4&   & 20.49 (0.08) &  19.83 (0.04)  &  19.64 (0.03) &  19.48 (0.09)    &  FTN\\
01/09/10  & 55441.37 &71.8&    &  20.54 (0.01)  & 19.84 (0.04)  &  19.71 (0.05)  &  19.56 (0.05)    &  LT\\
05/09/10  & 55445.37 &75.4&   &  20.64 (0.04)  & 19.89 (0.04)  &  19.76 (0.06)   & 19.65 (0.07)     & LT\\
07/09/10   &55447.80  &77.6& &  20.67 (0.03) & 19.92 (0.04)  &  19.82 (0.03)   & 19.78 (0.07)    & FTN\\
08/09/10   &55448.38  &78.2&   &  20.70 (0.03)  & 19.97 (0.04)  &  19.87 (0.04)   & 19.83 (0.08)    &  LT\\
13/09/10  & 55453.77 &83.1&   &  20.85 (0.05)  & 20.06 (0.05)   & 19.97 (0.05)  &  20.04 (0.09) &     FTN\\
20/09/10   &55460.36   &89.0&   & 21.01 (0.08)\tablenotemark{b}  & 20.21 (0.09)\tablenotemark{b}   &    &  20.21 (0.13)\tablenotemark{b}  &    LT\\
22/02/11  & 55615.21 &229.8&  &   & &  23.82 (0.31)\tablenotemark{b}  &    &    WHT\\
24/02/11   &55616.23   &231.2&   &   &23.92 (0.30)\tablenotemark{b} &    &    &    WHT\\
28/04/11   &55679.56   &288.3&   &  $>$21.98  & &    &    &    PS1 \\
\\
{\em Host}    \\
26/05/12 &   56073.00  &   646.0      &    &         23.64 (0.23)         &   22.01 (0.09)   &  &      & WHT \\
28/05/12 &   56075.00  &   648.0      &    &                  &           &        22.18 (0.15)        &    21.59 (0.15)  & TNG
\enddata
\tablenotetext{a}{phase with respect to the g-band maximum, corrected for time dilation}
\tablenotetext{b}{magnitudes after image subtraction using 646-648d images}
\tablecomments{PS1 = 1.8 m Pan-STARRS1; UVOT = Swift + UVOT; LT = 2.0 m Liverpool Telescope + RATCam; FTN = 2.0 m Faulkes Telescope North + MEROPE; WHT = 4.2 m William Herschel Telescope + ACAM.}
\end{deluxetable*}

\begin{deluxetable*}{lcccccccccc}
\tablecaption{Observed (non {\it K}-corrected) photometry of \xk\/ and associated errors in parentheses, plus host (see Tab.~\ref{table:10hgi}). \label{table:11xk}}
\tablehead{\colhead{Date}& \colhead{MJD}& \colhead{Phase\tablenotemark{a}} & \colhead{u}& \colhead{B} & \colhead{g}& \colhead{V} &\colhead{r}& \colhead{i} & \colhead{z} & \colhead{Tel.}}
\startdata
16/03/11    &  55637.00  &-43.3&    	&  & & $>$20.10   	 & & & & CSS\\
30/03/11	 &  55649.55 &-32.3& 	&   &  & 	 &$>$21.17   &  &	&PS1\\
30/03/11   &  55650.50 &-31.5&    &  &21.00 ( - ) &   & &   & & ATel\\
31/03/11	 &  55651.58 &-30.5& 	&   &  & 	 &   & 20.07 (0.06) &	&PS1\\
31/03/11	 &  55651.60 &-30.5& 	&   &  & 	 &   & 20.05 (0.06) &	&PS1\\
02/04/11	 &  55653.57 &-28.8& 	&   &  & 	 & 19.40 (0.05)  &  &	&PS1\\
02/04/11	 &  55653.58 &-28.8& 	&   &  & 	 &  19.41 (0.05) &  &	&PS1\\
06/04/11& 55657.99 &-24.9&  	&  &   & 18.61 (0.08)&	 &  & 	& CSS\\
14/04/11 &55665.98  &-17.9&   	&   & & 18.06 (0.07)&	 & &	& CSS\\
15/04/11 &55666.80 &-17.2&      & & 18.07 (0.08) &   &18.00  (0.06)   &  18.21  (0.08) &  &  PS1\\
25/04/11 &55677.00  &-8.3&   	&  &  & 17.65 (0.07)&	   &&	& CSS\\
05/05/11 &55686.50  &0.0&     &  &17.70  ( - ) & &    && 		  &  ATel\\
12/05/11 &55693.98  &6.5&   	&   &   & 17.75 (0.06)	&&  &	&CSS\\
13/05/11 &55694.99 &7.4&     & & 18.08 (0.08) &   &17.91 (0.06)    &  18.15 (0.06)   &  18.20 (0.09) &LT\\
13/05/11 &55695.54  &7.9& 18.34 (0.11)  &   &18.08 (0.04) &  & 17.92 (0.02)    &  18.10 (0.04) 	&18.13 (0.05) 	&FTN\\
15/05/11 &55695.92  &8.2&  &  18.26 (0.09) &   & 17.97 (0.08)  &    &    & 	   &   EKAR\\
14/05/11 &55695.92 &8.2&  &  & 18.15 (0.08) &	 &18.01 (0.10) &	 18.15 (0.09) 	&18.22 (0.12) 	&LT\\
14/05/11 &55695.95 &8.3&  &	& 18.16 (0.08) &	  &18.02 (0.04) &	 18.18 (0.04)&	18.18 (0.09)	    & LT\\
14/05/11 &55696.37 &8.6&  18.46 (0.20)	& &  &	  &&	  &	 	& UVOT\\ 
15/05/11 &55696.93  &9.1& 	& & 18.20 (0.14) &	 &18.01 (0.06) &	 18.28 (0.08) &	18.25 (0.20) 	  & LT\\
16/05/11 &55697.96  &10.0& 18.72 (0.17)  & & 18.17 (0.05) &	 &18.07 (0.06) &	 18.19 (0.03) 	&18.22 (0.04) 	    &LT\\
17/05/11 &55698.36  &10.4& 18.79 (0.24)  & & 18.15 (0.07)  &&	 18.06 (0.06) 	& 18.17 (0.04)& 	18.28 (0.05) 	     & FTN\\
17/05/11 &55698.95  &10.9& 18.95 (0.21)  &  &18.29 (0.05)	&&18.05 (0.03) &	 18.26 (0.04) 	&18.16 (0.07) 	  &    LT\\
18/05/11 &55700.04  &11.8& 	& 18.62 (0.13) & &18.11 (0.05) &	&         &	     & EKAR\\
18/05/11 &55700.05  &11.8& 	&   &	&&18.10 (0.08) 	& 18.25 (0.04) 	& 	 &  LT\\
19/05/11 &55700.90  &12.6& 	&  & &18.14 (0.05)  &	  & &  & EKAR\\
19/05/11 &55700.94  &12.6& 19.08 (0.10)  &  & 18.41 (0.02) &	&18.18 (0.02) &	 18.26 (0.02) &	18.35 (0.03) 	& LT\\
20/05/11 &55702.03 &13.6&  19.23 (0.10)  & & 18.46 (0.02) 	& &18.19 (0.02) &	 18.26 (0.03) &	18.43 (0.04)   &  LT\\
21/05/11 &55702.97  &14.4& &  &18.54 (0.02)  && 18.22 (0.02) &	 18.31 (0.02) &	18.37 (0.03)   &  LT\\
22/05/11 &55703.93  &15.2& 19.31 (0.07)  &  & 18.60 (0.02)  &	& 18.30 (0.03) 	& 18.34 (0.02) &	18.42 (0.03) 	& LT\\
23/05/11 &55704.92  &16.1& & 18.94 (0.04)& &18.35 (0.04) &	  &   &    & EKAR\\
23/05/11 &55704.93  &16.1& 19.33 (0.07)  & & 18.65 (0.02)& 	 &18.33 (0.02) 	& 18.42 (0.02) &	18.45 (0.03) 	& LT\\
27/05/11 &55708.95  &19.6& 	& 19.20 (0.04)  & &18.55 (0.03) &	  &   & 	     & EKAR\\
27/05/11 &55709.40  &20.0&   &	  & &   18.56 (0.07)	&  &  & 	&CSS\\
28/05/11 &55710.29  &20.8& 19.97 (0.03) &   &18.98 (0.04) & &18.51 (0.02) 	& 18.48 (0.03) &	18.56 (0.06)	 & FTN\\
30/05/11 &55710.37  &20.9& 20.08 (0.30)&    &	& &  	&  &	 	  & UVOT \\
31/05/11 &55712.94  &23.1& 20.08 (0.09)  &  &19.13 (0.02)& &18.60 (0.02) 	& 18.61 (0.02) &	18.62 (0.03) 	   &LT\\
31/05/11 &55712.97  &23.2& & 19.45 (0.03)  &  &18.73 (0.02) &	 &     & 	      &EKAR\\
03/06/11 &55715.93  &25.7& 20.25 (0.10) &	 & 19.30 (0.02)& & 18.77 (0.02) 	& 18.71 (0.02) &	18.70 (0.03) 	& LT\\
06/06/11 &55719.37  &28.8& 20.79 (0.41)& &  &	&   	&&    & UVOT \\
07/06/11 &55719.98  &29.3&   & 20.03 (0.09) & & &	&  &    & EKAR\\
09/06/11 &55721.93  &31.0& 21.69 (0.44)	 && 19.79 (0.05)& &19.09 (0.04) &	 19.06 (0.03)	&19.05 (0.06) &LT\\
12/06/11 &55724.50  &33.2&   	& &   & 19.25 (0.11)&	  &   &		   & CSS\\
17/06/11 &55730.00  &38.1&   & 20.55 (0.08) &  &19.61 (0.04) &	  &      &	 &   EKAR\\
24/06/11 &55736.75  &44.0&  22.05 (0.12) & 20.98 (0.06) &  &20.03 (0.04) &	19.81 (0.04)  &   19.76 (0.05)  &	 &   NTT\\
28/06/11 &55740.90  &47.6&& 21.24 (0.08)  && 20.33 (0.08) &	 &      &    &  EKAR\\
02/07/11 &55745.33  &51.5& &  & 21.40 (0.16)\tablenotemark{b}&  &20.42 (0.07)\tablenotemark{b} 	& 20.07 (0.07)\tablenotemark{b} &	19.94 (0.08)\tablenotemark{b}	& FTN\\
09/07/11 &55751.93  &57.2& 	&  & 21.95 (0.16)\tablenotemark{b} & 	 &20.99 (0.30)\tablenotemark{b} 	& 20.84 (0.17)\tablenotemark{b}& 	20.64 (0.25)\tablenotemark{b}&LT\\
13/07/11 &55755.93  &60.7&    & 21.52 (0.24)& &21.00 (0.27) &	   &   & 	  &   EKAR\\
23/07/11 &55765.94  &69.5& 	&  &22.15 (0.06)\tablenotemark{b} &	 &21.26 (0.06)\tablenotemark{b} 	& 21.02 (0.07)\tablenotemark{b} &	20.86 (0.14)\tablenotemark{b} &LT\\
24/07/11 &55766.89  &70.3& &   &22.18 (0.08)\tablenotemark{b} & &21.28 (0.07)\tablenotemark{b} 	& 21.06 (0.07)\tablenotemark{b} &	20.88 (0.14)\tablenotemark{b}& LT\\
25/07/11 &55767.90  &71.2& 	&  &22.20 (0.06)\tablenotemark{b} &	& 21.32 (0.06)\tablenotemark{b} 	& 21.08 (0.07)\tablenotemark{b} &	20.91 (0.12)\tablenotemark{b}& LT\\
03/08/11 &55776.89 &79.1& 	&& 22.40 (0.11)\tablenotemark{b} &	 &21.48 (0.10)\tablenotemark{b} 	& 21.24 (0.08)\tablenotemark{b} &	21.07 (0.15)\tablenotemark{b} &LT \\
21/12/11 &55917.74 &202.3 &	& & 23.95 (0.30)\tablenotemark{b} &	& 23.14 (0.24)\tablenotemark{b} & 22.50 (0.20)\tablenotemark{b}&	22.30 (0.23)\tablenotemark{b}& WHT  \\
\\
{\em Host} \\
26/05/12 & 56073.50 & 338.6 &                      &           &  21.18 (0.05) &	& 20.72 (0.04) &  20.47 (0.04) & 20.90 (0.10) & WHT\\
12/06/04             &   53169.22           &           & 21.54 (0.24) & 	 & 21.20 (0.08) & & 	 20.71 (0.08) & 	 20.59 (0.11) & 	 21.30 (0.67) & SDSS DR9
\enddata
\tablenotetext{a}{phase with respect to the g-band maximum, corrected for time dilation}
\tablenotetext{b}{magnitudes after image subtraction using 339d images}
\tablecomments{CSS = 0.7 m Catalina Schmidt Telescope; PS1 = 1.8 m Pan-STARRS1; LT = 2.0 m Liverpool Telescope + RATCam; FTN = 2.0 m Faulkes Telescope North + MEROPE; UVOT = Swift + UVOT; EKAR = 1.8 m Copernico Telescope + AFOSC; WHT = 4.2 m William Herschel Telescope + ACAM, NTT = 3.6m New Technology Telescope + EFOSC2, ATel = ATel 3344}
\end{deluxetable*}

\begin{deluxetable*}{lcccccccc}
\tablewidth{0pt}
\tablecaption{Observed (non {\it K}-corrected) photometry of \rks\/ and associated errors in parentheses, plus host (see Tab.~\ref{table:10hgi}).\label{table:11rks}}
\tablehead{\colhead{Date}& \colhead{MJD}& \colhead{Phase\tablenotemark{a}} & \colhead{u} & \colhead{g}& \colhead{r}& \colhead{i} & \colhead{z} & \colhead{Telescope}}
\startdata
21/12/11 &  55916.70    &	-13.4 &        &     &   19.9 (-)    &    &      &   ATel 3841\\
27/12/11  & 55922.70   & 	  -8.4  &   &        & 19.1 (-)     &       &    &  ATel 3841\\
30/12/11 &  55925.24  &  	 -6.3   & 20.49 (0.12)  &   &            &   &   & UVOT\\
01/01/12  & 55927.56    &	 -4.3  &  20.30 (0.11) &      &        &     &    &   UVOT\\      
02/01/12  & 55929.42	&  -2.8   &  &  19.26 (0.09)  &  19.03 (0.20)  &    &       &   LT \\
03/01/12  & 55930.42	& -1.9  &   &	  19.19 (0.06) 	&  18.94 (0.04) 	&   19.14 (0.04)   &  19.25 (0.27) &     LT \\
04/01/12  & 55931.34	 &  -1.1  &  &19.18 (0.04)  &  18.88 (0.03)   &  19.05 (0.08)   &  19.19 (0.24)   &   LT\\
05/01/12 &  55931.85    &	-0.7 &     20.46 (0.10)  &   &         &      &      &   UVOT\\
05/01/12  & 55932.72	& 0.0 &    &	  19.13 (0.05) 	&  18.87 (0.05) 	&   19.01 (0.03)    & 19.14 (0.21)   &   FTN \\
08/01/12 &  55935.73   & 2.5 &      	& 19.22 (0.04) &	  18.96 (0.02) 	  & 18.99 (0.04)    & 19.06 (0.06)    &  FTN\\
09/01/12  & 55936.40&	3.1 &      & 19.23 (0.03) &	  18.97 (0.03) 	&   19.00 (0.04)     &19.06 (0.08)    &  LT\\
10/01/12  & 55936.71   & 3.4	&     20.55 (0.10)  &   &    &      &   &   UVOT\\
12/01/12  & 55939.32  & 5.6	 &     & 19.30 (0.02) 	&  19.03 (0.06) 	&   18.97 (0.04)     &19.03 (0.05)   &   LT\\
14/01/12  & 55941.32&7.2 &	      & 19.37 (0.04) &	  19.04 (0.03) 	&   18.98 (0.05)  &   19.07 (0.28)   &   LT \\
15/01/12  & 55941.51   &7.4 	&     20.93 (0.16) &     &      &   &         &   UVOT\\
15/01/12  & 55942.35 &8.1	&       &  19.38 (0.02) 	&  19.04 (0.03) 	&   18.98 (0.04)    & 19.09 (0.09) &     LT\\
21/01/12 &  55948.75  &	13.5 &     	&  19.65 (0.04) 	 & 19.19 (0.02) 	 &  19.08 (0.04)&     19.10 (0.07) &     FTN\\
24/01/12  & 55951.73  &16.0	&      	&  19.79 (0.04) 	&  19.28 (0.03) 	  & 19.17 (0.05)	&    19.16 (0.06)  &    FTN \\
27/01/12  & 55954.44  &18.3	&      	&  19.93 (0.05) 	&  19.41 (0.06) 	&   19.22 (0.08)  &   19.17 (0.14)   &   LT \\
29/01/12  & 55956.84  &20.3	 &     	&   &	  19.46 (0.21) 	 &     &      &   FTN \\
04/02/12  & 55962.75  & 25.2   &	      	&  20.58 (0.14) &	  19.62 (0.04) 	  & 19.44 (0.03)    & 19.35 (0.03)  &    FTN \\
08/02/12  & 55966.74  &	28.6&      	& 20.90 (0.20)\tablenotemark{b} &	  20.04 (0.08)\tablenotemark{b} 	&   19.84 (0.13)\tablenotemark{b}   &  19.71 (0.16)\tablenotemark{b}  &    FTN \\
03/03/12  & 55990.34  &48.4	 &     	&  	&  20.82 (0.15)\tablenotemark{b} 	&   20.55 (0.44)\tablenotemark{b}  &     &    LT\\
26/06/12  & 56105.26  & 145.3	 &     	&  $>$22.9  	&  $>$22.4  	&   $>$22.1   &    &    WHT\\
\\
{\em Host}\\
22/09/12 &  56192.50 & 218.3      &             &   21.67 (0.07)                 &    20.83 (0.05)                &                     &                       & WHT \\
16/10/09  & 55121.47 &        &   21.51 (0.32) &	21.59 (0.11) &	 20.88 (0.10) & 20.71 (0.14) &	 20.03  (0.29) & SDSS DR9
\enddata
\tablenotetext{a}{phase with respect to the g-band maximum, corrected for time dilation}
\tablenotetext{b}{magnitudes after image subtraction using 218d images or SDSS iz images}
\tablecomments{Swift $u$-band data have been converted to SDSS magnitudes. UVOT = Swift +UVOT; LT = 2.0 m Liverpool Telescope +RATCam; FTN = 2.0 m Faulkes Telescope North +MEROPE; WHT = 4.2 m William Herschel Telescope +ACAM; GS = 8.1 m Gemini South +GMOS.}
\end{deluxetable*}

\begin{deluxetable*}{lcccccc}
\tablewidth{0pt}
\tablecaption{Observed (non {\it K}-corrected) NIR photometry of \rks\/ plus associated errors in parentheses.\label{table:11rksN}}
\tablehead{\colhead{Date}& \colhead{MJD}& \colhead{Phase\tablenotemark{a}} & \colhead{J} & \colhead{H}& \colhead{K}& \colhead{Telescope}}
\startdata
07/02/12 &  55965.84 &29.1 &	    &   	 &   18.86   (0.10)   &   NOT \\
08/02/12 &  55966.88  &	30.0 & 19.31 (0.10)  &    &    &   NOT\\
09/02/12  & 55967.83 &30.9 &	 19.32 (0.10)  &    &	     &   NOT \\
10/02/12  & 55968.84 &31.8 &	 19.42  (0.10)   &   19.41  (0.09)	  &    &   NOT
\enddata
\tablenotetext{a}{phase with respect to the g-band maximum, corrected for time dilation}
\tablecomments{NOT = 2.56 m Nordic Optical Telescope + NOTCam.}
\end{deluxetable*}

\begin{deluxetable*}{lccccccc}
\tablewidth{0pt}
\tablecaption{Observed (non {\it K}-corrected) photometry of \css\/ and associated errors in parentheses, plus host (see Tab.~\ref{table:10hgi}).\label{table:css}}
\tablehead{\colhead{Date}& \colhead{MJD}& \colhead{Phase\tablenotemark{a}} & \colhead{g}& \colhead{r}& \colhead{i} & \colhead{z} & \colhead{Telescope}}
\startdata
30/12/11 &  55925.54 &0.0&  18.60 (0.08)  & &    &    &   ATel 3873\\
25/01/12 & 55951.54  &20.9& 19.15 (0.08)  &  &     &  	 &     LT\\
21/02/12 & 55979.74  &43.5& 20.91 (0.08)  & 19.79 (0.06)  &    19.55 (0.05) &   19.40 (0.10) &     LT\\
22/02/12 & 55980.71 &44.3 & 20.91 (0.11)& 	19.83 (0.05)	&  19.52 (0.07) &	  19.31 (0.10)	 &     LT\\ 
23/02/12 & 55981.62  &45.0& 20.93 (0.05)	&  &    &     &   LT \\
24/02/12 & 55982.69  & 45.9&20.96 (0.11) &	19.86 (0.06) &	  19.61 (0.06) 	 & 19.41 (0.06)	&      LT\\
08/03/12 & 55995.71   & 56.4& &20.24 (0.08) 	 & 19.87 (0.10) 	&  19.64 (0.21)	 &     LT \\
10/03/12 & 55997.68 &57.9 & 21.40 (0.19) &	20.32 (0.11) 	&  19.95 (0.09) &	  19.71 (0.19)	  &    LT\\
12/03/12 & 55999.69 &59.6 & 21.44 (0.19) &	20.39 (0.10) 	&  19.97 (0.08) &	  19.78 (0.12)	   &   LT\\
13/03/13 & 56000.70 &60.4 & 21.45 (0.10) &	20.39 (0.05) 	&  19.99 (0.11) &	  19.81 (0.19)	&      LT\\
18/03/12 & 56005.67  &64.4 &21.54 (0.09) &	20.51 (0.06) &	  20.13 (0.09) 	&  19.97 (0.10)      &  LT \\
23/03/12 & 56010.65  &68.4& 21.69 (0.16) &	20.80 (0.10) &	  20.40 (0.15) 	&  20.25 (0.19)&	      LT\\
27/03/12 & 56014.66  &71.6 &21.95 (0.11) &	21.03 (0.12) &	  20.55 (0.18) 	&  20.52 (0.20)&	      LT\\
24/05/12 & 56073.15  &118.9 &23.14 (0.21)\tablenotemark{b}& 	22.07 (0.14)\tablenotemark{b} &	  21.88 (0.13)\tablenotemark{b} 	&  21.81 (0.15)\tablenotemark{b}&	      WHT \\
27/06/12 & 56106.04  &145.4 &$>$23.61&	$>$22.7&	 $>$22.2&  $>$22.0&	      WHT\\
\\
{\em Host} \\
27/06/12 & 56106.04  &145.4 &&	&	 &  $>$22.0&	      WHT\\
20/07/12  &  56128.94 &    163.8   & 24.24 (0.34)    & 23.94 (0.30)  & 22.56 (0.33) &   &  WHT\\
06/05/09  & 54987.25 &     	      & $>$22.2 &	 $>$22.2 & $>$21.3 &$>$20.5 & SDSS DR9
\enddata
\tablenotetext{a}{phase with respect to the g-band maximum, corrected for time dilation}
\tablenotetext{b}{magnitudes after image subtraction using 163d images and SDSS griz images}
\tablecomments{LT = 2.0 m Liverpool Telescope +RATCam; WHT = 4.2 m William Herschel Telescope +ACAM.}
\end{deluxetable*}

\begin{deluxetable*}{lccccccc}
\tablewidth{0pt}
\tablecaption{Observed (non {\it K}-corrected) photometry of \fo\/ and associated errors  in parentheses, plus host (see Tab.~\ref{table:10hgi}).\label{table:12fo}}
\tablehead{\colhead{Date}& \colhead{MJD}& \colhead{Phase\tablenotemark{a}}  & \colhead{g}& \colhead{r}& \colhead{i} & \colhead{z} & \colhead{Telescope}}
\startdata
31/12/11&	55927.56    &-11.8&   &    18.75 (0.09)  &        &    &   CSS\\
05/01/12&	55932.45    &-7.6&      &   &     &  18.99 (0.03) &    PS1\\
05/01/12&	55932.49   &-7.6&         &      &      &  18.96 (0.03)  &   PS1\\
14/01/12&	55941.43    &0.0&        &   17.98 (0.06)      &  &    &  CSS\\
19/01/12  &      55946.44 &4.3&       18.34 (0.01)   &  &	       &       &     PS1 \\    
21/01/12&	55948.52    &6.0&       &   18.11 (0.07)     &     &  &    CSS\\
04/02/12&	55961.54    &17.1&    	&      &   18.94 (0.02) &  	&    PS1\\
05/02/12&	55961.55    &17.1&       &    	 &    18.99 (0.02)   &	&  PS1\\
11/02/12&	55969.56    &23.9&      & 18.96 (0.10)     &    &  &   CSS\\
20/02/12&	55978.05    &31.2&    19.83 (0.03)   & 19.32 (0.09)  &   19.30 (0.07)  &  19.45 (0.12)   &  FTN\\
22/02/12&	55980.56     &33.3&         & 19.35 (0.10)    &    & &  CSS\\
23/02/12&	55981.02    &33.7&    20.00 (0.05) &   19.42 (0.04) &    &  &    FTN\\
29/02/12&	55987.48     &39.2&      & 19.74 (0.14)     &    & &    CSS\\
14/03/12&	56000.89     &50.6&   20.97 (0.07)  &  20.16 (0.07)   &  20.04 (0.08)  &  19.85 (0.09) &    FTN\\
19/03/12&	56005.86    &54.8&    21.11 (0.06)   & 20.25 (0.07)   &     &    &     FTN\\
20/03/12&56006.90      &55.7&  21.11 (0.07)  &  20.26 (0.05)&     20.16 (0.05)  &  20.06 (0.14)&     FTN\\
23/04/12&	56009.89      &58.3&  21.19\tablenotemark{b} (0.08)  &  20.35 (0.03)    &  &      &     FTN\\
24/03/12&	56010.88      &59.1&  21.20\tablenotemark{b} (0.09)   & 20.39 (0.08)   &  20.28 (0.12) &   20.18 (0.09)    & FTN\\
29/03/12&	56015.86     &63.3&   21.29\tablenotemark{b} (0.11)   & 20.47 (0.07)  &   20.40 (0.08)   & 20.30 (0.01)  &   FTN\\
26/05/12&	56074.02&113.2&            & 21.23 (0.20)\tablenotemark{b}  &   21.00 (0.22)\tablenotemark{b}   & 20.87 (0.20)\tablenotemark{b}   &  WHT\\
\\
{\em Host}\\
 10/02/13     & 56333.04 & 327.2  & 	 22.14  (0.17) &            21.57  (0.06) & 	 21.51 	 (0.08)& 	 21.54  (0.35)           & WHT\\
 03/02/05     & 53405.24 &   & 	 22.13  (0.08) &            21.46  (0.07) & 	 21.64 	 (0.11)& 	 21.50  (0.35)           & SDSS DR9
\enddata
\tablenotetext{a}{phase with respect to the g-band maximum, corrected for time dilation}
\tablenotetext{b}{magnitudes after image subtraction using 327d images}
\tablecomments{CSS = 0.7 m Catalina Schmidt Telescope; PS1 = 1.8 m Pan-STARRS1;  FTN = 2.0 m Faulkes Telescope North + MEROPE; WHT = 4.2 m William Herschel Telescope + ACAM.}
\end{deluxetable*}

\begin{deluxetable*}{lccccc}
\tablewidth{0pt}
\tablecaption{Observed (non {\it K}-corrected) UVOT ultraviolet AB magnitudes of SL-SNe Ic plus associated errors  in parentheses.\label{table:swift}}
\tablehead{\colhead{Date}& \colhead{MJD}& \colhead{Phase\tablenotemark{a}} & \colhead{uvw2}& \colhead{uvm2}& \colhead{uvw1}}
\startdata
\cutinhead{\paj\/}
13/07/10 & 55390.31 &25.8  &  21.93 (0.24)&  22.05 (0.27) & 20.97 (0.22) \\
18/07/10 & 55395.50   &30.5&  22.38 (0.43) & 22.40	 (0.44) & 21.77 (0.29)\\
\cutinhead{\xk\/}
14/05/11 & 55695.87  &8.6&   20.67 (0.12) & 20.42 (0.13) & 18.86 (0.12) \\
30/05/11 & 55711.92   &22.6&  21.90 (0.21) & 21.71 (0.17) & 21.09 (0.21) \\
06/06/11 & 55718.87   &28.7 & 22.01 (0.37) & 21.63 (0.27) & \\
07/06/11  &55719.73  & 29.6&  22.27 (0.41) & 22.14 (0.38)  &21.53 (0.27) \\
08/06/11 & 55720.87 & 30.5 &  22.43 (0.48) & 22.40 (0.49)  &21.67 (0.43) \\
08/03/12 & 55994.99 & 270.3 & $>$23.2 & $>$23.1 & $>$23.3 \\
\cutinhead{\rks\/}
30/12/11   &55925.24  & -6.2&   20.92 (0.12)   &21.12 (0.13)  & 20.42 (0.12) \\
01/01/12   &55927.06    & -4.6& 21.22 (0.14)  & 21.52 (0.15)   &20.45 (0.14)\\
05/01/12   &55931.35    &-0.8& 21.47 (0.14)  & 21.59 (0.15)  & 21.17 (0.18) \\
10/01/12  & 55936.21    & 3.6 & 21.54 (0.15)  & 21.63 (0.20)  & 21.28 (0.20) \\
15/01/12  & 55941.01    &7.8 & 22.00 (0.21)  & 21.68 (0.26)  & 21.42 (0.24)  \\
\cutinhead{\fo\/}
13/02/12 & 55971.71 & 25.3 && 22.14 (0.19) & 21.33 (0.38)\\
14/02/12 & 55971.07 & 25.6 &22.48 (0.26)& & 21.60 (0.26)
\enddata
\tablenotetext{a}{phases with respect to the g-band maxima, corrected for time dilation}
\end{deluxetable*}

\clearpage
\section{Sequence stars for \paj\/}\label{sec:sspaj}
\begin{figure}
\center
\includegraphics[width=8cm]{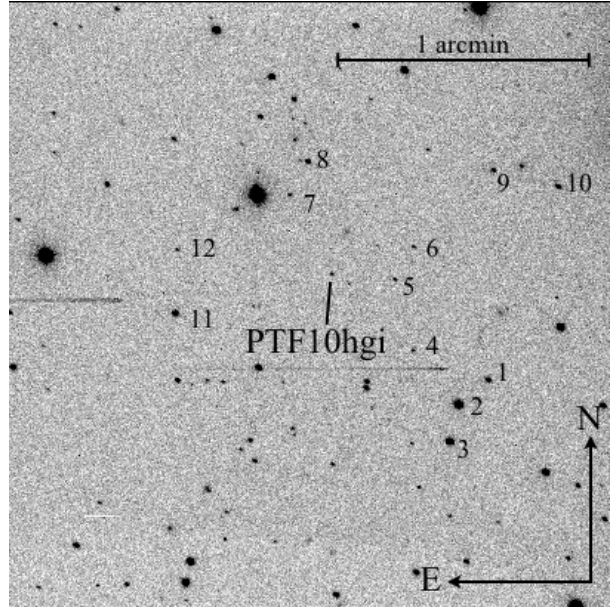}
\caption{{\it r}-band image of \paj\/ obtained with LT + RATCam on September 5th, 2010. The sequence of stars used to calibrate the optical magnitudes is indicated by numbers.} 
\label{fig:paj}
\end{figure}

Here we report the average magnitudes of the local sequence stars of \paj\/ used to calibrate the photometric zero points for non-photometric nights. They are reported in Tab.~\ref{table:ss10hgi} along with their r.m.s. (in brackets). Their positions are marked in Fig.\ref{fig:paj}, along with the SN position.

\begin{deluxetable}{cccccc}
\tablewidth{0pt}
\tablecaption{Magnitudes of the local sequence stars in the field of \paj\/ (cfr. Fig.~\ref{fig:paj}).\label{table:ss10hgi}}
\tablehead{\colhead{ID}& \colhead{g}& \colhead{r} & \colhead{i} & \colhead{z}}
\startdata
1 & 19.52 (0.05)& 18.04 (0.03) & 17.30 (0.03) & 17.00 (0.04)\\
2 & 16.80 (0.01)& 15.85 (0.02) & 15.40 (0.02) & 15.23 (0.02)\\
3 &17.44 (0.02)& 16.51 (0.02) & 16.12 (0.02) & 16.00 (0.02)\\
4& 21.27 (0.08)&19.99 (0.05) & 19.44  (0.05) & 19.13 (0.05)\\
5& 20.16 (0.05)& 19.35 (0.05) & 19.06  (0.04) &19.02 (0.05)\\
6& 20.60 (0.06)&19.97 (0.05) & 19.70  (0.05) & 19.56 (0.07)\\
7& 21.34 (0.08)& 19.89 (0.05) & 18.70 (0.04) & 18.24 (0.05)\\
8& 19.29 (0.04)&18.17 (0.04) & 17.66 (0.03) & 17.50 (0.05)\\
9& 19.64 (0.05)& 18.81 (0.04) & 18.37 (0.03) & 18.30 (0.05)\\
10& 19.26 (0.04)& 18.18 (0.04) & 17.76 (0.03) & 17.60 (0.05)\\
11& 17.94 (0.03)& 17.42 (0.03) & 17.24 (0.03)  & 17.20 (0.04)\\
12& 20.21 (0.06)& 19.76 (0.05) & 19.59  (0.05) & 19.61 (0.06)
\enddata
\end{deluxetable}

\clearpage 
\section{Spectral evolution}\label{sec:spbin}

\begin{figure*}[!hb]
\includegraphics[width=17cm]{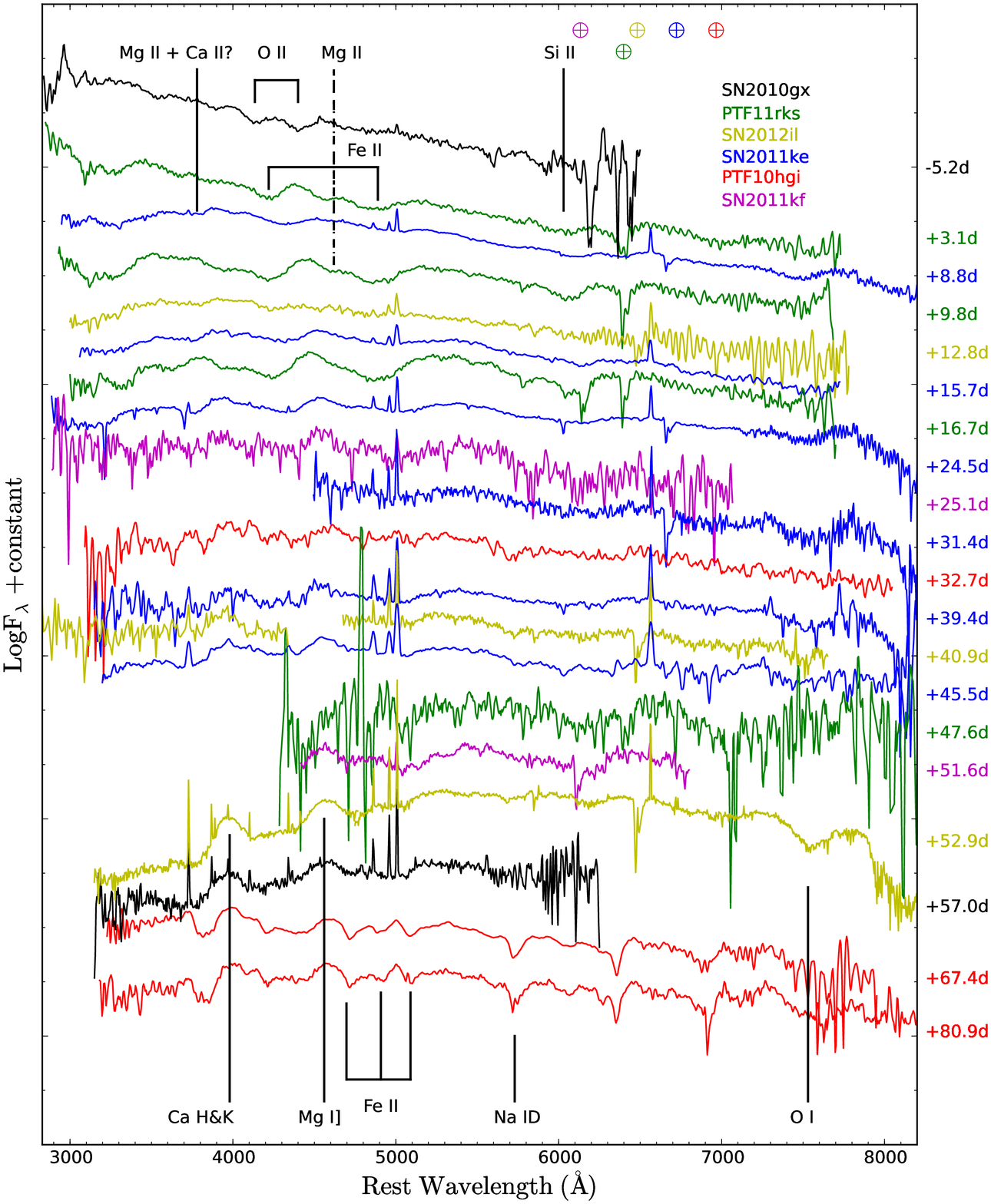}
\caption{Same figure as is Fig.~\ref{fig:spevol} but with all the spectra convolved with a factor of five and subsequently binned to a 5~\AA\/ scale. Spectra of \rks\/ are in green, \xk\/ in blue, \fo\/ in gold, \paj\/ in red, \css\/ in magenta and SN 2010gx \citep{pa10} in black. The phase of each spectrum relative to light curve peak in the rest frame is shown on the right. The spectra are corrected for extinction and reported in their rest frame. The most prominent features are labelled.} 
\label{fig:spevolb}
\end{figure*}

\clearpage

\section{Magnetar-powered light-curves}\label{sec:mag}
\citet{ar82} derived the solution for the bolometric light-curve of a homologously expanding ejecta subject to a total (absorbed) power $P(t)$ as
\begin{equation}
L_{\rm SN}{ (t)} = { e^{ -\left(t/\tau_{\rm m}\right)^2} \int_0^{ t/\tau_{\rm m}} P{ (t')} 2 \left(t'/\tau_{\rm m}\right) e^{\left(t'/\tau_{\rm m}\right)^2} \frac{dt'}{\tau_{\rm m}}~\mbox{erg s}^{-1}},
\end{equation}
where $\tau_{\rm m}$ is the diffusion time-scale parameter, which in the case of uniform density ($E_{\rm k} = \frac{3}{10} M_{\rm ej} V_{\rm ej}^2$) is
\begin{equation}
\tau_{\rm m} = \frac{1.05}{\left(\beta c\right)^{1/2}}\kappa^{1/2} M_{\rm ej}^{3/4} E_{\rm k}^{-1/4}~\mbox{s}~.
\label{eq:taum}
\end{equation}
The parameter $\beta$ has a typical value of 13.7 \citep{ar82}, which we use throughout. Apart from spherical symmetry and homology, the Arnett solution assumes a radiation pressure dominated gas, energy transport by diffusion, constant and grey opacity, and that the spatial distribution of energy input by the power source is proportional to the radiation energy density. The last assumption is for most cases the coarsest one, but the solutions are not critically dependent on deviations from it \citep{ar82}.

The power function $P(t)$ can generally be expressed as a sum of a set of source luminosities $L_i(t)$ multiplied by deposition functions $D_i(t)$:
\begin{equation}
P(t) = \sum L_{\rm i}(t)D_{\rm i}(t)~\mbox{erg s}^{-1}.
\end{equation}

For radioactivity from $^{56}$Ni and its daughter nucleus $^{56}$Co, the luminosity functions are
\begin{eqnarray}
L_{\rm 56Ni}(t) &=& 7.8\cdot 10^{43} \left(\frac{M_{\rm 56Ni}}{1~M_\odot}\right) e^{-t/\tau_{\rm 56Ni}}~\mbox{erg s}^{-1}, \\
L_{\rm 56Co}(t) &=& 1.4\cdot 10^{43}  \left(\frac{M_{\rm 56Ni}}{1~M_\odot}\right) \frac{e^{-t/\tau_{\rm 56Co}}- e^{-t/\tau_{\rm 56Ni}}}{1-\frac{\tau_{\rm 56Ni}}{\tau_{\rm 56Co}}}~\mbox{erg s}^{-1},  
\end{eqnarray}

where $M_{\rm 56Ni}$ is the amount of $^{56}$Ni formed in the explosion, and $\tau_{\rm 56Ni}$ and $\tau_{\rm 56Co}$ are the decay times of the isotopes (8.7 and 111 days, respectively).

For a magnetar with a 45-degree angle between the magnetic axis and spin axis, the dipole spin-down luminosity is \citep{og71,kb10}
\begin{equation}\label{eq:magnetar}
L_{\rm magnetar}(t) = 4.9\cdot 10^{46} B_{14}^2 P_{\rm ms}^{-4}\frac{1}{\left(1 + t/\tau_p\right)^2}~\mbox{erg s}^{-1},
\end{equation}
where $B_{14}$ is the magnetic field strength in $10^{14}$ G, $P_{\rm ms}$ is the initial spin period in ms, and the spin-down time-scale $\tau_{\rm p}$ is
\begin{equation}
\tau_p = 4.7~ B_{\rm 14}^{-2} P_{\rm ms}^2~\mbox{days}~.
\end{equation}

With only magnetar powering, the light-curve parameters are thus $\tau_m$, $B_{14}$ and $P_{\rm ms}$. For a given fit to $\tau_{\rm m}$, the ejecta mass is (from Eq. \ref{eq:taum})
\begin{equation}\label{eq:apa}
M_{ej} = 1~M_\odot \cdot \left(\frac{\tau_{\rm m}}{10~{\rm d}}\right)^{4/3}\left(\frac{\kappa}{0.1~\mbox{cm}^2 \mbox{g}^{-1}}\right)^{-2/3} \left(\frac{E_{\rm k}}{10^{51}~\mbox{erg}}\right)^{1/3}
\end{equation} 

The weak scaling with $\kappa$ and $E_k$ means that ejecta masses can be meaningfully estimated despite significant uncertainties in $\kappa$ and $E_{\rm k}$.

When the energy input $\int P(t) dt$ approaches or exceeds the explosion energy, accelerations will not be negligible and the homologous approximation becomes poor. A solar mass of $^{56}$Ni emits a total energy of $\int P(t) dt = 1.9\cdot 10^{50}$ erg (including the \co\/ decay), so for an explosion energy of $\sim 10^{51}$ erg, accelerations will not be important unless M(\ni\/)$ \gtrsim$ 5 $M_\odot$. A magnetar, on the other hand, has a rotation energy of $2\cdot 10^{52} P_{\rm ms}^{-2}$ erg, so for fast braking with efficient trapping, the ejecta can be accelerated to far beyond its explosive velocities. This is a major limitation of using the homologous expansion approximation, and the consequences for the light curve can only be investigated with numerical radiation hydrodynamics \citep[see][and reference therein for some examples]{bu09}. 

\subsection{The opacity}\label{sec:opa}
The opacity due to electron scattering is
\begin{equation}
\kappa_{\rm es} = 0.2 \left(\frac{\bar{Z}/\bar{A}}{0.5}\right)\left(\frac{x_{\rm e}}{\bar{Z}}\right)~\mbox{cm}^2~\mbox{g}^{-1}~,
\end{equation}                                                                                                                               where $\bar{Z}$ is the mean atomic charge, $\bar{A}$ is the mean atomic mass, and $x_{\rm e} = n_{\rm e}/n_{\rm nuclei}$ is the ionization fraction. For any chemical mixture excluding hydrogen, $\bar{Z}/\bar{A} \approx 0.5$, which we can assume since there is no trace of hydrogen lines in the spectra of these objects.                                                                                                                             
The radiation temperature in a radiation-pressure dominated uniform sphere of internal energy $E(t)$ is
\begin{equation}
T(t) = \left(\frac{E(t)}{a \frac{4\pi}{3} V_{\rm ej}^3 t^3}\right)^{1/4} \sim 10^5 K \left(\frac{E(t)}{10^{51}~\mbox{erg}}\right)^{-1/8}\left(\frac{M}{1~M_\odot}\right)^{3/8} \left(\frac{t}{10~{\rm d}}\right)^{-3/4}
\end{equation}

In LTE, and at relevant densities here, the ionization fraction of a pure oxygen plasma at $T=10^5$ K is $x_{\rm e} \sim 6$, for carbon it is $x_{\rm e} \sim 4$, and for Fe is $x_{\rm e} \sim 10$. It should thus be a reasonable approximation to use $x_{\rm e}/\bar{Z} = 0.5$. With this choice $\kappa_{\rm es} = 0.1$ cm$^2$ g$^{-1}$, comparable to the value $\kappa = 0.08$ that \citet{ar82} uses.

In rapidly expanding media, line opacity can become comparable to or exceed electron scattering opacity \citep{Karp1977}, which complicates analytical light curve modelling. A common simplistic approach to take line opacity into account is to put a minimum value on the opacity. In previous works, values ranging between 0.01 and 0.24 cm$^2$ g$^{-1}$ have been used \citep[e.g.,][]{Herzig1990, Swartz1991, Young2004, Bersten2011, Bersten2012}. 
The implication of line opacity is that $\kappa$ is likely to stay in the 0.01 - 0.2 cm$^{2}$ g$^{-1}$range throughout the evolution, and $\kappa = 0.1$ cm$^2$ g$^{-1}$ is probably not off by more than a factor of a few at any given time.
                                                                                                                                                        
\subsection{Deposition of magnetar radiation}\label{sec:dep}

The spectral and directional distribution of the magnetar radiation is highly uncertain. At early times, most radiation will undoubtedly be absorbed, but at later times, this is not certain \citep{ko13}. X-rays are efficiently absorbed for a long time by photoionization from inner shell electrons, but gamma rays see a smaller opacity and start escaping after a few weeks for typical ejecta parameters.

Here, we assume full trapping of the magnetar radiation. Most of the energy input occurs over a few weeks after explosion, and it is a reasonable assumption that most of the energy emerges as X-rays, as in the Crab pulsar for instance \citep{crab}.

\subsection{Photospheric velocities and temperatures}
To estimate the photospheric velocities and temperatures for a given solution for ejecta mass, we approximate the ejecta density structure with a core + envelope morphology. The velocity of the homogenous core (which contains almost all the mass to force consistency with the diffusion calculation) is determined from
\begin{equation}\label{eq:vel}
\frac{3}{10}M_{\rm ej} V_{\rm core}^2 = E_{\rm k}
\end{equation}
and its density is given by
\begin{equation}
\rho_{\rm core}(t) = \frac{M_{\rm ej}}{\frac{4\pi}{3} V_{\rm core}^3 t^3}~.
\end{equation}
Outside we attach an envelope with density profile
\begin{equation}
\rho(t,V) = \rho_{\rm core}(t)\left(\frac{V}{V_{\rm core}}\right)^{-\alpha}
\end{equation}

The optical depth of the envelope is 
\begin{eqnarray}
\tau_{\rm env} &=& \int_{R_{\rm core}}^\infty \rho_{\rm core}\left(\frac{r}{R_{\rm core}}\right)^{-\alpha} \kappa dr\\
   &=& \frac{\rho_{\rm core} R_{\rm core} \kappa}{\alpha - 1} \\
   &=& \frac{\tau_{\rm core}}{\alpha-1}
\end{eqnarray}
 
where 
\begin{equation}
\tau_{\rm core}(t) = \kappa \rho_{\rm core}(t)V_{\rm core} t~.
\end{equation}

One should in general distinguish between the photosphere and the thermalization layer, where the temperature of the continuum is determined. The position of these depend on the total opacity $\kappa$, and the absorption opacity $\kappa_{\rm a}$, respectively. However, due to our simple treatment such detail is unwarranted, and we approximate them to be the same, $R_*$. As long as the atmosphere is optically thick ($\tau_{\rm env} > 1$), $R_*(t)$ is found from solving

\begin{equation}
\int_{R_*(t)}^\infty \kappa \rho(r,t) dr = 1~,
\end{equation}

which gives

\begin{equation}
R_*(t) = R_{\rm core}(t)\left(\frac{\alpha-1}{\tau_{\rm core}(t)}\right)^{\frac{1}{1-\alpha}}~.
\end{equation}

If the photosphere has receded into the core, the expression is instead
\begin{equation}
R_*(t) = R_{\rm core}(t) - \frac{1 - \frac{\tau_{\rm core}}{\alpha-1}}{\kappa \rho_{\rm core}}
\end{equation}

Typical density profiles in Ib/c models show $\alpha \sim 10$ \citep[e.g. Fig. 1 in][]{kb10}, which is the value we use here.\\

From the photospheric radius $R_*$({\it t}) we retrieved the velocity at the photosphere ($V_{\rm phot}$)

\begin{equation}
V_{\rm phot}(t)=V_{\rm core}\frac{R_*(t)}{R_{\rm core}(t)}.
\end{equation}

The effective temperature is found from application of the black-body formula, using the model luminosities at corresponding times.

\subsection{Additional tests}\label{sec:test}
\begin{figure}
\includegraphics[width=1\linewidth]{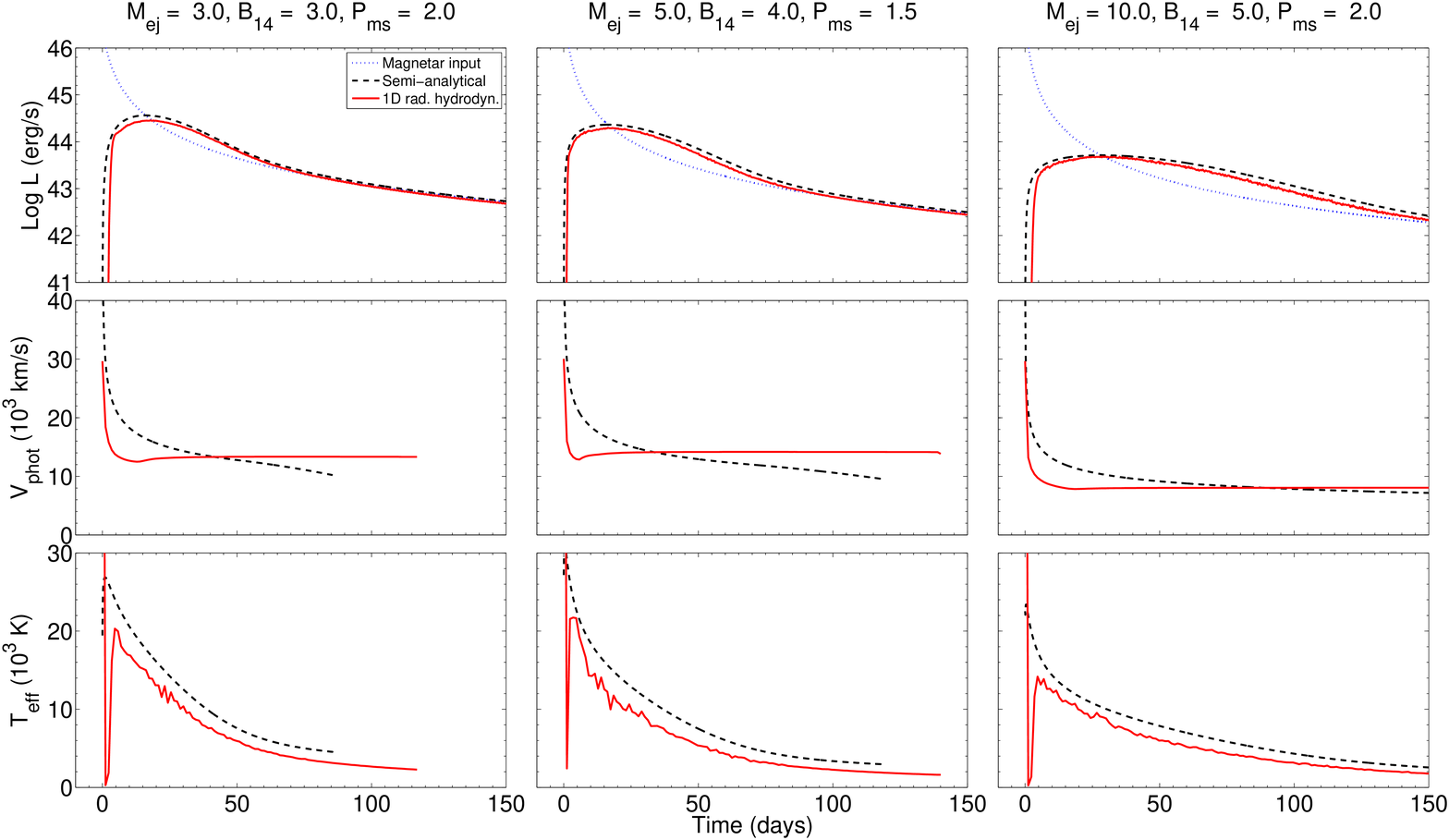}
\caption{Comparison of light curves, photospheric velocities, and photospheric temperatures from our semi-analytical model (black, dashed) and the 1D radiation hydrodynamical simulations of \citet{kb10} (red, solid). Also shown is the input magnetar energy (blue, dotted line).}
\label{fig:emag}
\end{figure}

To test the applicability of our approach, we perform three tests. The first one is to compare light-curves, photospheric velocities and temperatures to the radiation hydrodynamical simulations of \citet{kb10}. Fig.~\ref{fig:emag} shows three representative examples. For these comparisons, we use eq.~\ref{eq:apa2} to relate magnetar energy to ejecta kinetic energy, which shows itself a good approximation.
Tab.~\ref{table:comp} shows the derived parameters of $M_{\rm ej}$, $B_{14}$, $P_{\rm ms}$ and $\Delta t$ for a series of simulations in \citet{kb10}. The magnetar properties ($B_{14}$ and $P_{\rm ms}$) are reliably recovered to within 10-20\% of their input values, whereas the ejecta mass can vary, tending to be too low in the fits. It is, however, always within a factor 2 of the correct value.

\begin{deluxetable}{lccc||c}
\tablewidth{0pt}
\tablecaption{Best-fitting values of system parameters (after the slash) compared to the actual simulation values of \citet{kb10} (before the slash). Derived parameters on the righ.\label{table:comp}}
\tablehead{\colhead{$M_{\rm ej}$}& \colhead{$B_{14}$ }& \colhead{$P_{\rm ms}$} & \colhead{Shift (days)} & \colhead{$E^{\rm mag}$}}
\startdata
3.0/3.0 &  2.0/2.2 & 2.0/2.3 & $-1.57/0$ & 3.9 \\
3.0/3.1 &  3.0/3.2 & 2.0/2.3 & 0.17/0 & 4.0 \\
3.0/2.8 &  5.0/5.4 & 2.0/2.4 & $-0.038/0$ & 3.6 \\
\tableline
5.0/3.1 &  3.0/3.2 & 2.0/2.3 &   0.18/0   & 3.9 \\
5.0/3.0 &  4.0/4.3 & 2.0/2.4 & 7e$-4/0$ & 3.7 \\
5.0/3.0 &  5.0/5.4 & 2.0/2.4 & 0.18/0 & 3.7 \\
\tableline
7.0/4.1 &  2.0/2.1 & 1.5/1.8 & $-0.039/0$ & 6.6 \\
7.0/3.8 &  3.0/3.2 & 1.5/1.9 & $-4.9$e$-4$/0& 6.1 \\
\tableline
10/12.8  &  2.0/2.7 & 2.0/1.6 & $-5.0$/0 & 8.0 \\
10/9.2  &  3.0/3.5 & 2.0/2.2 & $-3.6$/0 & 4.3 \\
10/7.4  &  5.0/5.1 & 2.0/2.9 & 1.8e$-4$/0 & 2.4
\enddata
\end{deluxetable}

\begin{figure}[htb]
\centering
\includegraphics[width=8cm]{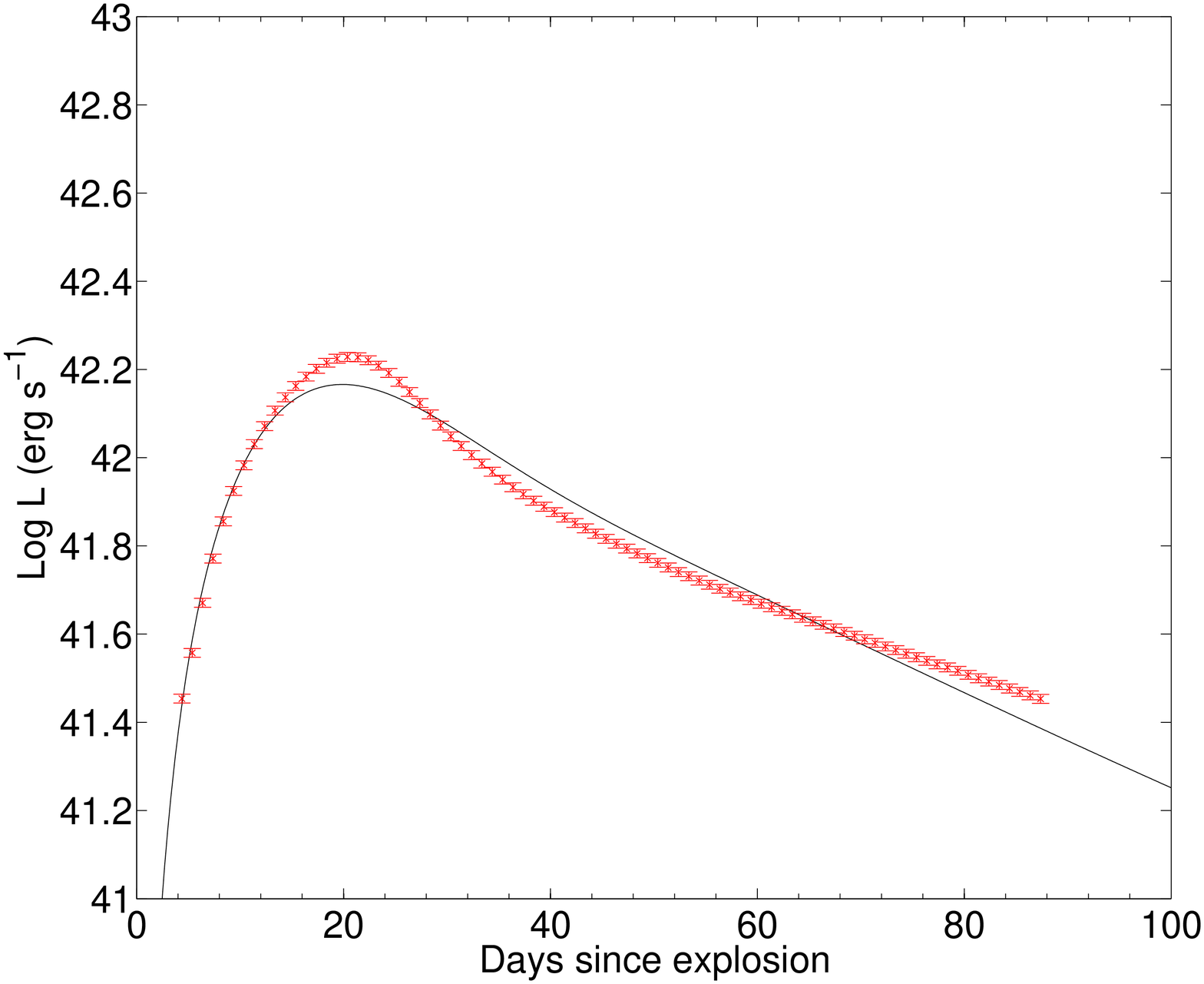}
\caption{Fit to the bolometric light curve of SN 2011dh (Ergon+, in prep). The fit values are $M_{\rm ej} = 2.2~M_\odot$  and $M_{\rm 56Ni} = 0.08~M_\odot$, for $E_{\rm k} = 10^{51}$ erg and $\kappa = 0.1$ cm$^{2}$ g$^{-1}$.}
\label{fig:2011dh}
\end{figure}

The second test is to derive parameters for the Type IIb SN 2011dh, using the approach described above (fixing $E_{\rm k}= 10^{51}$ erg, $\kappa = 0.1$, $\alpha = 10$). The best-fit is shown in Fig. \ref{fig:2011dh}. The recovered values $M_{\rm ej} = 2.2$ \M\/ and $M_{\rm 56Ni} = 0.08~M_\odot$ agree well with the results from \citet{Bersten2011}, who used radiation hydrodynamical simulations.

\begin{figure}[]
\centering
\includegraphics[width=10cm]{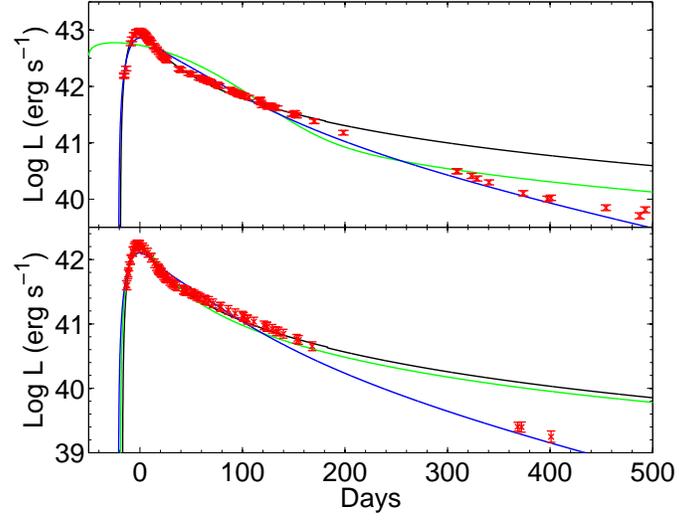}
\caption{Fit to the bolometric light curve of SNe 1998bw (top) and 2007gr (bottom). Magnetar fit of the first 200d is shown (black) along with the fit of the entire set (green) and the fit of radioactive decay (blue). Both models can fit the first 50-100d but are clearly differentiated at late times}
\label{fig:98bw}
\end{figure}

While for the third one we tried to test our code on Type Ic SNe 1998bw  \citep{ga98,mck99,so00,pat01} and 2007gr \citep{va08,hu09} to explore its limits.
We fitted the first 200d with the magnetar model (fixing $E_{\rm k}= 10^{51}$ erg, $\kappa = 0.1$, $\alpha = 10$), recovering $M_{\rm ej} = 1.8$ \M\/ and $B_{14} = 9.8$ G and $P_{\rm ms}=18.9$ ms for SN 1998bw and $M_{\rm ej} = 1.3$ \M\/ and $B_{14} = 22.9$ G and $P_{\rm ms}=47.6$ ms for SN 2007gr.
Instead, trying to fit the entire dataset, we recover $M_{\rm ej} = 19.9$ \M\/ and $B_{14} = 16.6$ G and $P_{\rm ms}=0.5$ ms for SN 1998bw and $M_{\rm ej} = 2.5$ \M\/ and $B_{14} = 26.0$ G and $P_{\rm ms}=42.6$ ms for SN 2007gr. The best fit to those data are shown in Fig.~\ref{fig:98bw} along with the fit of the radioactive decay. The applicability of the Arnett model to derive parameters for radioactivity-driven SNe has also been demonstrated by \citet{va11}.
The results imply that at those energies the magnetar contribution is more important after 150-200 d, highlighting the importance of obtaining late time data.

\subsection{\ni\/ fits parameters}\label{sec:nifit}

For the \ni\/ models we computed $\gamma$-ray trapping as in \citet{ar82}.
\begin{deluxetable*}{lcccccc}
\tablewidth{0pt}
\tablecaption{Best-fit parameters for \ni\/ modelling of the bolometric light curves and $\chi^2_{\rm red}$. \label{table:nifit}}
\tablehead{\colhead{Object}& \colhead{Shift}& \colhead{M(\ni)} & \colhead{$M_{\rm ej}$} & \colhead{$E$} &\colhead{$t_{\rm 0}$}& \colhead{$\chi^2_{\rm red}$} \\
\colhead{}& \colhead{(day)}& \colhead{(\M)} & \colhead{(\M)}& \colhead{($10^{51}$ erg)} & \colhead{(MJD)} & \colhead{} }
\startdata
\xk\/ 			& 36.0 & 3.0  & 6.1& 8.1  &  55650.65 & 35 \\
\css\/      		& 38.0 & 5.0 & 11& 15.0 &  55920.65 & 58  \\
\fo\/     		& 26.0 & 2.9 & 5.9 & 6.2   &  55918.56 & 193\\
\paj\/    		& 37.0 & 2.6 & 5.9 & 6.2   &  55322.78 & 1.2\\
\rks\/    		& 31.0 & 3.0 & 6.1  & 8.1   &  55912.11 & 53\\
SN 2010gx   	& 50.0 & 6.0 & 13.0 & 15.0  &  55269.22 & 860
\enddata
\end{deluxetable*}


\begin{thebibliography}{}
\bibitem[Abazajian et al.(2009)]{ab09} Abazajian, K.~N., 
Adelman-McCarthy, J.~K., Ag{\"u}eros, M.~A., et al.\ 2009, \apjs, 182, 543
\bibitem[Agnoletto et al.(2009)]{ag09} Agnoletto, I., 
Benetti, S., Cappellaro, E., et al.\ 2009, \apj, 691, 1348 
\bibitem[Alard(2000)]{al} Alard, C.\ 2000, \aaps, 144, 363
\bibitem[Ahn et al.(2012)]{ah12} Ahn, C.~P., Alexandroff, 
R., Allende Prieto, C., et al.\ 2012, \apjs, 203, 21
\bibitem[Arnett(1982)]{ar82} Arnett, W.~D.\ 1982, \apj, 253, 
785
\bibitem[Barbary et al.(2009)]{ba09} Barbary, K., Dawson, 
K.~S., Tokita, K., et al.\ 2009, \apj, 690, 1358
\bibitem[Barkov 
\& Komissarov(2011)]{bk11} Barkov, M.~V., \& Komissarov, S.~S.\ 2011, \mnras, 415, 944
\bibitem[Baron et al.(1996)]{ba96} Baron, E., Hauschildt, 
P.~H., Branch, D., Kirshner, R.~P., 
\& Filippenko, A.~V.\ 1996, \mnras, 279, 799
\bibitem[Berger et al.(2012)]{2012ApJ...755L..29B} Berger, E., Chornock, 
R., Lunnan, R., et al.\ 2012, \apjl, 755, L29
\bibitem[Bersten et al.(2012)]{Bersten2012} Bersten, M.~C., 
Benvenuto, O.~G., Nomoto, K., et al.\ 2012, \apj, 757, 31 
\bibitem[Bersten et al.(2011)]{Bersten2011} Bersten, M.~C., 
Benvenuto, O., \& Hamuy, M.\ 2011, \apj, 729, 61
\bibitem[Botticella et al.(2010)]{bot10} Botticella, M.~T., 
Trundle, C., Pastorello, A., et al.\ 2010, \apjl, 717, L52
\bibitem[Bouchet et 
al.(1989)]{bo89} Bouchet, P., Slezak, E., Le Bertre, T., Moneti, A., \& Manfroid, J.\ 1989, \aaps, 80, 379
\bibitem[Blinnikov 
\& Sorokina(2010)]{bl10} Blinnikov, S.~I., \& Sorokina, E.~I.\ 2010, arXiv:1009.4353
\bibitem[Bucciantini et al.(2009)]{bu09} Bucciantini, N., 
Quataert, E., Metzger, B.~D., et al.\ 2009, \mnras, 396, 2038
\bibitem[Cappellaro et 
al.(1997)]{ca97} Cappellaro, E., Mazzali, P.~A., Benetti, S., et al.\ 1997, \aap, 328, 203
\bibitem[Chandrasekhar 
\& Fermi(1953)]{ch53} Chandrasekhar, S., \& Fermi, E.\ 1953, \apj, 118, 116  
\bibitem[Chatzopoulos 
\& Wheeler(2012)]{cw12} Chatzopoulos, E., \& Wheeler, J.~C.\ 2012, \apj, 760, 154
\bibitem[Chatzopoulos et al.(2012)]{cw12a} Chatzopoulos, E., 
Wheeler, J.~C., \& Vinko, J.\ 2012, \apj, 746, 121 
\bibitem[Chatzopoulos et al.(2009)]{cw09} Chatzopoulos, E., 
Wheeler, J.~C., \& Vinko, J.\ 2009, \apj, 704, 1251
\bibitem[Chen et al.(2013)]{che12} Chen, T.-W., Smartt, 
S.~J., Bresolin, F., et al.\ 2013, \apjl, 763, L28
\bibitem[Chevalier 
\& Irwin(2011)]{ci11} Chevalier, R.~A., \& Irwin, C.~M.\ 2011, \apjl, 729, L6
\bibitem[Chomiuk et al.(2012)]{ch12} Chomiuk, L., Soderberg, 
A., Margutti, R., et al.\ 2012, The Astronomer's Telegram, 3931, 1
\bibitem[Chomiuk et al.(2011)]{ch11} Chomiuk, L., Chornock, 
R., Soderberg, A.~M., et al.\ 2011, \apj, 743, 114 
\bibitem[Dessart et al.(2012)]{2012MNRAS.426L..76D} Dessart, L., Hillier, 
D.~J., Waldman, R., Livne, E., \& Blondin, S.\ 2012, \mnras, 426, L76 
\bibitem[Dexter 
\& Kasen(2012)]{de12} Dexter, J., \& Kasen, D.\ 2012, arXiv:1210.7240
\bibitem[Drake et al.(2012)]{dr12} Drake, A.~J., Djorgovski, 
S.~G., Mahabal, A.~A., et al.\ 2012, The Astronomer's Telegram, 3873, 1
\bibitem[Drake et al.(2011)]{dr11} Drake, A.~J., Djorgovski, 
S.~G., Mahabal, A.~A., et al.\ 2011, The Astronomer's Telegram, 3343, 1
\bibitem[Drake et al.(2009)]{dr09} Drake, A.~J., Djorgovski, 
S.~G., Mahabal, A., et al.\ 2009, \apj, 696, 870
\bibitem[Drout et al.(2011)]{dro11} Drout, M.~R., Soderberg, 
A.~M., Gal-Yam, A., et al.\ 2011, \apj, 741, 97
\bibitem[Duncan 
\& Thompson(1992)]{du92} Duncan, R.~C., \& Thompson, C.\ 1992, \apjl, 392, L9
\bibitem[Eldridge et al.(2013)]{el13} Eldridge, J.~J., 
Fraser, M., Smartt, S.~J., Maund, J.~R., 
\& Crockett, R.~M.\ 2013, arXiv:1301.1975
\bibitem[Ensman 
\& Woosley(1988)]{Ensman1988} Ensman, L.~M., \& Woosley, S.~E.\ 1988, \apj, 333, 754 
\bibitem[Galama et al.(1998)]{ga98} Galama, T.~J., 
Vreeswijk, P.~M., van Paradijs, J., et al.\ 1998, \nat, 395, 670
\bibitem[Gal-Yam(2012)]{2012Sci...337..927G} Gal-Yam, A.\ 2012, Science, 
337, 927 
\bibitem[Gal-Yam et al.(2009)]{gy09} Gal-Yam, A., Mazzali, 
P., Ofek, E.~O., et al.\ 2009, \nat, 462, 624
\bibitem[Gezari et al.(2012)]{2012Natur.485..217G} Gezari, S., Chornock, 
R., Rest, A., et al.\ 2012, \nat, 485, 217 
\bibitem[Gezari et al.(2010)]{ge10} Gezari, S., Rest, A., 
Huber, M.~E., et al.\ 2010, \apjl, 720, L77
\bibitem[Gezari et al.(2009)]{ge09} Gezari, S., Halpern, 
J.~P., Grupe, D., et al.\ 2009, \apj, 690, 1313
\bibitem[Ginzburg 
\& Balberg(2012)]{gb12} Ginzburg, S., \& Balberg, S.\ 2012, \apj, 757, 178
\bibitem[Hadjiyska et al.(2012)]{ha12} Hadjiyska, E., 
Rabinowitz, D., Baltay, C., et al.\ 2012, IAU Symposium, 285, 324
\bibitem[Hamuy(2003)]{h03a} Hamuy, M.\ 2003, \apj, 582, 905
\bibitem[Hamuy et al.(1988)]{ha88} Hamuy, M., Suntzeff, 
N.~B., Gonzalez, R., \& Martin, G.\ 1988, \aj, 95, 63
\bibitem[Herzig et 
al.(1990)]{Herzig1990} Herzig, K., El Eid, M.~F., Fricke, K.~J., \& Langer, N.\ 1990, \aap, 233, 462 
\bibitem[Hodapp et al.(2004)]{PS1_optics} Hodapp, K.~W., Siegmund, 
W.~A., Kaiser, N., Chambers, K.~C., Laux, U., Morgan, J., 
\& Mannery, E.\ 2004, \procspie, 5489, 667 
\bibitem[Hunter et 
al.(2009)]{hu09} Hunter, D.~J., Valenti, S., Kotak, R., et al.\ 2009, \aap, 508, 371
\bibitem[Kaiser et al.(2010)]{PS1_system} Kaiser, N., et al. \ 2010, \procspie, 7733,  12K.
\bibitem[Karp et al.(1977)]{Karp1977} Karp, A.~H., Lasher, G., 
Chan, K.~L., \& Salpeter, E.~E.\ 1977, \apj, 214, 161
\bibitem[Kasen 
\& Bildsten(2010)]{kb10} Kasen, D., \& Bildsten, L.\ 2010, \apj, 717, 245
\bibitem[Kotera et al.(2013)]{ko13} Kotera, K., Phinney, 
E.~S., \& Olinto, A.~V.\ 2013, arXiv:1304.5326
\bibitem[Law et al.(2009)]{la09} Law, N.~M., Kulkarni, 
S.~R., Dekany, R.~G., et al.\ 2009, \pasp, 121, 1395
\bibitem[Leloudas et al.(2012)]{le12} Leloudas, G., Chatzopoulos, E., Dilday, B., et al.\ 2012, \aap, 541, A129 
\bibitem[Magnier et al.(2013)]{ps1_photladder} Magnier E.A., et
  al. 2013, ApJ, in press
  \bibitem[Magnier et al.(2008)]{ma08} Magnier, E.~A., Liu, 
M., Monet, D.~G., \& Chambers, K.~C.\ 2008, IAU Symposium, 248, 553 
\bibitem[Magnier(2007)]{ma07} Magnier, E.\ 2007, The Future 
of Photometric, Spectrophotometric and Polarimetric Standardization, 364, 
153 
\bibitem[Mahabal 
\& Drake(2010)]{ma10} Mahabal, A.~A., \& Drake, A.~J.\ 2010, The Astronomer's Telegram, 2508, 1
\bibitem[Margutti et al.(2012)]{ma12} Margutti, R., 
Soderberg, A., Chomiuk, L., et al.\ 2012, The Astronomer's Telegram, 3925, 
1
\bibitem[McKenzie 
\& Schaefer(1999)]{mck99} McKenzie, E.~H., \& Schaefer, B.~E.\ 1999, \pasp, 111, 964
\bibitem[Meynet et 
al.(2011)]{me11} Meynet, G., Eggenberger, P., \& Maeder, A.\ 2011, \aap, 525, L11
\bibitem[Moriya 
\& Tominaga(2012)]{mo12} Moriya, T.~J., \& Tominaga, N.\ 2012, \apj, 747, 118
\bibitem[Neill et al.(2011)]{2011ApJ...727...15N} Neill, J.~D., Sullivan, 
M., Gal-Yam, A., et al.\ 2011, \apj, 727, 15 
\bibitem[Ofek et al.(2007)]{of07} Ofek, E.~O., Cameron, 
P.~B., Kasliwal, M.~M., et al.\ 2007, \apjl, 659, L13
\bibitem[Ostriker 
\& Gunn(1971)]{og71} Ostriker, J.~P., \& Gunn, J.~E.\ 1971, \apjl, 164, L95
\bibitem[Pastorello et al.(2010)]{pa10} Pastorello, A., 
Smartt, S.~J., Botticella, M.~T., et al.\ 2010, \apjl, 724, L16 
\bibitem[Patat et al.(2001)]{pat01} Patat, F., Cappellaro, 
E., Danziger, J., et al.\ 2001, \apj, 555, 900
\bibitem[Poole et al.(2008)]{poo08} Poole, T.~S., Breeveld, 
A.~A., Page, M.~J., et al.\ 2008, \mnras, 383, 627 
\bibitem[Prieto et al.(2012)]{pr12} Prieto, J.~L., Drake, 
A.~J., Mahabal, A.~A., et al.\ 2012, The Astronomer's Telegram, 3883, 1
\bibitem[Quimby et al.(2013)]{qu13} Quimby, R.~M., Yuan, F., 
Akerlof, C., \& Wheeler, J.~C.\ 2013, \mnras, 431, 912
\bibitem[Quimby et al.(2011a)]{qu11} Quimby, R.~M., Kulkarni, 
S.~R., Kasliwal, M.~M., et al.\ 2011a, \nat, 474, 487
\bibitem[Quimby et al.(2011b)]{qu11b} Quimby, R.~M., Gal-Yam, 
A., Arcavi, I., et al.\ 2011b, The Astronomer's Telegram, 3841, 1 
\bibitem[Quimby et al.(2011c)]{qu11c} Quimby, R.~M., 
Sternberg, A., \& Matheson, T.\ 2011c, The Astronomer's Telegram, 3344, 1 
\bibitem[Quimby et al.(2010)]{qu10} Quimby, R.~M., Kulkarni, 
S., Ofek, E., et al.\ 2010, The Astronomer's Telegram, 2740, 1
\bibitem[Quimby et al.(2007)]{qu07} Quimby, R.~M., Aldering, 
G., Wheeler, J.~C., et al.\ 2007, \apjl, 668, L99
\bibitem[Quimby et al.(2005)]{qu05} Quimby, R.~M., Castro, 
F., Gerardy, C.~L., et al.\ 2005, Bulletin of the American Astronomical 
Society, 37, \#171.02
\bibitem[Rau et al.(2009)]{rau09} Rau, A., Kulkarni, S.~R., 
Law, N.~M., et al.\ 2009, \pasp, 121, 1334 
\bibitem[Schlegel et al.(1998)]{sc98} Schlegel, D.~J., 
Finkbeiner, D.~P., \& Davis, M.\ 1998, \apj, 500, 525
\bibitem[Shigeyama et al.(1990)]{Shigeyama1990_2} Shigeyama, T., 
Nomoto, K., Tsujimoto, T., \& Hashimoto, M.-A.\ 1990, \apjl, 361, L23
\bibitem[Shigeyama 
\& Nomoto(1990)]{Shigeyama1990} Shigeyama, T., \& Nomoto, K.\ 1990, \apj, 360, 242 
\bibitem[Smartt et al.(2012)]{sma12} Smartt, S.~J., Wright, 
D., Valenti, S., et al.\ 2012, The Astronomer's Telegram, 3918, 1 
\bibitem[Smartt et al.(2011)]{sma11} Smartt, S.~J., Valenti, 
S., Magill, L., et al.\ 2011, The Astronomer's Telegram, 3351, 1
\bibitem[Smith et al.(2007)]{sm07} Smith, N., Li, W., Foley, 
R.~J., et al.\ 2007, \apj, 666, 1116
\bibitem[Smith 
\& McCray(2007)]{smc07} Smith, N., \& McCray, R.\ 2007, \apjl, 671, L17 
\bibitem[Sollerman et al.(2000)]{so00} Sollerman, J., Kozma, 
C., Fransson, C., et al.\ 2000, \apjl, 537, L127
\bibitem[Swartz et al.(1991)]{Swartz1991} Swartz, D.~A., Wheeler, 
J.~C., \& Harkness, R.~P.\ 1991, \apj, 374, 266 
\bibitem[Taubenberger et al.(2006)]{ta06} Taubenberger, S., 
Pastorello, A., Mazzali, P.~A., et al.\ 2006, \mnras, 371, 1459
\bibitem[Thompson et al.(2004)]{th04} Thompson, T.~A., 
Chang, P., \& Quataert, E.\ 2004, \apj, 611, 380
\bibitem[Tonry \& Onaka(2009)]{PS1_GPCA} Tonry, J., \& Onaka, P.\ 2009, Advanced Maui Optical and Space Surveillance Technologies Conference, 
Proceedings of the Advanced Maui Optical and Space Surveillance Technologies Conference, Ed.: S. Ryan, p.E40.   
\bibitem[Tonry et al.(2012)]{2012ApJ...745...42T} Tonry, J.~L., Stubbs, 
C.~W., Kilic, M., et al.\ 2012, \apj, 745, 42 
\bibitem[Tonry et al.(2012)]{2012ApJ...750...99T} Tonry, J.~L., Stubbs, 
C.~W., Lykke, K.~R., et al.\ 2012, \apj, 750, 99 
\bibitem[Umeda 
\& Nomoto(2008)]{um08} Umeda, H., \& Nomoto, K.\ 2008, \apj, 673, 1014 
\bibitem[Usov(1992)]{us92} Usov, V.~V.\ 1992, \nat, 357, 472 
\bibitem[Valenti et al.(2008)]{va08} Valenti, S., 
Elias-Rosa, N., Taubenberger, S., et al.\ 2008, \apjl, 673, L155
\bibitem[Valenti et al.(2010)]{2010ATel.2773....1V} Valenti, S., Smartt, 
S., Young, D., et al.\ 2010, The Astronomer's Telegram, 2773, 1 
\bibitem[Valenti et al.(2011)]{va11} Valenti, S., Fraser, 
M., Benetti, S., et al.\ 2011, \mnras, 416, 3138
\bibitem[Weisskopf et al.(2000)]{crab} Weisskopf, M.~C., 
Hester, J.~J., Tennant, A.~F., et al.\ 2000, \apjl, 536, L81
\bibitem[Wheeler et al.(2000)]{wh00} Wheeler, J.~C., Yi, I., 
H{\"o}flich, P., \& Wang, L.\ 2000, \apj, 537, 810
\bibitem[Woods 
\& Thompson(2006)]{wt06} Woods, P.~M., \& Thompson, C.\ 2006, Compact stellar X-ray sources, 547
\bibitem[Woosley(2010)]{wo10} Woosley, S.~E.\ 2010, \apjl, 
719, L204
\bibitem[Woosley et al.(2007)]{wo07} Woosley, S.~E., 
Blinnikov, S., \& Heger, A.\ 2007, \nat, 450, 390
\bibitem[Woosley et al.(1989)]{wo89} Woosley, S.~E., 
Hartmann, D., \& Pinto, P.~A.\ 1989, \apj, 346, 395
\bibitem[York et al.(2000)]{yo00} York, D.~G., Adelman, J., 
Anderson, J.~E., Jr., et al.\ 2000, \aj, 120, 1579
\bibitem[Young et 
al.(2010)]{yo10} Young, D.~R., Smartt, S.~J., Valenti, S., et al.\ 2010, \aap, 512, A70
\bibitem[Young(2004)]{Young2004} Young, T.~R.\ 2004, \apj, 617, 
1233

\end{thebibliography}
\end{document}